\documentclass[twocolumn,superscriptaddress,aps,floatfix,nofootinbib]{revtex4}
\usepackage{graphicx}
\usepackage{dcolumn}
\usepackage{amsmath}
\input epsf

\begin{document}

\title{\textbf{Statistical theory of the continuous double auction}}

\author{Eric Smith\footnote{Corresponding author: desmith@santafe.edu}}
\affiliation{Santa Fe Institute, 1399 Hyde Park Rd., Santa Fe NM 87501}

\author{J. Doyne Farmer\footnote{McKinsey Professor}}
\affiliation{Santa Fe Institute, 1399 Hyde Park Rd., Santa Fe NM 87501}

\author{L\'{a}szl\'{o} Gillemot}
\affiliation{Santa Fe Institute, 1399 Hyde Park Rd., Santa Fe NM 87501}

\author{Supriya Krishnamurthy}
\affiliation{Santa Fe Institute, 1399 Hyde Park Rd., Santa Fe NM 87501}

\date{\today}

\begin{abstract}

Most modern financial markets use a continuous double auction
mechanism to store and match orders and facilitate trading.  In this
paper we develop a microscopic dynamical statistical model for the
continuous double auction under the assumption of IID random order
flow, and analyze it using simulation, dimensional analysis, and
theoretical tools based on mean field approximations.  The model makes
testable predictions for basic properties of markets, such as price
volatility, the depth of stored supply and demand vs. price, the
bid-ask spread, the price impact function, and the time and
probability of filling orders.  These predictions are based on
properties of order flow and the limit order book, such as share
volume of market and limit orders, cancellations, typical order size,
and tick size. Because these quantities can all be measured directly
there are no free parameters.  We show that the order size, which can
be cast as a nondimensional granularity parameter, is in most cases a
more significant determinant of market behavior than tick size.  We
also provide an explanation for the observed highly concave nature of
the price impact function. On a broader level,
this work suggests how stochastic models based on
zero-intelligence agents may be useful to probe the structure of
market institutions.  Like the model of perfect rationality, a
stochastic-zero intelligence model can be used to make strong
predictions based on a compact set of assumptions, even if these
assumptions are not fully believable.

\end{abstract}
\maketitle

\tableofcontents

\section{Introduction}

This section provides background and motivation, a description of the
model, and some historical context for work in this area.
Section~\ref{overview} gives an overview of the phenomenology of the
model, explaining how dimensional analysis applies in this context,
and presenting a summary of numerical results. Section~\ref{analysis}
develops an analytic treatment of model, explaining some of the
numerical findings of Section~\ref{overview}.  We conclude in Section
\ref{summa} with a discussion of how the model may be enhanced to
bring it closer to real-life markets, and some comments comparing the
approach taken here to standard models based on information arrival
and valuation.

\subsection{Motivation}
\label{motivation}

In this paper we analyze the continuous double auction trading
mechanism under the assumption of random order flow, developing a
model introduced in \cite{Daniels01}.  This analysis produces
quantitative predictions about the most basic properties of markets,
such as volatility, depth of stored supply and demand, the bid-ask
spread, the price impact, and probability and time to fill.  These
predictions are based on the rate at which orders flow into the
market, and other parameters of the market, such as order size and
tick size.  The predictions are falsifiable with no free parameters.
This extends the original random walk model of Bachelier
\cite{Bachelier00} by providing a basis for the diffusion rate of
prices.  The model also provides a possible explanation for the highly
concave nature of the price impact function. Even though some of the
assumptions of the model are too simple to be literally true, the
model provides a foundation onto which more realistic assumptions may
easily be added.

The model demonstrates the importance of financial institutions in
setting prices, and how solving a necessary economic function such as
providing liquidity can have unanticipated side-effects.  In a world
of imperfect rationality and imperfect information, the task of demand
storage necessarily causes persistence.  Under perfect rationality all
traders would instantly update their orders with the arrival of each
piece of new information, but this is clearly not true for real
markets.  The limit order book, which is the queue used for storing
unexecuted orders, has long memory when there are persistent orders.
It can be regarded as a device for storing supply and demand, somewhat
like a capacitor is a device for storing charge.  We show that even
under completely random IID order flow, the price process displays
anomalous diffusion and interesting temporal structure.  The converse
is also interesting: For prices to be effectively random, incoming
order flow must be non-random, in just the right way to compensate for
the persistence.  (See the remarks in Section~\ref{valuation}.)

This work is also of interest from a fundamental point of view because
it suggests an alternative approach to doing economics.  The
assumption of perfect rationality has been popular in economics
because it provides a parsimonious model that makes strong
predictions.  In the spirit of Gode and Sunder \cite{Gode93}, we show
that the opposite extreme of zero intelligence random behavior
provides another reference model that also makes very strong
predictions.  Like perfect rationality, zero intelligence is an
extreme simplification that is obviously not literally true.  But as
we show here, it provides a useful tool for probing the behavior of
financial institutions.  The resulting model may easily be extended by
introducing simple boundedly rational behaviors.  We also differ from
standard treatments in that we do not attempt to understand the
properties of prices from fundamental assumptions about utility.
Rather, we split the problem in two.  We attempt to understand how
prices depend on order flow rates, leaving the problem of what
determines these order flow rates for the future.

One of our main results concerns the average price impact function.
The liquidity for executing a market order can be characterized by a
price impact function $\Delta p=\phi(\omega,\tau,t)$.  $\Delta p$ is
the shift in the logarithm of the price at time $t+\tau$ caused by a
market order of size $\omega$ placed at time $t$.  Understanding price
impact is important for practical reasons such as minimizing
transaction costs, and also because it is closely related to an excess
demand function\footnote{In financial models it is common to define an excess demand
function as demand minus supply; when the context is clear the
modifier ``excess'' is dropped, so that demand refers to both supply
and demand.}, providing a natural starting point for theories of
statistical or dynamical properties of markets \cite{Farmer98,
Bouchaud98}.  A naive argument predicts that the price impact $\phi
(\omega)$ should increase at least linearly.  This argument goes as
follows: Fractional price changes should not depend on the scale of
price. Suppose buying a single share raises the price by a factor
$k>1$. If $k$ is constant, buying $\omega$ shares in succession should
raise it by $k^{\omega}$. Thus, if buying $\omega$ shares all at once
affects the price at least as much as buying them one at a time, the
ratio of prices before and after impact should increase at least
exponentially.  Taking logarithms implies that the price impact as we
have defined it above should increase at least linearly.\footnote{
This has practical implications.  It is common practice to break up
orders in order to reduce losses due to market impact. With a
sufficiently concave market impact function, in contrast, it is
cheaper to execute an order all at once.}

In contrast, from empirical studies $\phi(\omega)$ for buy orders
appears to be concave
\cite{Hausman92,Farmer96,Torre97,Kempf98,Plerou01,Lillo02}.  
Lillo {\it et al.} have shown for that for stocks in the NYSE
the concave behavior of the price impact is quite consistent across
different stocks \cite{Lillo02}. Our model produces concave price impact functions
that are in qualitative agreement with these results.

Our work also demonstrates the value of physics techniques for economic
problems.  Our analysis makes extensive use of dimensional analysis,
the solution of a master equation through a generating functional, and
a mean field approach that is commonly used to analyze non-equilibrium
reaction-diffusion systems and evaporation-deposition problems.

\subsection{Background: The continuous double auction}
\label{background}

Most modern financial markets operate continuously.  The mismatch
between buyers and sellers that typically exists at any given instant
is solved via an order-based market with two basic kinds of
orders. Impatient traders submit
\textit{market orders}, which are requests to buy or sell a given
number of shares immediately at the best available price. More patient
traders submit \textit{limit orders}, or {\it quotes} which also state
a limit price, corresponding to the worst allowable price for the
transaction.  (Note that the word ``quote'' can be used either to
refer to the limit price or to the limit order itself.) Limit orders
often fail to result in an immediate transaction, and are stored in a
queue called the \textit{limit order book}. Buy limit orders are
called \textit{bids}, and sell limit orders are called
\textit{offers }or \textit{asks}. We use the logarithmic price $a(t)$ to denote the
position of the best (lowest) offer and $b(t)$ for the position the 
best (highest) bid. These are also called the {\it inside quotes}.  There is
typically a non-zero price gap between them, called the
\textit{spread} $s(t)=a(t)-b(t)$.  Prices are not continuous, but
rather have discrete quanta called {\it ticks}.  Throughout this
paper, all prices will be expressed as logarithms, and to avoid
endless repetition, the word {\it price} will mean the logarithm of
the price.  The minimum interval that prices change on is the {\it
tick size} $dp$ (also defined on a logarithmic scale; note this is not
true for real markets).  Note that $dp$ is {\it not} necessarily
infinitesimal.

\begin{figure}[ptb]
  \begin{center} 
  \includegraphics[scale=0.5]{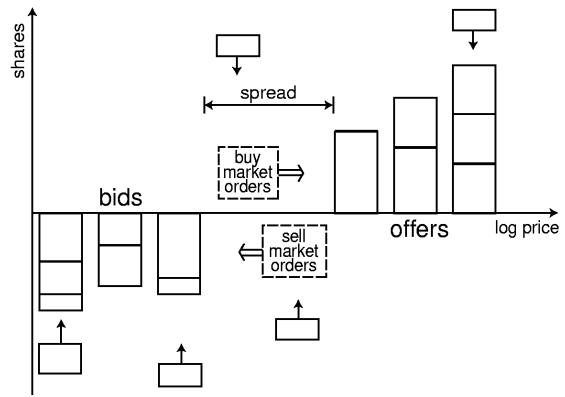}
  \caption{
  A schematic illustration of the continuous double auction
  mechanism and our model of it.  Limit orders are stored in the limit
  order book.  We adopt the arbitrary convention that buy orders are
  negative and sell orders are positive.  As a market order arrives,
  it has transactions with limit orders of the opposite sign, in order
  of price (first) and time of arrival (second).  The best quotes at
  prices $a(t)$ or $b(t)$ move whenever an incoming market order has 
  sufficient size
  to fully deplete the stored volume at $a(t)$ or $b(t)$.  Our model assumes
  that market order arrival, limit order arrival, and limit order cancellation
  follow a Poisson process. New offers (sell limit orders) can be placed at 
  any price greater than the best bid, and are shown here as ``raining down''
  on the price axis.  Similarly, new bids (buy limit orders) can be
  placed at any price less than the best offer.  Bids and
  offers that fall inside the spread become the new best bids and offers.  
  All prices in this model are logarithmic.}
  \label{bookschematic}
  \end{center}
\end{figure}

As market orders arrive they are matched against limit orders of the
opposite sign in order of first price and then arrival time, as shown in
Fig.~\ref{bookschematic}.
Because orders are placed for varying numbers of shares, matching is
not necessarily one-to-one. For example, suppose the best offer is for
200 shares at \$60 and the the next best is for 300 shares at \$60.25;
a buy market order for 250 shares buys 200 shares at \$60 and 50
shares at \$60.25, moving the best offer $a(t)$ from \$60 to \$60.25.
A high density of limit orders per price results in high
\textit{liquidity} for market orders, i.e., it decreases the price
movement when a market order is placed.  Let $n(p,t)$ be the stored
density of limit order volume at price $p$, which we will call the
{\it depth profile} of the limit order book at any given time $t$.
The total stored limit order volume at price level $p$ is $n(p,t)
dp$.  For unit order size the shift in the best ask $a(t)$ produced by
a buy market order is given by solving the equation
\begin{equation}
\omega = \sum_{p=a(t)}^{p\prime} n(p,t) dp
\label{volumeConstraint}
\end{equation}
for $p\prime$.  The shift in the best ask $p\prime - a(t)$, where
is the instantaneous price impact for buy market orders.
A similar statement applies for sell market orders, where
the price impact can be defined in terms of the shift in the best bid.
(Alternatively, it is also possible to define the price impact in
terms of the change in the midpoint price).

We will refer to a buy limit order whose limit price is greater than
the best ask, or a sell limit order whose limit price is less than the
best bid, as a {\it crossing limit order} or {\it marketable limit
order}.  Such limit orders result in immediate transactions, with at
least part of the order immediately executed.

\subsection{The model}
\label{model}

This model introduced in reference \cite{Daniels01}, is designed to be
as analytically tractable as possible while capturing key features of
the continuous double auction.  All the order flows are modeled as
Poisson processes.  We assume that market orders arrive in chunks of
$\sigma$ shares, at a rate of $\mu$ shares per unit time. The market
order may be a `buy' order or a `sell' order with equal probability.
(Thus the rate at which buy orders or sell orders arrive individually
is $\mu/2$.)  Limit orders arrive in chunks of $\sigma$ shares as well, at a
rate $\alpha$ shares per unit price and per unit time for buy orders
and also for sell orders.  Offers are placed with uniform probability
at integer multiples of a tick size $dp$ in the range of
price $b(t)<p<\infty$, and similarly for bids on
$-\infty<p<a(t)$.  When a market order arrives it causes a
transaction; under the assumption of constant order size, a buy market
order removes an offer at price $a(t)$, and if it was the
last offer at that price, moves the best ask up to the next occupied
price tick.  Similarly, a sell market order removes a bid at
price $b(t)$, and if it is the last bid at that price,
moves the best bid down to the next occupied price tick.  In addition,
limit orders may also be removed spontaneously by being canceled or by
expiring, even without a transaction having taken place.  We model
this by letting them be removed randomly with constant probability
$\delta$ per unit time.

While the assumption of limit order placement over an infinite
interval is clearly unrealistic, it provides a tractable boundary
condition for modeling the behavior of the limit order book near the
midpoint price $m(t)=(a(t)+b(t))/2$, which is the region
of interest since it is where transactions occur.  Limit orders far
from the midpoint are usually canceled before they are executed (we
demonstrate this later in Fig.~\ref{effprof}), and so far from the
midpoint, limit order arrival and cancellation have a steady state
behavior characterized by a simple Poisson distribution.  Although
under the limit order placement process the total number of orders
placed per unit time is infinite, the order placement per unit price
interval is bounded and thus the assumption of an infinite interval
creates no problems.  Indeed, it guarantees that there are always an
infinite number of limit orders of both signs stored in the book, so
that the bid and ask are always well-defined and the book never
empties.  (Under other assumptions about limit order placement this is
not necessarily true, as we later demonstrate in
Fig. \ref{fig:powerlaw}.)  We are also
considering versions of the model involving more realistic order
placement functions; see the discussion in
Section~\ref{subsec:futurework}.

In this model, to keep things simple, we are using the conceptual
simplification of {\it effective market orders} and {\it effective
limit orders}.  When a crossing limit order is placed part of it may
be executed immediately.  The effect of this part on the price is
indistinguishable from that of a market order of the same size.
Similarly, given that this market order has been placed, the remaining
part is equivalent to a non-crossing limit order of the same size.
Thus a crossing limit order can be modeled as an effective market
order followed by an effective (non-crossing) limit order.\footnote{In
assigning independently random distributions for the two events, our
model neglects the correlation between market and limit order arrival
induced by crossing limit orders.} Working in terms of effective
market and limit orders affects data analysis: The effective market
order arrival rate $\mu$ combines both pure market orders and the
immediately executed components of crossing limit orders, and
similarly the limit order arrival rate $\alpha$ corresponds only to
the components of limit orders that are not executed immediately.
This is consistent with the boundary conditions for the order
placement process, since an offer with $p\leq b(t)$ or a bid with
$p\geq a(t)$ would result in an immediate transaction, and thus would
be effectively the same as a market order. Defining the order
placement process with these boundary conditions realistically allows
limit orders to be placed anywhere inside the spread.

Another simplification of this model is the use of logarithmic prices,
both for the order placement process and for the tick size $dp$.  This
has the important advantage that it ensures that prices are always
positive. In real markets price ticks are linear, and the use of
logarithmic price ticks is an approximation that makes both the
calculations and the simulation more convenient. We find that the
limit $dp \rightarrow 0$, where tick size is irrelevant, is a
good approximation for many purposes.  We find that tick size is less
important than other parameters of the problem, which provides some
justification for the approximation of logarithmic price ticks.

Assuming a constant probability for cancellation is clearly
\textit{ad hoc}, but in simulations we find that other assumptions with
well-defined timescales, such as constant duration time, give similar
results.  For our analytic model we use a constant order size
$\sigma$.  In simulations we also use variable order size,
e.g. half-normal distributions with standard deviation $\sqrt{\pi /
2}\sigma$, which ensures that the mean value remains $\sigma$.  
As long as these distributions have thin tails, the
differences do not qualitatively affect most of the results reported
here, except in a trivial way.  As discussed in Section
\ref{subsec:futurework}, decay processes without well-defined
characteristic times and size distributions with power law tails give
qualitatively different results and will be treated elsewhere.

Even though this model is simply defined, the time evolution is not
trivial.  One can think of the dynamics as being composed of three
parts: (1) the buy market order/sell limit order interaction, which
determines the best ask; (2) the sell market order/buy limit order
interaction, which determines the best bid; and (3) the random
cancellation process.  Processes (1) and (2) determine each others'
boundary conditions.  That is, process (1) determines the best ask,
which sets the boundary condition for limit order placement in process
(2), and process (2) determines the best bid, which determines the
boundary conditions for limit order placement in process (1).  Thus
processes (1) and (2) are strongly coupled.  It is this coupling that
causes the bid and ask to remain close to each other, and guarantees
that the spread $s(t) = a(t) - b(t)$ is a stationary random variable,
even though the bid and ask are not.  It is the coupling of these
processes through their boundary conditions that provides the
nonlinear feedback that makes the price process complex.

\subsection{Summary of prior work}

There are two independent lines of prior work, one in the financial
economics literature, and the other in the physics literature.  The
models in the economics literature are directed toward empirical
analysis, and treat the order process as static. In contrast, the
models in the physics literature are conceptual toy models, but they
allow the order process to react to changes in prices, and are thus
fully dynamic.  Our model bridges this gap.  This is explained in more
detail below.

The first model of this type that we are aware of was due to Mendelson
\cite{Mendelson82}, who modeled random order placement with periodic
clearing.  This was developed along different directions by Cohen
{\it et al.} \cite{Cohen85}, who used techniques from queuing theory, but
assumed only one price level and addressed the issue of time priority
at that level (motivated by the existence of a specialist who
effectively pinned prices to make them stationary).  Domowitz and Wang
\cite{Domowitz94} and Bollerslev {\it et al.} \cite{Bollerslev97} further
developed this to allow more general order placement processes that
depend on prices, but without solving the full dynamical problem.
This allows them to get a stationary solution for prices. In
contrast, in our model the prices that emerge make a random walk, and
so are much more realistic.  In order to get a solution for the depth
of the order book we have to go into price coordinates that co-move
with the random walk.  Dealing with the feedback between order
placement and prices makes the problem much more difficult, but it is
key for getting reasonable results.

The models in the physics literature incorporate price dynamics, but
have tended to be conceptual toy models designed to understand the
anomalous diffusion properties of prices.  This line of work begins
with a paper by Bak {\it et al.} \cite{Bak96} which was developed by Eliezer
and Kogan \cite{Eliezer98} and by Tang \cite{Tang}. They assume that
limit orders are placed at a fixed distance from the midpoint, and
that the limit prices of these orders are then randomly shuffled until
they result in transactions.  It is the random shuffling that causes
price diffusion.  This assumption, which we feel is unrealistic, was
made to take advantage of the analogy to a standard reaction-diffusion
model in the physics literature.  Maslov \cite{Maslov00} introduced an
alterative model that was solved analytically in the mean-field limit
by Slanina \cite{Slanina01}.  Each order is randomly chosen to be
either a buy or a sell, and either a limit order or a market order.
If a limit order, it is randomly placed within a fixed distance of the
current price.  This again gives rise to anomalous price diffusion.  A
model allowing limit orders with Poisson order cancellation was
proposed by Challet and Stinchcombe \cite{Challet01}.  Iori and
Chiarella \cite{Iori01} have numerically studied a model including
fundamentalists and technical traders. 

The model studied in this paper was introduced by Daniels {\it et al.} \cite{Daniels01}.
This adds to the literature by introducing a model that treats the
feedback between order placement and price movement, while having
enough realism so that the parameters can be tested against real data.
The prior models in the physics literature have tended to focus
primarily on the anomalous diffusion of prices.  While interesting and
important for refining risk calculations, this is a second-order
effect.  In contrast, we focus on the first order effects of primary
interest to market participants, such as the bid-ask spread,
volatility, depth profile, price impact, and the probability and time
to fill an order.  We demonstrate how dimensional analysis becomes a
useful tool in an economic setting, and develop mean field theories in
a context that is more challenging than that of the toy models of
previous work.

Subsequent to reference \cite{Daniels01}, Bouchaud et
al. \cite{Bouchaud02} demonstrated that, under the assumption that
prices execute a random walk, by introducing an additional free
parameter they can derive a simple equation for the depth profile.  In
this paper we show how to do this from first principles without
introducing a free parameter.  

\section{Overview of predictions of the model}
\label{overview}

In this section we give an overview of the phenomenology of the model.
Because this model has five parameters, understanding all their
effects would generally be a complicated problem in and of itself.
This task is greatly simplified by the use of dimensional analysis,
which reduces the number of independent parameters from five to two.  Thus,
before we can even review the results, we need to first explain how
dimensional analysis applies in this setting.  One of the surprising
aspects of this model is that one can derive several powerful results
using the simple technique of dimensional analysis alone.

Unless otherwise mentioned the results presented in this section
are based on simulations.  These results are compared to
theoretical predictions in Section~\ref{analysis}.

\subsection{Dimensional analysis}

Because dimensional analysis is not commonly used in economics we
first present a brief review.  For more details see Bridgman
\cite{Bridgman22}.  

Dimensional analysis is a technique that is commonly used in physics
and engineering to reduce the number of independent degrees of freedom
by taking advantage of the constraints imposed by dimensionality.  For
sufficiently constrained problems it can be used to guess the answer
to a problem without doing a full analysis.  The idea is to write down
all the factors that a given phenomenon can depend on, and then find
the combination that has the correct dimensions.  For example,
consider the problem of the period of a pendulum: The period $T$ has
dimensions of {\it time}.  Obvious candidates that it might depend on
are the mass of the bob $m$ (which has units of {\it mass}), the length
$l$ (which has units of {\it distance}), and the acceleration of
gravity $g$ (which has units of $distance/time^2$).  There is only one
way to combine these to produce something with dimensions of {\it
time}, i.e. $T \sim \sqrt{l/g}$.  This determines the correct formula
for the period of a pendulum up to a constant.  Note that it makes it
clear that the period does not depend on the mass, a result that is
not obvious {\it a priori}.  We were lucky in this problem because
there were three parameters and three dimensions, with a unique
combination of the parameters having the right dimensions; in general
dimensional analysis can only be used to reduce the number of free
parameters through the constraints imposed by their dimensions.

For this problem the three fundamental dimensions in the model are
{\it shares}, {\it price}, and {\it time}.  Note that by {\it price},
we mean the logarithm of price; as long as we are consistent, this
does not create problems with the dimensional analysis.  There are
five parameters: three rate constants and two discreteness parameters.
The {\it order flow rates} are $\mu$, the market order arrival rate,
with dimensions of {\em shares per time}; $\alpha$, the limit order
arrival rate per unit price, with dimensions of {\em shares per price
per time}; and $\delta$, the rate of limit order decays, with
dimensions of {\em 1/time}.  These play a role similar to rate
constants in physical problems.  The two {\it discreteness parameters}
are the price tick size $dp$, with dimensions of {\em price}, and the
order size $\sigma$, with dimensions of {\em shares}.  This is
summarized in table~\ref{tab:parameters}.
\begin{table}
  \begin{tabular}[c]{lll}%
    {\bf Parameter} &  {\bf Description} &  {\bf Dimensions} \\
    $\alpha$ & 
      limit order rate &
      $shares/(price~~time)$\\
    $\mu$ & 
      market order rate & 
      $shares/time$\\
    $\delta$ & 
      order cancellation rate &
      $1/time$\\
    $dp$ & 
      tick size & 
      $price$\\
    $\sigma$ & 
      characteristic order size & 
      $shares$
  \end{tabular}
  \caption{
    The five parameters that characterize this model.  $\alpha$, $\mu$,
    and $\delta$ are order flow rates, and $dp$ and $\sigma$ are
    discreteness parameters.
  \label{tab:parameters}
  }
\end{table}

Dimensional analysis can be used to reduce the number of relevant
parameters. Because there are five parameters and three dimensions
({\em price, shares, time}), and because in this case the
dimensionality of the parameters is sufficiently rich, the dimensional
relationships reduce the degrees of freedom, so that all the
properties of the limit-order book can be described by functions of
two parameters.  It is useful to construct these two parameters 
so that they are nondimensional.

We perform the dimensional reduction of the model by guessing that the
effect of the order flow rates is primary to that of the discreteness
parameters.  This leads us to construct nondimensional units based on
the order flow parameters alone, and take nondimensionalized versions
of the discreteness parameters as the independent parameters whose effects
remain to be understood.  As we will see, this is justified by the fact
that many of the properties of the model depend only weakly on the
discreteness parameters.  We can thus understand much of the richness
of the phenomenology of the model through dimensional analysis alone.

There are three order flow rates and three fundamental dimensions.  If
we temporarily ignore the discreteness parameters, there are unique
combinations of the order flow rates with units of shares, price, and
time.  These define a characteristic number of shares $N_c = \mu / 2
\delta$, a characteristic price interval $p_c = {\mu} / 2 {\alpha}$,
and a characteristic timescale $t_c = 1/ \delta$.  This is summarized
in table~\ref{discreteParams}.
\begin{table}
 \begin{tabular}[c]{lll}%
 {\bf Parameter} & {\bf Description} &  {\bf Expression} \\ 
$N_c$ & characteristic number of shares & $\mu / 2 \delta$ \\
$p_c$ & characteristic price interval & $\mu / 2 \alpha$ \\
$t_c$ & characteristic time & $1/\delta$ \\
$dp/p_c$ & nondimensional tick size & $2 \alpha dp / \mu $\\ 
$\epsilon$ & nondimensional order size & $2 \delta \sigma / \mu$\\ 
\end{tabular}
  \caption{
    Important characteristic scales and nondimensional quantities.  We 
    summarize the characteristic share size, price and times defined by the 
    order flow rates, as well as the two nondimensional scale 
    parameters $dp/p_c$ and $\epsilon$
    that characterize the effect of finite tick size and order
    size.  Dimensional analysis makes it clear that all the properties
    of the limit order book can be characterized in terms of functions of these
    two parameters.
  \label{discreteParams}
  } 
\end{table}
The factors of two occur because we have defined the market order rate
for either a buy or a sell order to be $\mu/2$.  We can thus express
everything in the model in nondimensional terms by dividing by $N_c$,
$p_c$, or $t_c$ as appropriate, e.g. to measure shares in
nondimensional units $\hat{N} = N/N_c$, or to measure price in
nondimensional units $\hat{p} = p/p_c$.

\begin{figure}[ptb]
  \begin{center} 
    \includegraphics[scale=0.37]{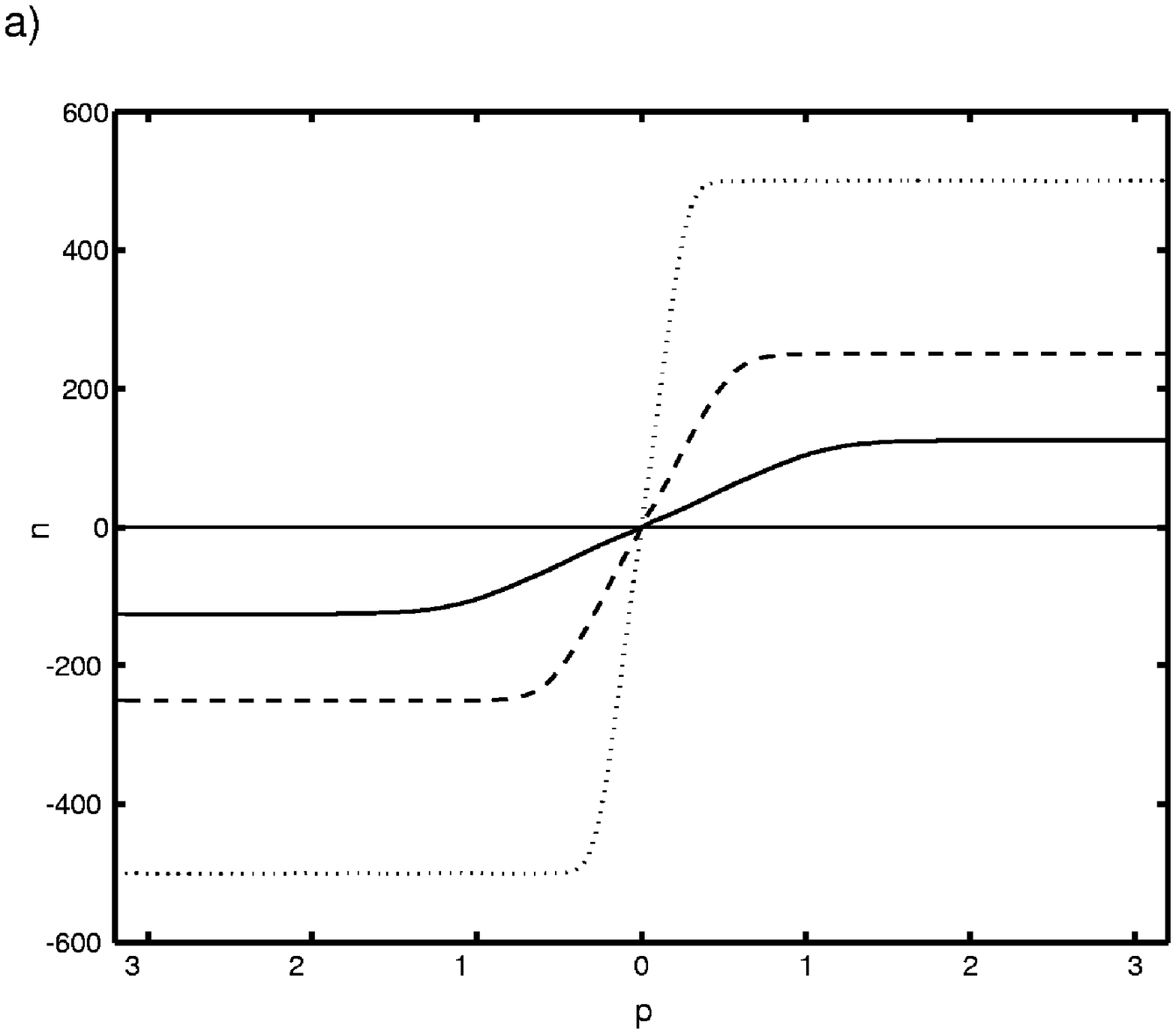}
    \includegraphics[scale=0.37]{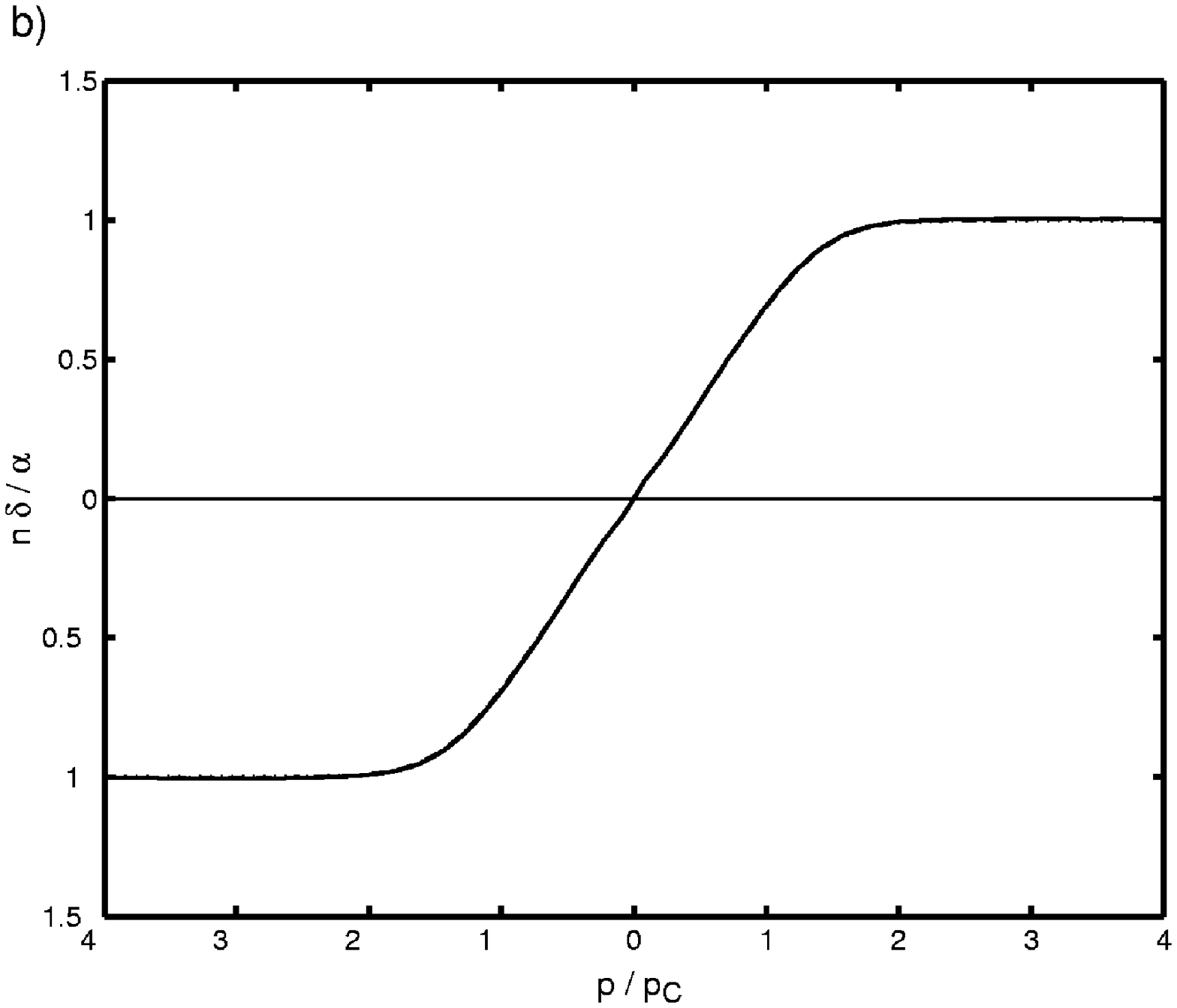}
  \caption{The usefulness of nondimensional units.
    (a) We show the average depth profile for three different parameter sets.
    The parameters $\alpha = 0.5$, $\sigma = 1$, and $dp = 0$ are held 
    constant, while $\delta$ and $\mu$ are varied.  The line types are:
    (dotted) $\delta = 0.001$, $\mu = 0.2$; (dashed) $\delta = 0.002$, 
    $\mu = 0.4$ and (solid) $\delta = 0.004$, $\mu = 0.8$.
    (b) is the same, but plotted in nondimensional units.  The horizontal 
    axis has units of {\it price}, and so has nondimensional units
    $\hat{p} = p/pc = 2 \alpha p /\mu $.  The vertical axis has
    units of $n$ {\it shares/price}, and so has nondimensional units 
    $\hat{n} = n p_c/N_c = n \delta/\alpha$.
    Because we have chosen the parameters to keep the nondimensional order 
    size $\epsilon$ constant, the collapse is perfect.  
    Varying the tick size has little effect on the results other than making 
    them discrete.
  \label{dimDepth}
  }
    \end{center}
\end{figure}

The value of using nondimensional units is illustrated in
Fig.~\ref{dimDepth}.
Fig.~\ref{dimDepth}(a) shows the average depth profile for three
different values of $\mu$ and $\delta$ with the other parameters held
fixed.  When we plot these results in dimensional units the results
look quite different.  However, when we plot them in terms of
nondimensional units, as shown in Fig.~\ref{dimDepth}(b), the results
are indistinguishable.  As explained below, because we have kept the
nondimensional order size fixed, the collapse is perfect.  Thus, the
problem of understanding the behavior of this model is reduced to
studying the effect of tick size and order size.

To understand the effect of tick size and order size it is useful to
do so in nondimensional terms. The nondimensional scale parameter
based on tick size is constructed by dividing by the characteristic
price, i.e. $dp/p_c = 2 {\alpha} dp / {\mu} $.  The theoretical
analysis and the simulations show that there is a sensible continuum
limit as the tick size $dp \rightarrow 0$, in the sense that there is
non-zero price diffusion and a finite spread.  Furthermore, the
dependence on tick size is weak, and for many purposes the limit $dp
\rightarrow 0$ approximates the case of finite tick size fairly
well. As we will see, working in this limit is essential for getting
tractable analytic results.

A nondimensional scale parameter based on order size is constructed by
dividing the typical order size (which is measured in shares) by the
characteristic number of shares $N_c$, i.e. $\epsilon \equiv
\sigma/N_c = 2 \delta \sigma / {\mu}$.  $\epsilon$ characterizes the 
``chunkiness'' of the orders stored in the limit order book.  As we
will see, $\epsilon$ is an important determinant of liquidity, and it
is a particularly important determinant of volatility. In the
continuum limit $\epsilon \rightarrow 0$ there is no price
diffusion. This is because price diffusion can occur only if there is
a finite probability for price levels outside the spread to be empty,
thus allowing the best bid or ask to make a persistent shift.  If we
let $\epsilon \rightarrow 0$ while the average depth is held fixed the
number of individual orders becomes infinite, and the probability that
spontaneous decays or market orders can create gaps outside the spread
becomes zero.  This is verified in simulations.  Thus the limit
$\epsilon \rightarrow 0$ is always a poor approximation to a real
market. $\epsilon$ is a more important parameter than the tick size
$dp/p_c$.  In the mean field analysis in Section~\ref{analysis}, we
let $dp/p_c \rightarrow 0$, reducing the number of independent parameters 
from two to one, and in many cases find that this is a good approximation.

The order size $\sigma$ can be thought of as the order {\it
granularity}.  Just as the properties of a beach with fine sand are
quite different from that of one populated by fist-sized boulders, a
market with many small orders behaves quite differently from one with
a few large orders.  $N_c$ provides the scale against which the order
size is measured, and $\epsilon$ characterizes the granularity in
relative terms.  Alternatively, $1/\epsilon$ can be thought of as the
annihilation rate from market orders expressed in units of the size of
spontaneous decays. Note that in nondimensional units the number of
shares can also be written $\hat{N} = N/N_c = N \epsilon /
\sigma$.

The construction of the nondimensional granularity parameter
illustrates the importance of including a spontaneous decay process in
this model.  If $\delta = 0$ (which implies $\epsilon = 0$) there is
no spontaneous decay of orders, and depending on the relative values
of $\mu$ and $\alpha$, generically either the depth of orders will
accumulate without bound or the spread will become infinite.  As long
as $\delta > 0$, in contrast, this is not a problem.

For some purposes the effects of varying tick size and order size are
fairly small, and we can derive approximate formulas using dimensional
analysis based only on the order flow rates.  For example, in
table~\ref{tab:dim_scaling} we give dimensional scaling formulas for
the average spread, the market order liquidity (as measured by the average 
slope of the depth profile near the midpoint), the volatility, and the
asymptotic depth (defined below).  Because these estimates neglect the
effects of discreteness, they are only approximations of the true
behavior of the model, which do a better job of explaining some
properties than others. Our numerical and analytical results show that
some quantities also depend on the granularity parameter $\epsilon$
and to a weaker extent on the tick size $dp/p_c$.  Nonetheless, the
dimensional estimates based on order flow alone provide a good
starting point for understanding market behavior.
\begin{table}
  \begin{tabular} [c]{lll}%
  {\bf Quantity} & {\bf Dimensions} & {\bf Scaling relation}\\
  Asymptotic depth & $shares/price$ & $d \sim \alpha/\delta$\\ 
  Spread & $price$ & $s \sim \mu/\alpha$\\ 
  Slope of depth profile & $shares/price^{2}$ & 
      $\lambda \sim \alpha^{2}/\mu\delta = d/s$\\
  Price diffusion rate & $price^{2}/time$ & 
      $D_0 \sim \mu^{2}\delta/{\alpha^{2}}$\\ 
 \end{tabular}
  \caption{
  Estimates from dimensional analysis for the scaling of a few market 
  properties based on order flow rates alone. $\alpha$ is the limit order 
  density rate, $\mu$ is the market order rate, and $\delta$ is the 
  spontaneous limit order removal rate. These estimates are constructed
  by taking the combinations of these three rates that have the proper units.
  They neglect the dependence on on the order granularity $\epsilon$ and
  the nondimensional tick size $dp/p_c$.  More accurate relations from
  simulation and theory are given in table \ref{tab:params_scaling}.
  \label{tab:dim_scaling} 
  }
\end{table}
A comparison to more precise formulas derived from theory and simulations
is given in table \ref{tab:params_scaling}.
\begin{table}
  \begin{tabular} [c]{lll}%
  {\bf Quantity} & {\bf Scaling relation} & {\bf Figure}\\
  Asymptotic depth & $d = \alpha/\delta$ & \ref{epsDepth}\\ 
  Spread & $s = (\mu/\alpha) f(\epsilon, dp/p_c)$ & \ref{meanSpread}, 
      \ref{spread_compare}\\ 
  Slope of depth profile & 
      $\lambda = (\alpha^{2}/\mu \delta) g(\epsilon, dp/p_c)$ &
      \ref{epsDepth}, \ref{fig:better_wrong_ind_eps_0p2} -
      \ref{fig:better_wrong_ind_eps_0p02}\\
  Price diffusion ($\tau \rightarrow 0$) & 
      $D_0 = (\mu^{2}\delta/{\alpha^{2}})\epsilon^{-0.5}$ &
      \ref{epsVarVsTau}, \ref{finiteTicks}(c)\\ 
  Price diffusion ($\tau \rightarrow \infty$) & 
      $D_\infty = (\mu^{2}\delta /{\alpha^{2}})\epsilon^{0.5}$ &
      \ref{epsVarVsTau}, \ref{finiteTicks}(c)\\ 
 \end{tabular}
  \caption{
  The dependence of market properties on model parameters based on simulation
  and theory, with the relevant figure numbers.
  These formulas include corrections for order granularity $\epsilon$ and 
  finite tick size $dp/p_c$. The formula for asymptotic depth from dimensional
  analysis in table \ref{tab:dim_scaling} is exact with zero tick size. The 
  expression
  for the mean spread is modified by a function of $\epsilon$ and $dp/p_c$,
  though the dependence on them is fairly weak.  
  For the liquidity $\lambda$, corresponding to the slope
  of the depth profile near the origin, the dimensional estimate must be 
  modified because the depth profile is no longer linear (mainly depending 
  on $\epsilon$) and so the slope depends on price.  
  The formulas for the volatility are empirical 
  estimates from simulations. The dimensional estimate for the volatility from 
  Table~\ref{tab:dim_scaling} is modified by a factor of
  $\epsilon^{-0.5}$ for the early time price diffusion rate and a factor of
  $\epsilon^{0.5}$ for the late time price diffusion rate.
  \label{tab:params_scaling} 
  }
\end{table}

An approximate formula for the mean spread can be derived by noting
that it has dimensions of $\mbox{\textit{price}}$, and the unique
combination of order flow rates with these dimensions is $\mu/\alpha$.
While the dimensions indicate the scaling of the spread, they cannot
determine multiplicative factors of order unity.  A more intuitive
argument can be made by noting that inside the spread removal due to
cancellation is dominated by removal due to market orders.  Thus the
total limit order placement rate inside the spread, for either buy or
sell limit orders $\alpha s$, must equal the order removal rate $\mu /
2$, which implies that spread is $s = \mu/2 \alpha$. As we will see
later, this argument can be generalized and made more precise within
our mean-field analysis which then also predicts the observed
dependence on the granularity parameter $\epsilon$.  However this
dependence is rather weak and only causes a variation of roughly a
factor of two for $\epsilon < 1$ (see Figs.~\ref{meanSpread} and
\ref{spread_compare}), and the factor of $1/2$ derived above is a good
first approximation.  Note that this prediction of the mean spread is
just the characteristic price $p_c$.

It is also easy to derive the mean {\it asymptotic depth}, which is
the density of shares far away from the midpoint.  The asymptotic
depth is an artificial construct of our assumption of order placement
over an infinite interval; it should be regarded as providing a simple
boundary condition so that we can study the behavior near the midpoint
price.  The mean asymptotic depth has dimensions of
$\mbox{\textit{shares}}/\mbox{\textit{price}}$, and is therefore given
by $\alpha/\delta$.  Furthermore, because removal by market orders is
insignificant in this regime, it is determined by the balance between
order placement and decay, and far from the midpoint the depth at any
given price is Poisson distributed.  This result is exact.

The average slope of the depth profile near the midpoint is an
important determinant of liquidity, since it affects the expected
price response when a market order arrives. The slope has dimensions of
$\mbox{\textit{shares}}/{\mbox{\textit{price}}}^{2}$, which implies
that in terms of the order flow rates it scales roughly as
$\alpha^{2}/\mu\delta$.  This is also the ratio of the asymptotic
depth to the spread.  As we will see later, this is a good
approximation when $\epsilon \sim 0.01$, but for smaller values of
$\epsilon$ the depth profile is not linear near the midpoint, 
and this approximation
fails.

The last two entries in table~\ref{tab:params_scaling} are empirical
estimates for the price diffusion rate $D$, which is proportional to
the square of the volatility.  That is, for normal diffusion, starting
from a point at $t=0$, the variance $v$ after time $t$ is $v = Dt$.
The volatility at any given timescale $t$ is the square root of the
variance at timescale $t$.  The estimate for the diffusion rate based
on dimensional analysis in terms of the order flow rates alone is
$\mu^{2}\delta/{\alpha^{2}}$. However, simulations show that short
time diffusion is much faster than long time diffusion, due to
negative autocorrelations in the price process, as shown in
Fig.~\ref{epsVarVsTau}. The initial and the asymptotic diffusion rates
appear to obey the scaling
relationships given in table~\ref{tab:params_scaling}. Though our
mean-field theory is not able to predict this functional form, the
fact that early and late time diffusion rates are different can be
understood within the framework of our analysis, as described in
Sec.~\ref{subsec:ren_diff}.  Anomalous diffusion of this type implies
negative autocorrelations in midpoint prices.  Note that we use the
term ``anomalous diffusion'' to imply that the diffusion rate is
different on short and long timescales. We do not use this term in the
sense that it is normally used in the physics literature, i.e. that
the long-time diffusion is proportional to $t^\gamma$ with $\gamma
\neq 1$ (for long times $\gamma = 1$ in our case).

\subsection{Varying the granularity parameter $\epsilon$}
\label{subsec:epsi_ranges}

We first investigate the effect of varying the order granularity $\epsilon$
in the limit $dp \rightarrow 0$.  As we will see, the granularity has an
important effect on most of the properties of the model, and particularly
on depth, price impact, and price diffusion.  The behavior 
can be divided into three regimes, roughly as follows:

\begin{itemize}

\item
{\bf Large $\bf \epsilon$, i.e. $\epsilon \gtrsim 0.1$}.  This
corresponds to a large accumulation of orders at the best bid and ask,
nearly linear market impact, and roughly equal short and long time price
diffusion rates.  This is the regime where the mean-field
approximation used in the theoretical analysis works best.

\item
{\bf Medium $\bf \epsilon$ i.e. $\epsilon \sim 0.01$}.  In this range
the accumulation of orders at the best bid and ask is small and near
the midpoint price the depth profile increases nearly linearly with
price.  As a result, as a crude approximation the price impact
increases as roughly the square root of order size.

\item
{\bf Small $\bf \epsilon$ i.e. $\epsilon \lesssim 0.001$}.  The
accumulation of orders at the best bid and ask is very small, and near
the midpoint the depth profile is a convex function of price.  The
price impact is very concave.  The short time price diffusion rate is
much greater than the long time price diffusion rate.

\end{itemize}

Since the results for bids are symmetric with those for offers about
$p=0$, for convenience we only show the results for offers,
i.e. buy market orders and sell limit orders.  In this sub-section
prices are measured relative to the midpoint, and simulations are in the
continuum limit where the tick size $dp \rightarrow 0$.  The results
in this section are from numerical simulations.  Also, bear in mind
that far from the midpoint the predictions of this model are not valid
due to the unrealistic assumption of an order placement process with
an infinite domain.  Thus the results are potentially relevant to real
markets only when the price $p$ is at most a few times as large as the
characteristic price $p_c$.

\subsubsection{Depth profile}

\begin{figure}[ptb]
  \begin{center}
    \includegraphics[scale=0.37]{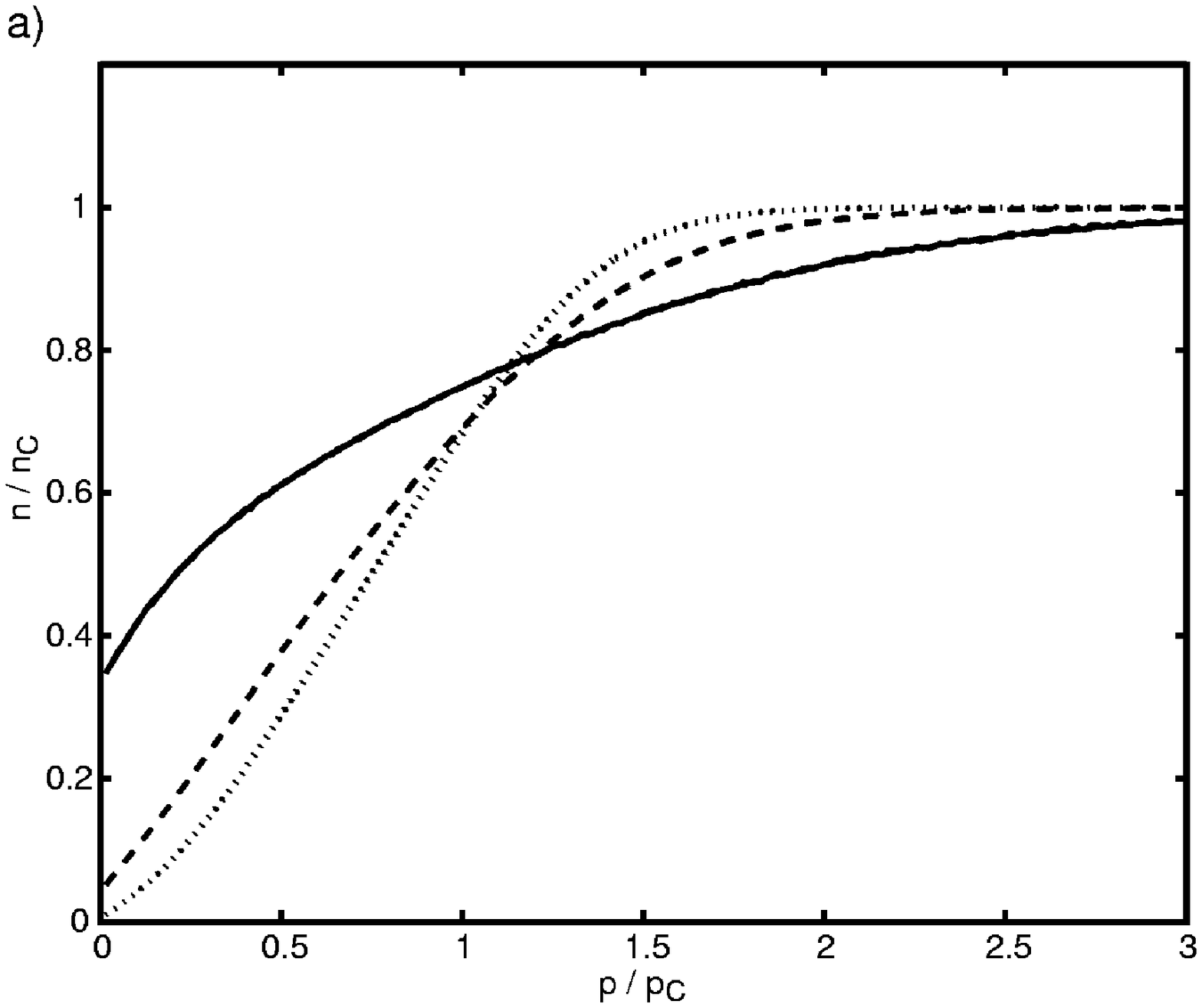}
    \includegraphics[scale=0.37]{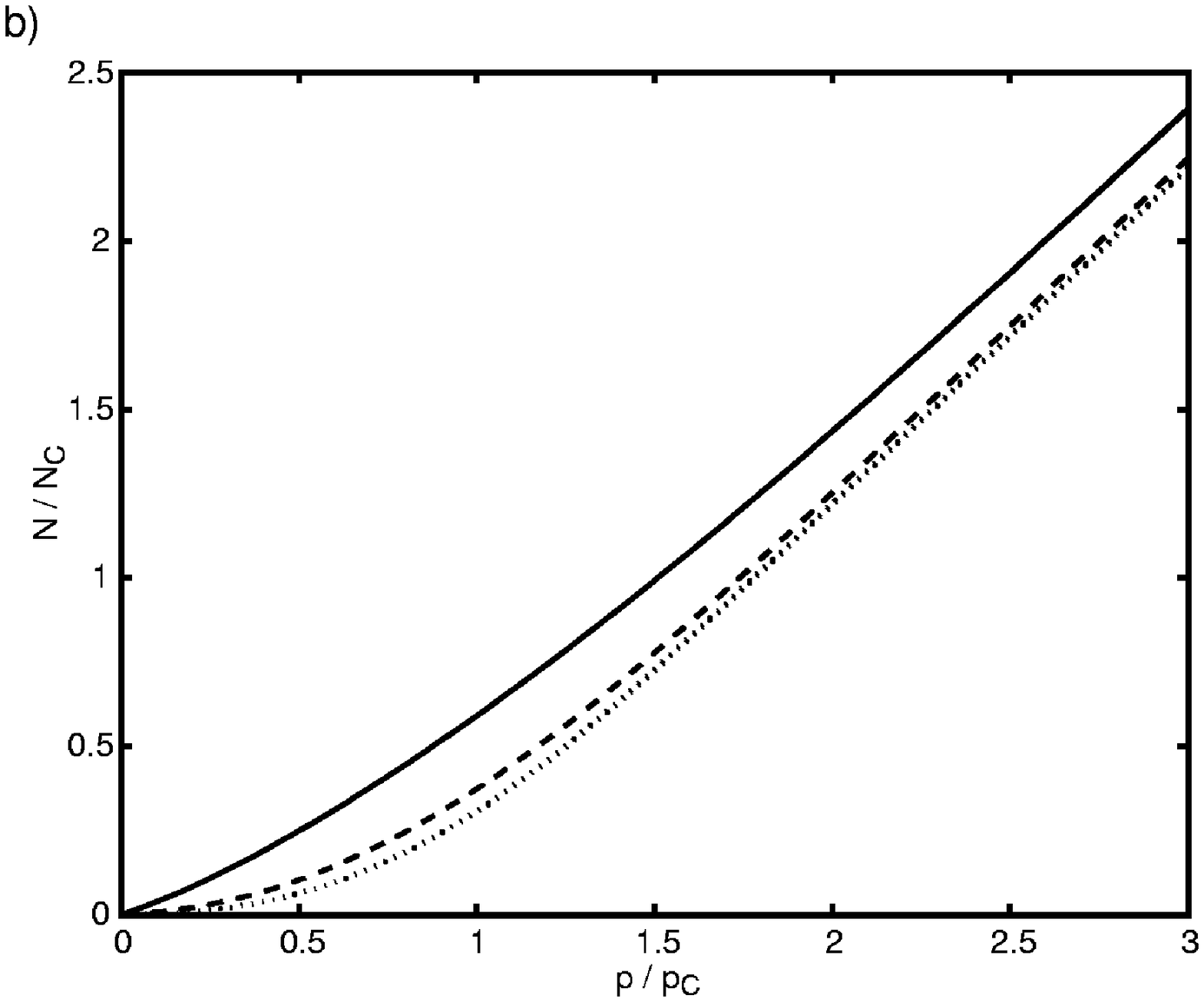}
    \caption{ 
      The mean depth profile and cumulative depth
      versus $\hat{p} = p/p_c = 2 \alpha p / \mu$.  The origin $p/p_c = 0$ 
      corresponds to the midpoint. (a) is the average depth
      profile $n$ in nondimensional coordinates 
      $\hat{n} = n p_c/N_c = n \delta / \alpha$.  (b) is nondimensional 
      cumulative depth $N(p)/N_c$. We show three different values of the 
      nondimensional granularity parameter: $\epsilon = 
      0.2$ (solid), $\epsilon = 0.02$ (dash), $\epsilon = 0.002$
      (dot), all with tick size $dp = 0$.  
    \label{epsDepth} 
    }
  \end{center}
\end{figure}

\begin{figure}[ptb]
  \begin{center} 
  \includegraphics[scale=0.37]{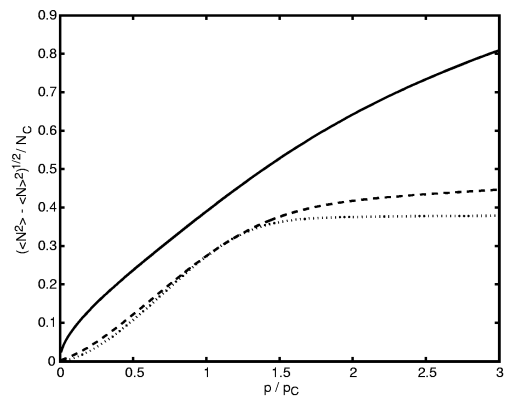}
  \caption{
    Standard deviation of the nondimensionalized cumulative depth
    versus nondimensional price, corresponding to
    Fig.~(\ref{epsDepth}). 
  \label{epsVarDepth} 
  }
  \end{center}
\end{figure}

The {\em mean depth profile}, i.e. the average number of shares per price
interval, and the mean cumulative depth profile are shown in
Fig.~\ref{epsDepth}, and the standard deviation of the cumulative profile is
shown in Fig.~\ref{epsVarDepth}.  Since the depth profile has units of
{\it shares/price}, nondimensional units of depth profile are $\hat{n}
= n p_c/N_c = n \delta / \alpha$.
The cumulative depth profile at any given time $t$ is defined as
\begin{equation}
N(p,t) = \sum_{\tilde{p}=0}^p n(\tilde{p},t) dp .
\label{cumDepth}
\end{equation}
This has units of shares and so in nondimensional terms is 
$\hat{N}(p) = N(p)/N_c = 2\delta N(p)/\mu = N(p) \epsilon / \sigma$.

In the high $\epsilon$ regime the annihilation rate due to market
orders is low (relative to $\delta \sigma$), and there is a
significant accumulation of orders at the best ask, so that the
average depth is much greater than zero at the midpoint.  The mean
depth profile is a concave function of price.  In the medium
$\epsilon$ regime the market order removal rate increases, depleting
the average depth near the best ask, and the profile is nearly linear
over the range $p / p_c \le 1$.  In the small $\epsilon$ regime the
market order removal rate increases even further, making the average
depth near the ask very close to zero, and the profile is a convex
function over the range $p / p_c \le 1$.

The standard deviation of the depth profile is shown in Fig.~\ref{epsVarDepth}.
We see that the standard deviation of the cumulative depth is
comparable to the mean depth, and that as $\epsilon$ increases, near
the midpoint there is a similar transition from convex to concave
behavior.

\begin{figure}[ptb]
  \begin{center}
    \includegraphics[scale=0.37]{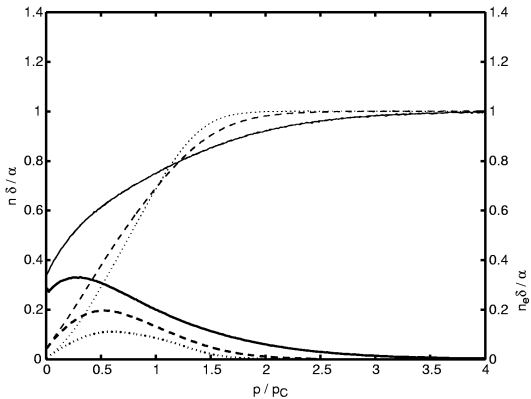}
    \caption{A comparison between the depth profiles and the 
      effective depth profiles as defined in the text, for different
      values of $\epsilon$. Heavy lines refer to the effective
      depth profiles $n_e$ and the light lines correspond to
      the depth profiles.
    \label{effprof} 
    }
  \end{center}
\end{figure}

The uniform order placement process seems at first glance 
one of the most unrealistic assumptions of our model, leading to
depth profiles with a finite asymptotic depth (which
also implies that there are an infinite number of orders in the book).
However, orders far away from the spread in the asymptotic 
region almost never get executed and thus do not affect the
market dynamics. To demonstrate this in Fig.~\ref{effprof} 
we show the comparison between the limit-order depth profile and the
depth $n_e$ of only those orders which eventually get
executed.\footnote{Note that the ratio $n_e/n$  is not the
same as the probability of filling orders (Fig. \ref{probfill})
because in that case the price $p/p_c$ refers to the distance of the
order from the midpoint at the time when it was placed.}
The density $n_e$ of executed orders decreases rapidly as a function
of the distance from the mid-price. Therefore we expect that near the
midpoint our results should be similar to alternative order placement
processes, as long as they also lead to an exponentially decaying profile
of executed orders (which is what we observe above).  However, to
understand the behavior further away from the midpoint we are also
working on enhancements that include more realistic order placement
processes grounded on empirical measurements of market data, as
summarized in section \ref{subsec:futurework}.

\subsubsection{Liquidity for market orders: The price impact function}

In this sub-section we study the {\it instantaneous price impact}
function $\phi(t, \omega, \tau \rightarrow 0)$.  This is defined as
the (logarithm of the) midpoint price shift immediately after the arrival
of a market order in the absence of any other events.  This should be
distinguished from the asymptotic price impact $\phi(t, \omega, \tau
\rightarrow \infty)$, which describes the permanent price shift.
While the permanent price shift is clearly very important, we do not
study it here.  The reader should bear in mind that all prices $p$, $a(t)$,
etc. are logarithmic.

The price impact function provides a measure of the liquidity for
executing market orders. (The liquidity for limit orders, in contrast,
is given by the probability of execution, studied in section
\ref{limitLiquidity}).  At any given time $t$, the 
instantaneous ($\tau = 0$) price impact function is the inverse of the
cumulative depth profile.  This follows immediately
from equations (\ref{volumeConstraint}) and (\ref{cumDepth}), which in
the limit $dp \rightarrow 0$ can be replaced by the continuum
transaction equation:
\begin{equation}
\omega= N(p,t) = \int_{0}^{p}n(\tilde{p},t)d\tilde{p}
\label{integral}
\end{equation}
This equation makes it clear that at any fixed $t$ the price impact can
be regarded as the inverse of the cumulative depth profile $N(p,t)$.
When the fluctuations are sufficiently small we can replace $n(p,t)$
by its mean value $n(p) = \langle n(p,t) \rangle$.  In general, however, the
fluctuations can be large, and the average of the inverse is not equal
to the inverse of the average.  There are corrections based on higher
order moments of the depth profile, as given in the moment expansion
derived in Appendix
\ref{momentExpansion}.  Nonetheless, the inverse of the mean
cumulative depth provides a qualitative approximation that gives
insight into the behavior of the price impact function.  (Note that
everything becomes much simpler using medians, since the median of the
cumulative price impact function is exactly the inverse of the median
price impact, as derived in Appendix \ref{momentExpansion}).

\begin{figure}[ptb]
  \begin{center} 
    \includegraphics[scale=0.37]{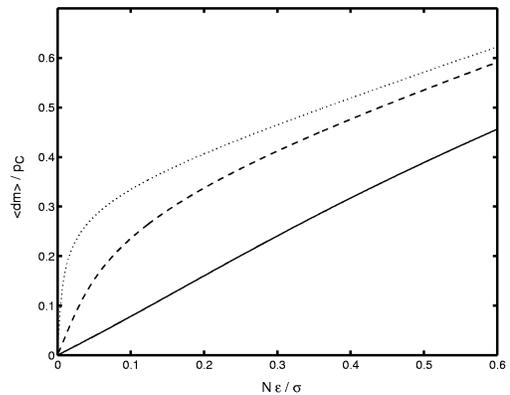}
  \caption{
    The average price impact corresponding to the results in
    Fig.~(\ref{epsDepth}). The average instantaneous movement of the
    nondimensional mid-price, $\langle dm \rangle/ p_c$ caused by an order 
    of size $N/N_c = N \epsilon/\sigma$.  $\epsilon = 0.2$ (solid),
    $\epsilon = 0.02$ (dash), $\epsilon = 0.002$ (dot).  
  \label{epsImpact}
  }
  \end{center}
\end{figure}
\begin{figure}[ptb]
  \begin{center}
  \includegraphics[scale=0.37]{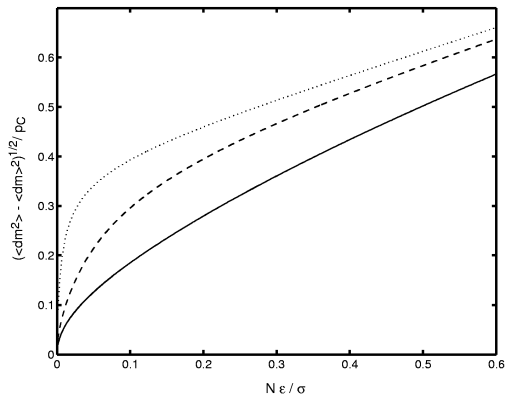}
  \caption{
    The standard deviation of the instantaneous price impact $dm / p_c$
    corresponding to the means in Fig.~\ref{epsImpact}, as a function
    of normalized order size $\epsilon N / \sigma$.  $\epsilon = 0.2$
    (solid), $\epsilon = 0.02$ (dash), $\epsilon = 0.002$ (dot).
  \label{epsVarImpact}
  }
 \end{center}
\end{figure}
Mean price impact functions are shown in Fig.~\ref{epsImpact} and the
standard deviation of the price impact is shown in
Fig.~\ref{epsVarImpact}. The price impact exhibits very large fluctuations 
for all values of $\epsilon$: The 
standard deviation has the same order of magnitude as the mean or even greater for 
small $N\epsilon / \sigma$ values.
Note that these are actually {\it virtual
price impact} functions.  That is, to explore the behavior of the
instantaneous price impact for a wide range of order sizes, we
periodically compute the price impact that an order of a given size
would have caused at that instant, if it had been submitted.  We have
checked that real price impact curves are the same, but they require a much
longer time to accumulate reasonable statistics.

\begin{figure}[ptb]
  \begin{center}
\includegraphics[scale=0.37]{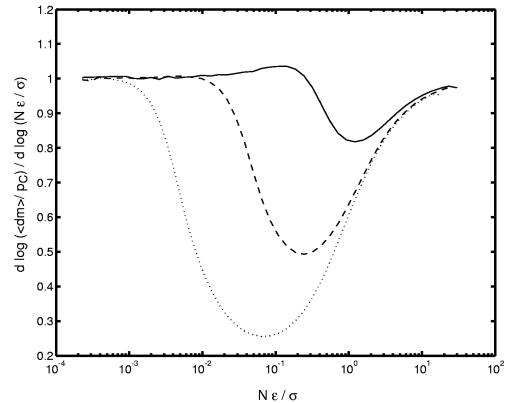}
  \caption{
    Derivative of the nondimensional mean mid-price movement,
    with respect to logarithm of the nondimensional order size 
    $N/N_c = N \epsilon/\sigma$, obtained from the price impact curves in 
    Fig.~\ref{epsImpact}.
  \label{epsImpactSlope}
  }
 \end{center}
\end{figure}

One of the interesting results in Fig.~\ref{epsImpact} is the scale of
the price impact.  The price impact is measured relative to the
characteristic price scale $p_c$, which as we have mentioned earlier
is roughly equal to the mean spread.  As we will argue in relation to
Fig.~\ref{epsImpactSlope}, the range of nondimensional shares shown on
the horizontal axis spans the range of reasonable order sizes.  This
figure demonstrates that throughout this range the price is the order
of magnitude (and typically less than) the mean spread size.

Due to the accumulation of orders at the ask in the large $\epsilon$
regime, for small $p$ the mean price impact is roughly linear.  This
follows from equation (\ref{integral}) under the assumption that
$n(p)$ is constant.  In the medium $\epsilon$ regime, under the
assumption that the variance in depth can be neglected, the mean price
impact should increase as roughly $\omega^{1/2}$.  This follows from
equation (\ref{integral}) under the assumption that $n(p)$ is linearly
increasing and $n(0) \approx 0$.  (Note that we see this as a crude
approximation, but there can be substantial corrections caused by the
variance of the depth profile).  Finally, in the small $\epsilon$
regime the price impact is highly concave, increasing much slower than
$\omega^{1/2}$.  This follows because $n(0) \approx 0$ and the depth
profile $n(p)$ is convex.

To get a better feel for the functional form of the price impact
function, in Fig.~\ref{epsImpactSlope} we numerically differentiate it
versus log order size, and plot the result as a function of the
appropriately scaled order size.  (Note that because our prices are
logarithmic, the vertical axis already incorporates the logarithm).
If we were to fit a local power law approximation to the function at
each price, this corresponds to the exponent of that power law near
that price.  Notice that the exponent is almost always less than one,
so that the price impact is almost always concave.  Making the
assumption that the effect of the variance of the depth is not too
large, so that equation (\ref{integral}) is a good assumption, the
behavior of this figure can be understood as follows: For $N/N_c
\approx 0$ the price impact is dominated by $n(0)$ (the constant term
in the average depth profile) and so the logarithmic slope of the
price impact is always near to one.  As $N/N_c$ increases, the
logarithmic slope is driven by the shape of the average depth profile,
which is linear or convex for smaller $\epsilon$, resulting in concave
price impact.  For large values of $N/N_c$, we reach the asymptotic
region where the depth profile is flat (and where our model is invalid
by design).  Of course, there can be deviations to this behavior
caused by the fact that the mean of the inverse depth profile is not
in general the inverse of the mean, i.e.  $\langle N^{-1}(p) \rangle
\neq \langle N(p) \rangle^{-1}$ (see App.~\ref{momentExpansion}).

To compare to real data, note that $N/N_c = N \epsilon/\sigma$.
$N/\sigma$ is just the order size in shares in relation to the average
order size, so by definition it has a typical value of one.  For the
London Stock Exchange, we have found that typical values of $\epsilon$
are in the range $0.001 - 0.1$.  For a typical range of order sizes
from $100 - 100,000$ shares, with an average size of $10,000$ shares,
the meaningful range for $N/N_c$ is therefore roughly $10^{-5}$ to
$1$.  In this range, for small values of $\epsilon$ the exponent can
reach values as low as $0.2$.  This offers a possible explanation for
the previously mysterious concave nature of the price impact function,
and contradicts the linear increase in price impact based on the
naive argument presented in the introduction.

\subsubsection{Spread}
\label{spreadSection}

\begin{figure}[ptb]
  \begin{center} 
  \includegraphics[scale=0.37]{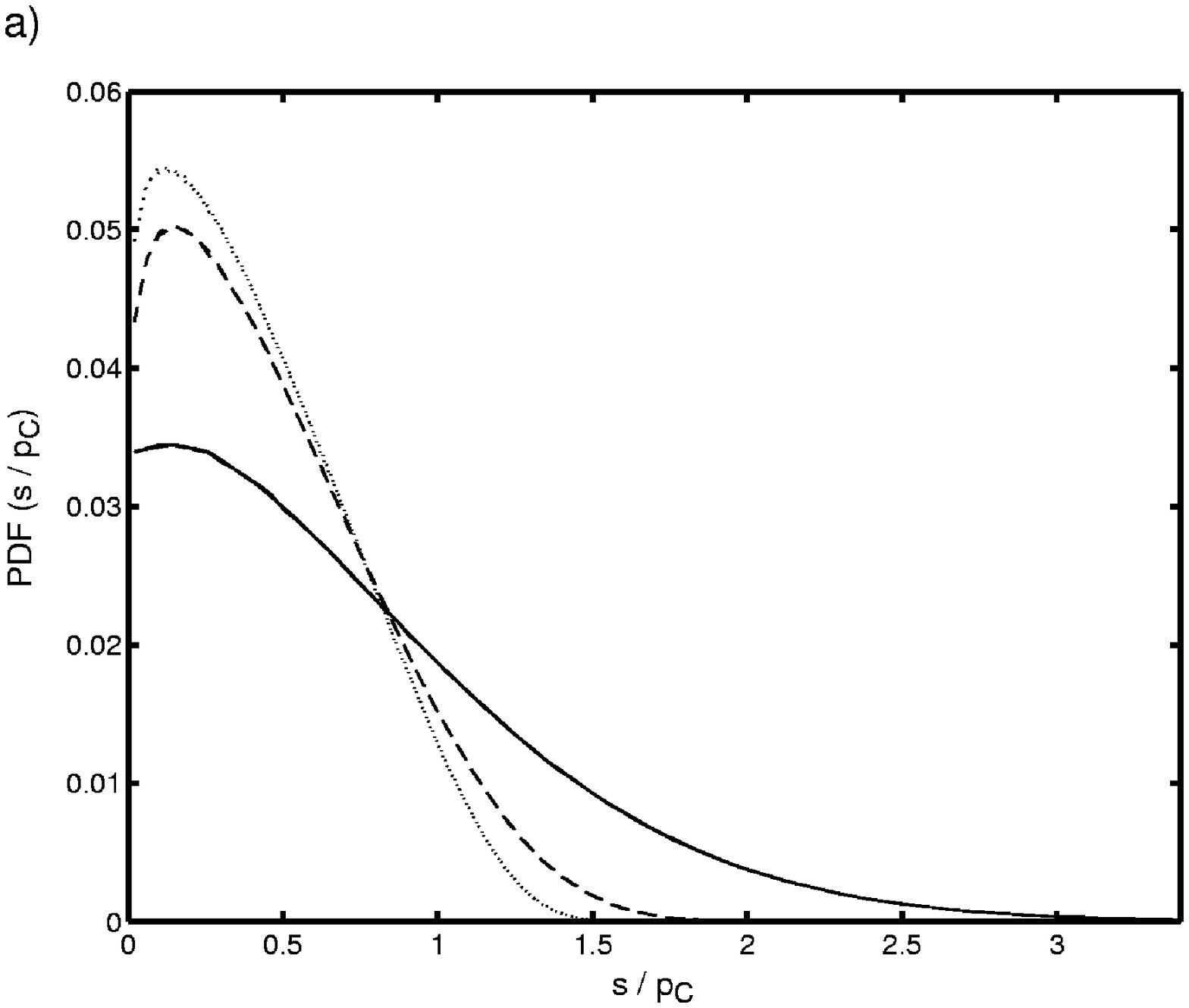}
  \includegraphics[scale=0.37]{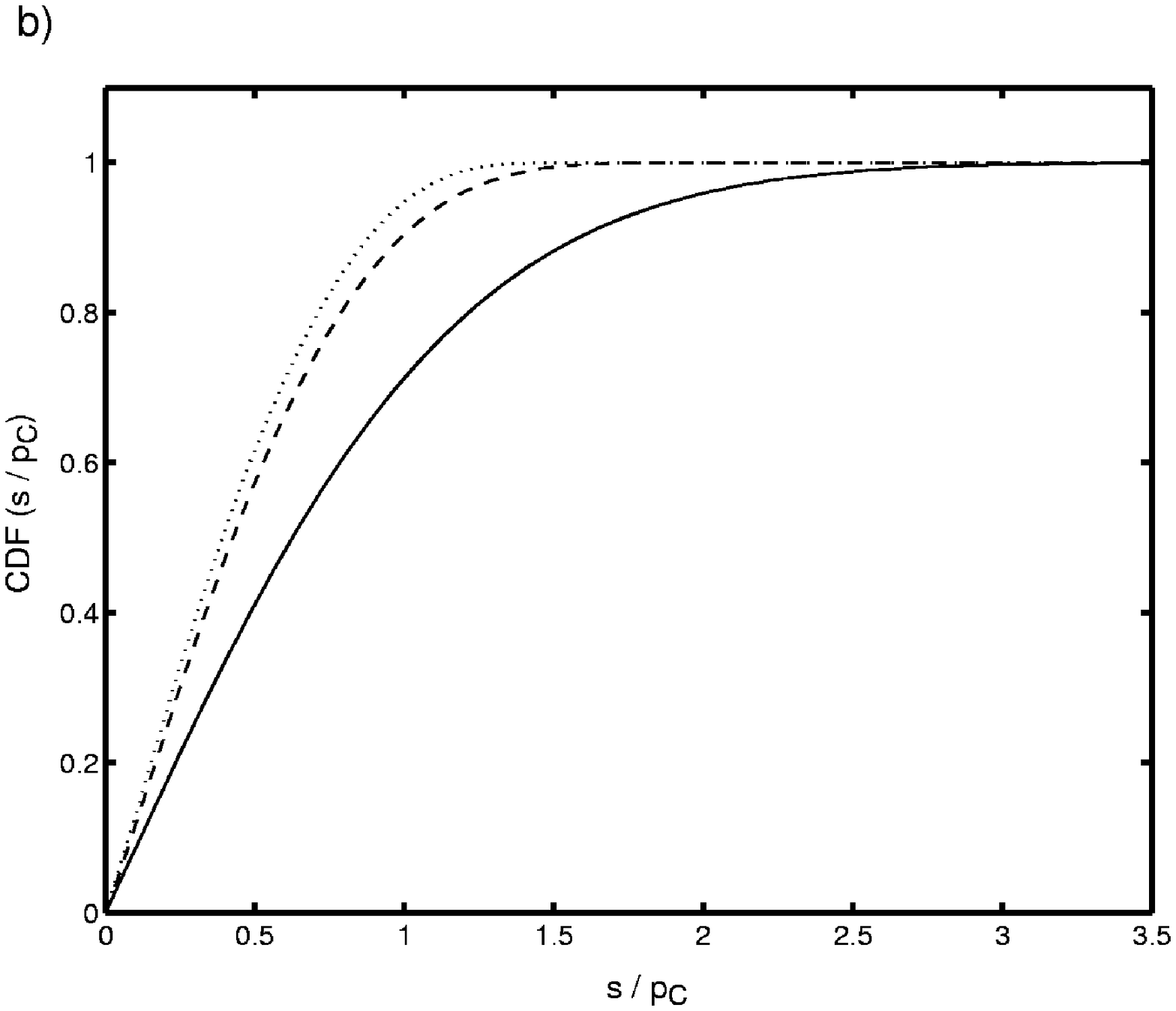}
  \caption{
    The probability density function (a), and cumulative distribution
    function (b) of the nondimensionalized bid-ask spread $s / p_c$,
    corresponding to the results in Fig.~(\ref{epsDepth}).
    $\epsilon = 0.2$ (solid), $\epsilon = 0.02$ (dash), $\epsilon =
    0.002$ (dot).
  \label{epsSpread} 
  }  
  \end{center}
\end{figure}

The probability density of the spread is shown in Fig.~\ref{epsSpread}.
This shows that the probability density is substantial at $s/p_c = 0$.
(Remember that this is in the limit $dp \rightarrow 0$).  The
probability density reaches a maximum at a value of the
spread approximately $0.2 p_c$, and then decays.  It
might seem surprising at first that it decays more slowly for large
$\epsilon$, where there is a large accumulation of orders at the ask.
However, it should be borne in mind that the characteristic price $p_c
= \mu/\alpha$ depends on $\epsilon$.  Since $\epsilon =
2\delta\sigma/\mu$, by eliminating $\mu$ this can be written $p_c = 2
\sigma \delta / (\alpha
\epsilon)$.  Thus, holding the other parameters fixed, large
$\epsilon$ corresponds to small $p_c$, and vice versa.  So in fact,
the spread is very small for large $\epsilon$, and large for small
$\epsilon$, as expected.  The figure just shows the small corrections
to the large effects predicted by the dimensional scaling relations.

For large $\epsilon$ the probability density of the spread decays
roughly exponentially moving away from the midpoint.  This is because
for large $\epsilon$ the fluctuations around the mean depth are
roughly independent.  Thus the probability for a market order to
penetrate to a given price level is roughly the probability that all
the ticks smaller than this price level contain no orders, which gives
rise to an exponential decay.  This is no longer true for small
$\epsilon$.  Note that for small $\epsilon$ the probability
distribution of the spread becomes insensitive to $\epsilon$, i.e.
the nondimensionalized distribution for $\epsilon = 0.02$ is nearly
the same as that for $\epsilon = 0.002$.

\begin{figure}[ptb]
  \begin{center} 
  \includegraphics[scale=0.37]{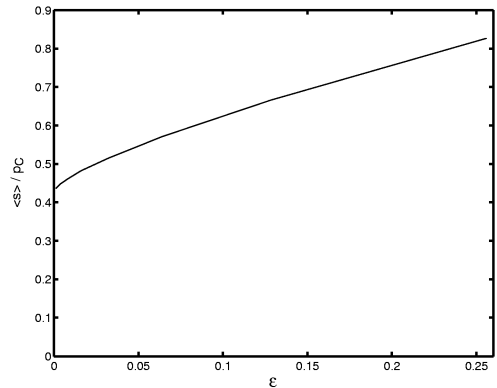}
  \caption{
     The mean value of the spread in nondimensional units $\hat{s} = s
     / p_c$ as a function of $\epsilon$.  This demonstrates that the
     spread only depends weakly on $\epsilon$, indicating that the
     prediction from dimensional analysis given in 
     table (\ref{tab:dim_scaling}) is a reasonable approximation.
  \label{meanSpread}.
  }
  \end{center}
\end{figure}

It is apparent from Fig.~\ref{epsSpread} that in nondimensional units
the mean spread increases with $\epsilon$.  This is confirmed in
Fig.~\ref{meanSpread}, which displays the mean value of the spread as
a function of $\epsilon$.
The mean spread increases monotonically with $\epsilon$.  It depends
on $\epsilon$ as roughly a constant (equal to approximately 0.45 in
nondimensional coordinates) plus a linear term whose slope is rather
small.  We believe that for most financial instruments $\epsilon <
0.3$.  Thus the variation in the spread caused by varying $\epsilon$
in the range $0 < \epsilon < 0.3$ is not large, and the 
dimensional analysis based only on rate parameters given in table
\ref{tab:params_scaling} is a good approximation.  
We get an accurate prediction of the $\epsilon$ dependence across the
full range of $\epsilon$ from the Independent Interval Approximation
technique derived in section ~\ref{subsec:IIA}, as shown in
Fig.~\ref{spread_compare}.

\subsubsection{Volatility and price diffusion
\label{subsec:longterm}}

The price diffusion rate, which is proportional to the square of the
volatility, is important for determining risk and is a property of
central interest.  From dimensional analysis in terms of the order
flow rates the price diffusion rate has units of $price^2/time$, and
so must scale as $\mu^{2}\delta/{\alpha}^2$.  We can also make a crude
argument for this as follows: The dimensional estimate of the spread (see Table
\ref{tab:params_scaling}) is $\mu / 2{\alpha}$. Let this be the 
characteristic step size of a random walk, and let the step frequency
be the characteristic time $1/\delta$ (which is the average lifetime
for a share to be canceled).  This argument also gives the above
estimate for the diffusion rate.  However, this is not correct in the presence
of negative autocorrelations in the step sizes.  The numerical results
make it clear that there are important $\epsilon$-dependent
corrections to this result, as demonstrated below.

\begin{figure}[ptb] 
  \begin{center} %
  \includegraphics[scale=0.37]{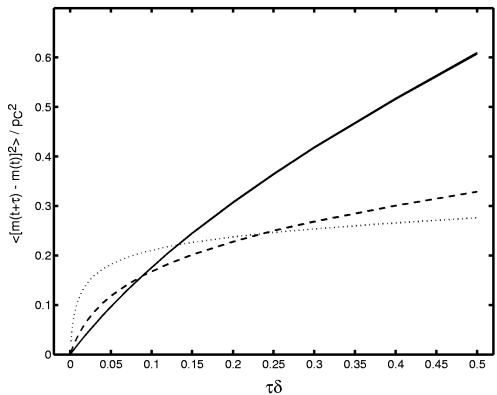}
    \caption{ 
      The variance of the change in the nondimensionalized midpoint
      price versus the nondimensional time delay interval $\tau \delta$.
      For a pure random walk this would be a straight line whose slope is 
      the diffusion 
      rate, which is proportional to the square of the volatility.  The 
      fact that the slope is steeper for short times comes from the nontrivial
      temporal persistence of the order book.  The three cases
      correspond to Fig.~\ref{epsDepth}: $\epsilon = 0.2$ (solid),
      $\epsilon = 0.02$ (dash), $\epsilon = 0.002$ (dot).
    \label{epsVarVsTau}
    }
    \end{center}
\end{figure}

In Fig.~\ref{epsVarVsTau} we plot simulation results for the
variance of the change in the midpoint price at timescale $\tau$,
$\mbox{Var} \left( m \left( t+\tau \right) - m \left( t \right)
\right)$.  The slope is the diffusion rate, which at any fixed timescale is 
proportional to the square of the volatility. It appears that there
are at least two timescales involved, with a faster diffusion rate for
short timescales and a slower diffusion rate for long timescales. Such
anomalous diffusion is not predicted by mean-field analysis.
Simulation results show that the diffusion rate is correctly described
by the product of the estimate from dimensional analysis based on
order flow parameters alone, $\mu^{2}\delta/{\alpha}^2$, and a
$\tau$-dependent power of the nondimensional granularity parameter
$\epsilon = 2 \delta\sigma/\mu$, as summarized in table
\ref{tab:params_scaling}.  We cannot currently explain why this power
is $-1/2$ for short term diffusion and $1/2$ for long-term diffusion.
However, a qualitative understanding can be gained based on the
conservation law we derive in Section~\ref{subsec:frames_marginals}.  A
discussion of how this relates to price diffusion is given in
Section~\ref{subsec:ren_diff}.

Note that the temporal structure in the diffusion process also implies
non-zero autocorrelations of the midpoint price $m(t)$.  This
corresponds to weak negative autocorrelations in price differences
$m(t) - m(t-1)$ that persist for timescales until the variance
vs. $\tau$ becomes a straight line. The timescale depends on
parameters, but is typically the order of 50 market order arrival
times.  This temporal structure implies that there exists an arbitrage
opportunity which, when exploited, would make prices more random and
the structure of the order flow non-random.

\subsubsection{Liquidity for limit orders: Probability and time to fill.}
\label{limitLiquidity}

\begin{figure}[ptb]
  \begin{center}
    \includegraphics[scale=0.37]{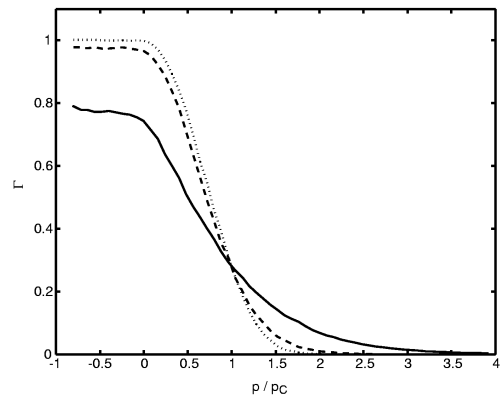}
    \caption{The probability $\Gamma$ for filling a limit order
      placed at a price $p/p_c$ where $p$ is calculated from the
      instantaneous mid-price at the time of placement. The three cases
      correspond to Fig.~\ref{epsDepth}: $\epsilon = 0.2$ (solid),
      $\epsilon = 0.02$ (dash), $\epsilon = 0.002$ (dot).
    \label{probfill} 
    }
 \end{center}
\end{figure}

The liquidity for limit orders depends on the probability that they
will be filled, and the time to be filled.  This obviously depends on
price: Limit orders close to the current transaction prices are more
likely to be filled quickly, while those far away have a lower
likelihood to be filled.  Fig. \ref{probfill} plots the probability
$\Gamma$ of a limit order being filled versus the nondimensionalized
price at which it was placed (as with all the figures in this section,
this is shown in the midpoint-price centered
frame). Fig. \ref{probfill} shows that in nondimensional coordinates
the probability of filling close to the bid for sell limit orders (or
the ask for buy limit orders) decreases as $\epsilon$ increases. For
large $\epsilon$, this is less than $1$ even for negative prices.
This says that even for sell orders that are placed close to the best
bid there is a significant chance that the offer is deleted before
being executed.  This is not true for smaller values of $\epsilon$,
where $\Gamma(0) \approx 1$.  Far away from the spread the fill
probabilities as a function of $\epsilon$ are reversed, i.e. the
probability for filling limit orders increases as $\epsilon$
increases. The crossover point where the fill probabilities are
roughly the same occurs at $p \approx p_c$.  This is consistent with
the depth profile in Fig.  \ref{epsDepth} which also shows that depth
profiles for different values of $\epsilon$ cross at about $p \sim p_c
$.

\begin{figure}[ptb]
  \begin{center}
    \includegraphics[scale=0.37]{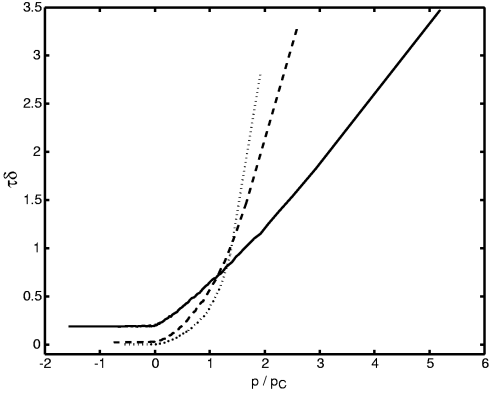}
    \caption{ The average time $\tau$ nondimensionalized
      by the rate $\delta$,  to fill a limit order
      placed at a distance $p/p_c$ from the
      instantaneous mid-price.
    \label{timefill} 
    }
  \end{center}
\end{figure}

Similarly Fig \ref{timefill} shows the average time $\tau$
taken to fill an order placed at a distance $p$ from
the instantaneous mid-price. 
Again we see that though the average time is larger
at larger values of $\epsilon$ for small $ p/p_c$, this
behaviour reverses at $ p \sim p_c$.

\subsection{Varying tick size $dp/p_c$}

\begin{figure}[ptb]
  \begin{center} 
  \includegraphics[scale=0.35]{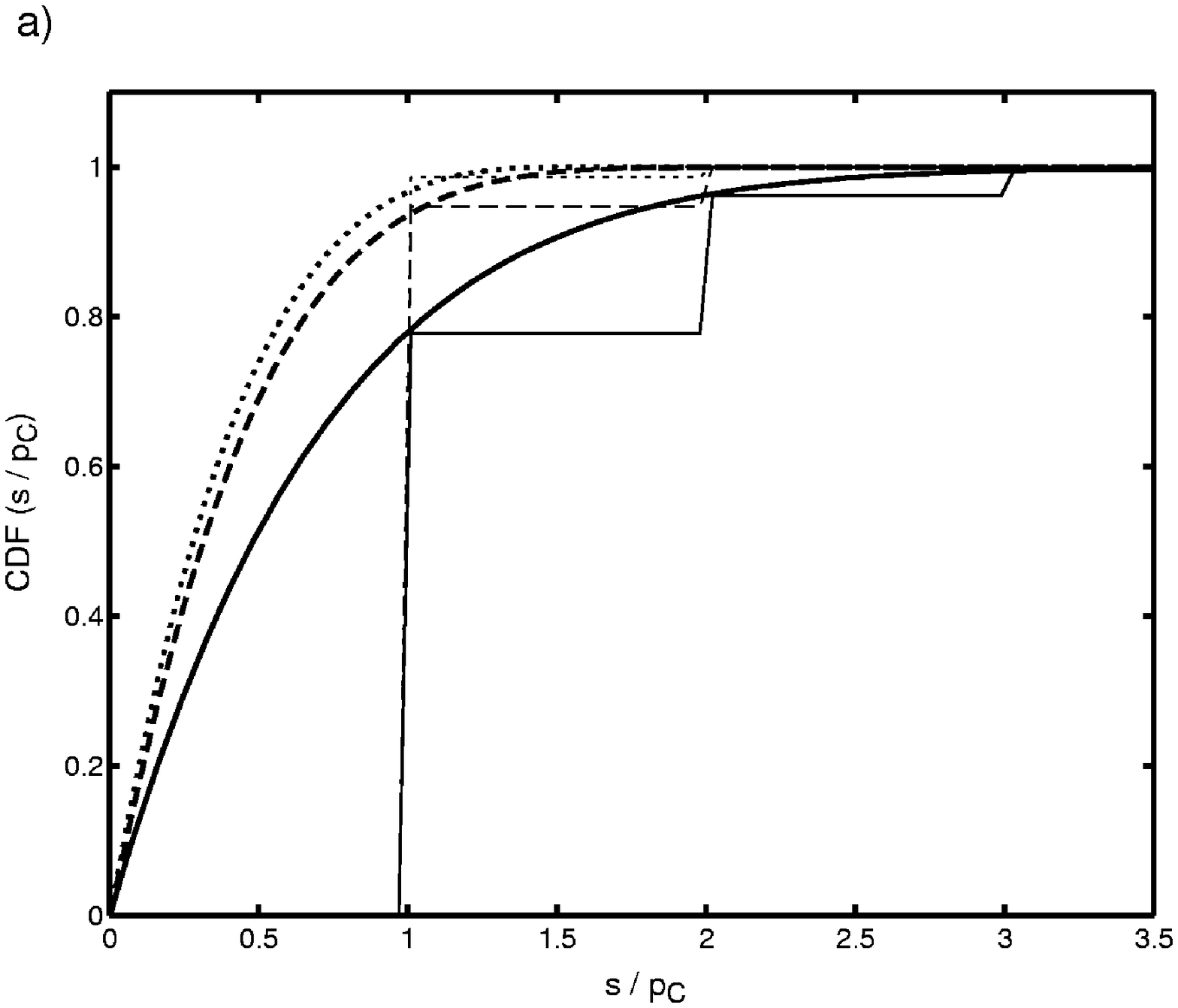}
  \includegraphics[scale=0.35]{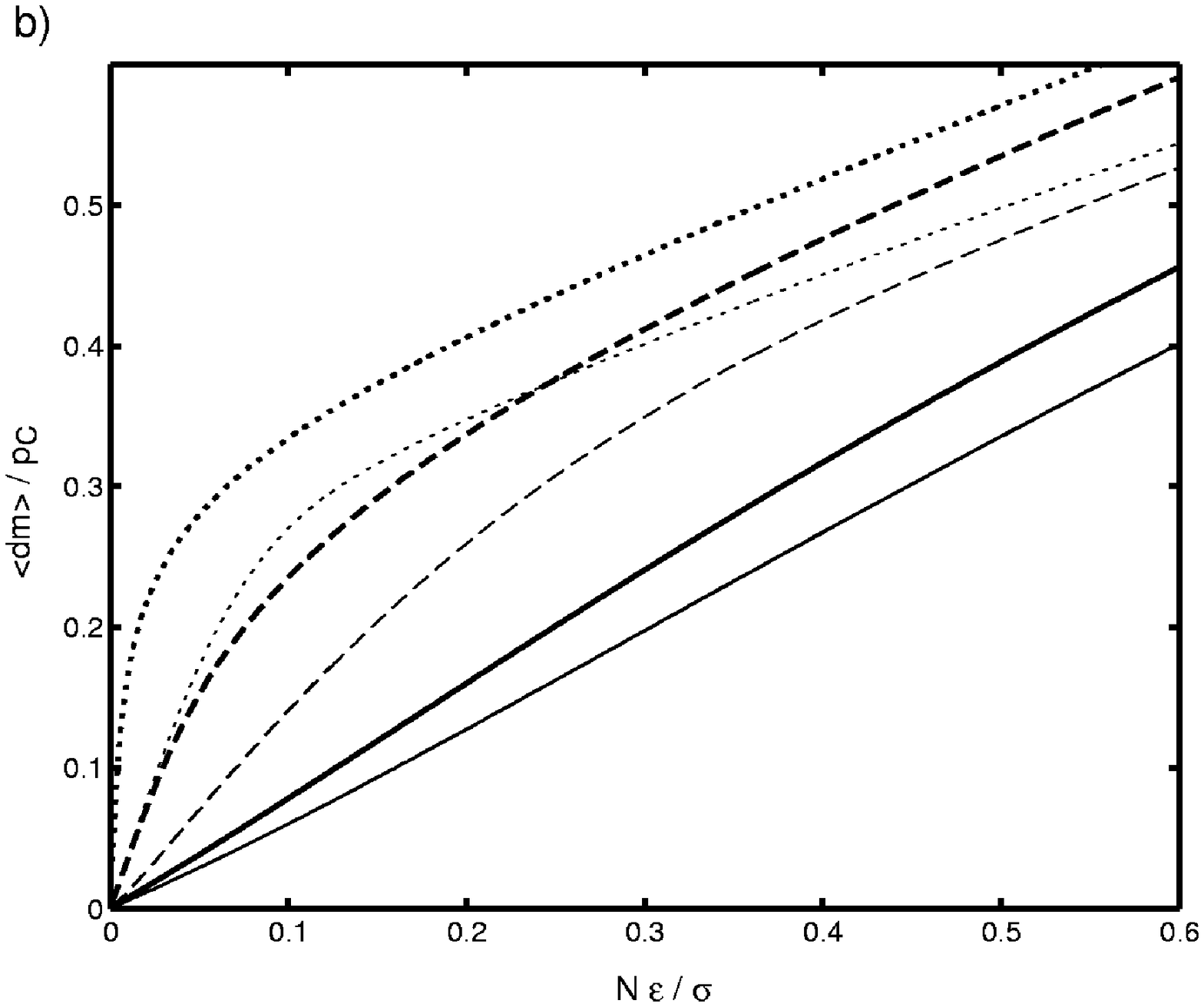}
  \includegraphics[scale=0.35]{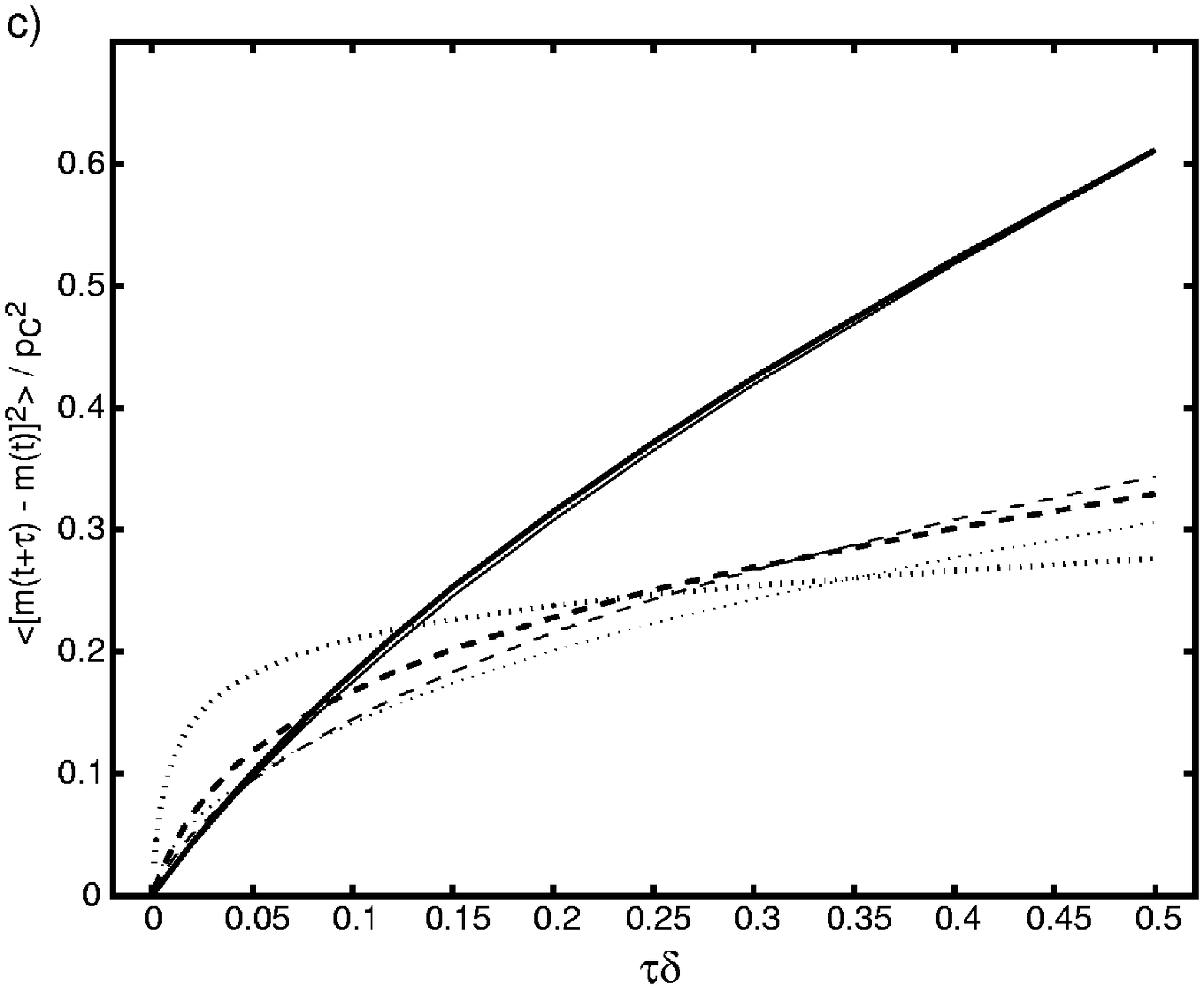}
  \caption{
    Dependence of market properties on tick size.  Heavy lines are $dp
    / p_c \rightarrow 0$; light lines are $dp / p_c = 1$.  Cases
    correspond to Fig.~\ref{epsDepth}, with $\epsilon = 0.2$ (solid),
    $\epsilon = 0.02$ (dash), $\epsilon = 0.002$ (dot).  (a) is the
    cumulative distribution function for the nondimensionalized
    spread.  (b) is instantaneous nondimensionalized price impact,
    (c) is diffusion of the nondimensionalized midpoint shift,
    corresponding to Fig.~\ref{epsVarVsTau}.  
  \label{finiteTicks} 
  }  
  \end{center}
\end{figure}

The dependence on discrete tick size $dp / p_c$, of the cumulative
distribution function for the spread, instantaneous price impact, and
mid-price diffusion, are shown in Fig.~\ref{finiteTicks}.  We chose an
unrealistically large value of the tick size, with $dp/p_c = 1$, to
show that, even with very coarse ticks, the qualitative changes in
behavior are typically relatively minor.

Fig.~\ref{finiteTicks}(a) shows the cumulative density function of the
spread, comparing $dp/p_c = 0$ and $dp/p_c = 1$.  It is apparent from
this figure that the spread distribution for coarse ticks
``effectively integrates'' the distribution in the limit $dp
\rightarrow 0$.  That is, at
integer tick values the mean cumulative depth profiles roughly match,
and in between integer tick values, for coarse ticks the probability
is smaller.  This happens for the obvious reason that coarse ticks
quantize the possible values of the spread, and place a lower
limit of one tick on the value the spread can take.  The shift in the
mean spread from this effect is not shown, but it is consistent with
this result; there is a constant offset of roughly $1/2$ tick.

The alteration in the price impact is shown in
Fig.~\ref{finiteTicks}(b).  Unlike the spread distribution, the average price
impact varies continuously. Even though the tick size is quantized, we
are averaging over many events and the probability of a price impact
of each tick size is a continuous function of the order size.  
Large tick size consistently lowers the price impact.  
The price impact rises more slowly for small $p$, but
is then similar except for a downward translation.

The effect of coarse ticks is less trivial for mid-price diffusion, as
shown in Fig.~\ref{finiteTicks}(c).  At $\epsilon = 0.002$, coarse
ticks remove most of the rapid short-term volatility of the midpoint,
which in the continuous-price case arises from price fluctuations
smaller than $dp/p_c = 1$.  This lessens the negative autocorrelation
of midpoint price returns, and reduces the anomalous diffusion.  At
$\epsilon = 0.2$, where both early volatility and late negative
autocorrelation are smaller, coarse ticks have less effect.  The net
result is that the mid-price diffusion becomes less sensitive to the
value of $\epsilon$ as tick size increases, and there is less
anomalous price diffusion.

\section{Theoretical analysis}
\label{analysis}

\subsection{Summary of analytic methods}

We have investigated this model analytically using two approaches.
The first one is based on a master equation, given in
Section~\ref{subsec:ME}.  This approach works best in the midpoint
centered frame. Here we attempt to solve directly for the average
number of shares at each price tick as a function of price.  The
midpoint price makes a random walk with a nonstationary distribution.
Thus the key to finding a stationary analytic solution for the average
depth is to use comoving price coordinates, which are centered on a
reference point near the center of the book, such as the midpoint or
the best bid.  In the first approximation, fluctuations about the mean
depth at adjacent prices are treated as independent.  This allows us
to replace the distribution over depth profiles with a simpler
probability density over occupation numbers $n$ at each $p$ and $t$.
We can take a continuum limit by letting the tick size $dp$ become
infinitesimal.  With finite order flow rates, this gives vanishing
probability for the existence of more than one order at any tick as
$dp \rightarrow 0$.  This is described in detail in
section~\ref{subsec:ctm_limit}.  With this approach we are able to
test the relevance of correlations as a function of the parameter
$\epsilon$ as well as predict the functional dependence of the
cumulative distribution of the spread on the depth profile. It is seen
that correlations are negligible for large values of $\epsilon
(\epsilon \sim 0.2)$ while they are very important for small values
$(\epsilon \sim 0.002)$.

Our second analytic approach which we term the {\it Independent
Interval Approximation (IIA)} is most easily carried out in the
bid-centered frame and is described in section ~\ref{subsec:IIA}. This
approach uses a different representation, in which the solution is
expressed in terms of the empty intervals between non-empty price
ticks. The system is characterized at any instant of time by a set of
intervals $\{...x_{-1},x_0,x_1,x_2...\}$ where for example $x_0$ is
the distance between the bid and the ask (the spread), $x_{-1}$ is the
distance between the second buy limit order and the bid and so on (see Fig.
 ~\ref{fig:price_bins_label}).
Equations are written for how a given interval varies in time. Changes
to adjacent intervals are related, giving us an infinite set of
coupled non-linear equations. However using a mean-field approximation
we are able to solve the equations, albeit only numerically.  Besides
predicting how the various intervals (for example the spread) vary
with the parameters, this approach also predicts the depth profiles as
a function of the parameters. The predictions from the {\it IIA} are
compared to data from numerical simulations, in Section~\ref{subsec:xsimulations}. They match very well for large $\epsilon$ 
and less well for smaller values of $\epsilon$.
The {\it IIA}  can also be modified to
incorporate various extensions to the model, as mentioned
in Section~\ref{subsec:xsimulations}.

In both approaches, we use a mean field approximation to get a
solution.  The approximation basically lies in assuming that
fluctuations in adjacent intervals (which might be adjacent price
ranges in the master equation approach or adjacent empty intervals in
the IIA) are independent. Also, both approaches
are most easily tractable only in the continuum limit
$dp \rightarrow 0$, when every tick has at most
only one order. They may however be extended to
general tick size as well. This is explained in the
appendix for the Master Equation approach.

Because correlations are important for small $\epsilon$,
both methods  work well mostly in the large
$\epsilon$ limit, though qualitative aspects of small $\epsilon$
behavior may also be gleaned from them.
Unfortunately, at least based on our preliminary
investigation of London Stock Exchange data, it seems that it is this 
small $\epsilon$ limit that real
markets may tend more towards. So our
approximate solutions may not be as useful as we would like.
Nonetheless, they do provide some conceptual insights into what
determines depth and price impact.  

In particular, we find that the shape of the mean depth profile
depends on a single parameter $\epsilon$, and that the relative sizes
of its first few derivatives account for both the order
size-dependence of the market impact, and the renormalization of the
midpoint diffusivity.  A higher relative rate of market versus limit
orders depletes the center of the book, though less than the classical
estimate predicts.  This leads to more concave impact (explaining
Fig.~\ref{epsImpactSlope}) and faster short-term diffusivity.
However, the orders pile up more quickly (versus classically
nondimensionalized price) with distance from the midpoint, causing the
rapid early diffusion to suffer larger mean reversion. These are the
effects shown in Fig.~\ref{epsVarVsTau}.  We will elaborate on the
above remarks in the following sections, however, the qualitative
relation of impact to midpoint autocorrelation supplies a potential
interpretation of data, which may be more robust than details of the
model assumptions or its quantitative results.

Both of the treatments described above are approximations.  We can
derive an exact global conservation law of order placement and removal
whose consequences we elaborate in section
\ref{subsec:frames_marginals}.  This conservation law must be
respected in any sensible analysis of the model, giving us a check on
the approximations.  It also provides some insight into the anomalous
diffusion properties of this model.

\subsection{Characterizing limit-order books: dual coordinates}
\label{subsec:defns}

We begin with the assumption of a {\em price space}.  Price is a
dimensional quantity, and the space is divided into bins of length
$dp$ representing the ticks, which may be finite or infinitesimal.
Prices are then discrete or continuous-valued, respectively.

\begin{figure}[t]
\epsfxsize=2.75in 
\epsfbox{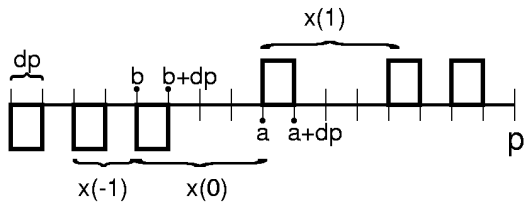}
\caption{
  The price space and order profile.  $n \left( p , t \right)$ has
  been chosen to be $0$ or $\pm 1$, a restriction that will be
  convenient later.  Price bins are labeled by their lower boundary
  price, and intervals $x \left( N \right)$ will be defined below.  
  \label{fig:price_bins_label}
}
\end{figure}

Statistical properties of interest are computed from temporal
sequences or ensembles of {\em limit-order book configurations}.  If
$n$ is the variable used to denote the number of shares from limit
orders in some bin $\left( p, p+dp \right)$ at the beginning $t$ of an
elementary time interval, a configuration is specified by a function
$n \left( p , t \right)$.  It is convenient to take $n$ positive for
sell limit orders, and negative for buy limit orders.  Because the
model dynamics precludes crossing limit orders, there is in general a
highest instantaneous buy limit-order price, called the bid $b\left( t
\right)$, and a lowest sell limit-order price, the ask $a\left( t
\right)$, with $b\left( t \right) < a\left( t \right)$ always.  The
{\em midpoint price}, defined as $m\left( t \right) \equiv \left[
a\left( t \right) + b\left( t \right) \right] / 2$, may or may not be
the price of any actual bin, if prices are discrete  ($m\left( t
\right)$ may be a half-integer multiple of $dp$).  These quantities
are diagrammed in Fig.~\ref{fig:price_bins_label}.  

\begin{figure}[t]
\includegraphics[height=2in, width=2.5in]{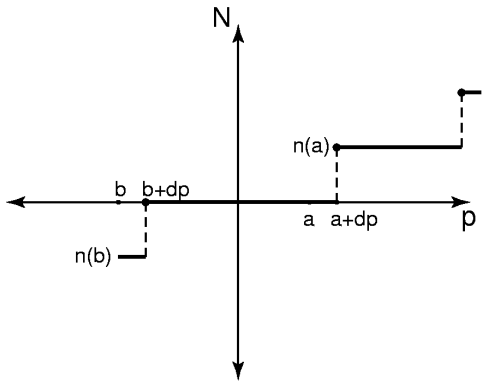}
\caption{
  The accumulated order number $N \left( p,t \right)$.  $N \left( a,t
  \right) \equiv 0$, because contributions from all bins cancel in the
  two sums.  $N$ remains zero down to $b \left( t \right) +dp$,
  because there are no uncanceled, nonzero terms.  $N \left( b,t
  \right)$ becomes negative, because the second sum in
  Eq.~(\ref{eq:cumulative_define}) now contains $n \left( b,t
  \right)$, not canceled by the first.
  \label{fig:N_of_p_graph}
}
\end{figure}

An equivalent specification of a limit-order book configuration is
given by the cumulative order count
\begin{equation}
  N \left( p,t \right) \equiv 
  \sum_{-\infty}^{p-dp} \left| n \left( p,t \right) \right| - 
  \sum_{-\infty}^{a-dp} \left| n \left( p,t \right) \right| , 
\label{eq:cumulative_define}
\end{equation}
where $-\infty$ denotes the lower boundary of the price space, whose
exact value must not affect the results.  (Because by definition there
are no orders between the bid and ask, the bid could equivalently have
been used as the origin of summation.  Because price bins will be
indexed here by their lower boundaries, though, it is convenient here
to use the ask.)  The absolute values have been placed so that $N$,
like $n$, is negative in the range of buy orders and positive in the
range of sells.  The construction of $N \left( p,t \right)$ is
diagrammed in Fig.~\ref{fig:N_of_p_graph}.

\begin{figure}[t]

\includegraphics[height=2in, width=2.5in]{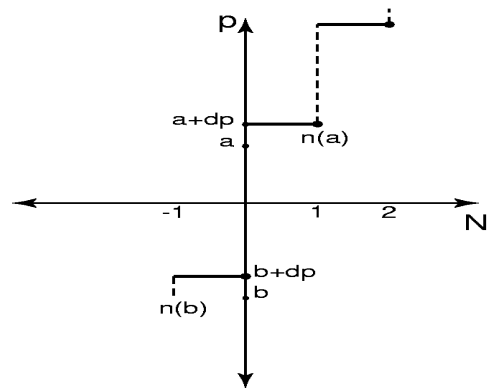}

\caption{
  The inverse function $p \left( N,t \right)$.  The function is in
  general defined only on discrete values of $N$, so this domain is
  only invariant when order size is fixed, a convenience that will be
  assumed below.  Between the discrete domain, and the definition of
  $p$ as a maximum, the inverse function effectively interpolates
  between vertices of the reflected image of $N \left( p,t \right)$,
  as shown by the dotted line.
  \label{fig:p_of_N_graph}
}
\end{figure}

In many cases of either sparse orders or infinitesimal $dp$, with
fixed order size (which we may as well define to be one share) there
will be either zero or one share in any single bin, and
Eq.~(\ref{eq:cumulative_define}) will be invertible to an equivalent
specification of the limit-order book configuration
\begin{equation}
  p \left( N,t \right) \equiv 
  \mbox{max}
  \left\{ 
    p
  \mid 
    N \left( p,t \right) = N 
  \right\} , 
\label{eq:p_of_N_def}
\end{equation}
shown in Fig.~\ref{fig:p_of_N_graph}.  (Strictly, the inversion may be
performed for any distribution of order sizes, but the resulting
function is intrinsically discrete, so its domain is only invariant
when order size is fixed.  To give $p \left( N,t \right)$ the
convenient properties of a well-defined function on an invariant
domain, this will be assumed below.)

With definition~(\ref{eq:p_of_N_def}), $p \left( 0,t \right) \equiv
a\left( t \right)$, $p \left( -1,t \right) \equiv b\left( t \right)$,
and one can define the intervals between orders as
\begin{equation}
  x \left( N,t \right) \equiv 
  p \left( N,t \right) -
  p \left( N-1,t \right) .
\label{eq:x_of_N_def}
\end{equation}
Thus $x \left( 0,t \right) = a\left( t \right) - b\left( t \right)$,
the instantaneous bid-ask spread.  The lowest values of $x \left( N,t
\right)$ bracketing the spread are shown in
Fig.~\ref{fig:price_bins_label}.  For symmetric order-placement rules,
probability distributions over configurations will be symmetric under
either $n \left( p,t \right) \rightarrow -n \left( -p,t \right)$, or
$x \left( N,t \right) \rightarrow x \left( -N,t \right)$.  Coordinates
$N$ and $p$ furnish a dual description of configurations, and $n$ and
$x$ are their associated differences.  The Master Equation approach of 
section ~\ref{subsec:ME} assumes independent fluctuation in $n$ while
the Independent Interval
Approximation of Sec.~\ref{subsec:IIA} assumes independent fluctuation in
$x$ (In this section,
it will be convenient to
abbreviate $x \left( N,t \right) \equiv x_N \left( t \right)$).  

\subsection{Frames and marginals}
\label{subsec:frames_marginals}

The $x\left( N,t \right)$ specification of limit-order book
configurations has the property that its distribution is stationary
under the dynamics considered here.  The same is not true for $p
\left( N,t \right)$ or $n \left( p,t \right)$ directly, because bid,
midpoint, and ask prices undergo a random walk, with a renormalized 
diffusion coefficient.
Stationary distributions for $n$-variables can be obtained in {\em
co-moving frames}, of which there are several natural choices.

The {\em bid-centered configuration} is defined as 
\begin{equation}
  n_b \left( p,t \right) \equiv 
  n \left( p-b\left( t \right) ,t \right) . 
\label{eq:bid_cent_def}
\end{equation}
If an appropriate rounding convention is adopted in the case of
discrete prices, a {\em midpoint-centered configuration} can also be
defined, as
\begin{equation}
  n_m \left( p,t \right) \equiv 
  n \left( p-m\left( t \right) ,t \right) . 
\label{eq:mid_cent_def}
\end{equation}
The midpoint-centered configuration has qualitative differences from
the bid-centered configuration, which will be explored below.  Both
give useful insights to the order distribution and diffusion
processes.  The ask-centered configuration, $n_a \left( p,t \right)$,
need not be considered if order placement and removal are symmetric,
because it is a mirror image of $n_b \left( p,t \right)$.

The {\em spread} is defined as the difference $s \left( t \right)
\equiv a \left( t \right) - b \left( t \right)$, and is the value of
the ask in bid-centered coordinates.  In midpoint-centered
coordinates, the ask appears at $s \left( t \right) / 2$.

The configurations $n_b$ and $n_m$ are dynamically correlated over
short time intervals, but evolve ergodically in periods longer than
finite characteristic correlation times.  Marginal probability
distributions for these can therefore be computed as time averages,
either as functions on the whole price space, or at discrete sets of
prices.  Their marginal mean values at a single price $p$ will be
denoted $\left< n_b \left( p \right) \right>$, $\left< n_m \left( p
\right) \right>$, respectively.  

These means are subject to global balance constraints, between total
order placement and removal in the price space.  
Because all limit orders are placed above the bid, the bid-centered
configuration obeys a simple balance relation: 
\begin{equation}
  \frac{\mu}{2} = 
  \sum_{p = b+dp}^{\infty}
  \left(
    \alpha - \delta 
    \left< n_b \left( p \right) \right>
  \right) . 
\label{eq:n_b_balance_dim}
\end{equation}
Eq.~(\ref{eq:n_b_balance_dim}) says that buy market orders must
account, on average, for the difference between all limit orders
placed, and all decays.  After passing to nondimensional coordinates
below, this will imply an inverse relation between corrections to the
classical estimate for diffusivity at early and late times, discussed
in Sec.~\ref{subsec:ren_diff}.  In addition, this conservation law
plays an important role in the analysis and determination of the
$x(N,t)$'s, as we will see later in the text.

The midpoint-centered averages satisfy a different constraint: 
\begin{equation}
  \frac{\mu}{2} = 
  \alpha 
  \frac{
    \left< s \right> 
  }{
    2 
  } + 
  \sum_{p = b+dp}^{\infty}
  \left(
    \alpha - \delta 
    \left< n_m \left( p \right) \right>
  \right) . 
\label{eq:n_m_balance_dim}
\end{equation}
Market orders in Eq.~(\ref{eq:n_m_balance_dim}) account not only for
the excess of limit order placement over evaporation at prices above
the midpoint, but also the ``excess'' orders placed between $b \left(
t \right)$ and $m \left( t \right)$.  Since these always lead to
midpoint shifts, they ultimately appear at positive comoving
coordinates, altering the shape of $\left< n_m \left( p \right)
\right>$ relative to $\left< n_b \left( p \right) \right>$.  Their
rate of arrival is $\alpha \left< m-b
\right> = \alpha \left< s \right> / 2$.  These results are also
confirmed in simulations.

\subsection{Factorization tests}
\label{subsec:fact_tests}

Whether in the bid-centered frame or the midpoint centered frame,
the probability distribution function for the entire
configuration $n \left( {p} \right)$
is too difficult a problem to solve in its entirety.
However, an approximate master equation can be formed for $n$
independently at each $p$ if all joint probabilities factor into
independent marginals, as
\begin{equation}
  \mbox{Pr}
  \left(
    {
      \left\{
        n \left( p_i \right)
      \right\}
    }_i
  \right) = 
  \prod_i
  \mbox{Pr}
  \left(
    n \left( p_i \right)
  \right) , 
\label{eq:factorization}
\end{equation}
where $\mbox{Pr}$ denotes, for instance, a probability density for
$n$ orders in some interval around $p$.  

Whenever orders are sufficiently sparse that the expected number in
any price bin is simply the probability that the bin is occupied (up
to a constant of proportionality), the independence assumption implies
a relation between the cumulative distribution for the spread
of the ask and the mean density profile.  In units where the order
size is one, the relation is 
\begin{equation}
  \mbox{Pr}
  \left(
    s/2 < p
  \right) = 
  1 - 
  \exp
  \left( - 
    \sum_{p' = b+dp}^{p - dp}
      \left< n_m \left( p' \right) \right>
  \right)  .  
\label{eq:factor_test}
\end{equation}

\begin{figure}[t]
\epsfxsize=2.5in 
\epsfbox{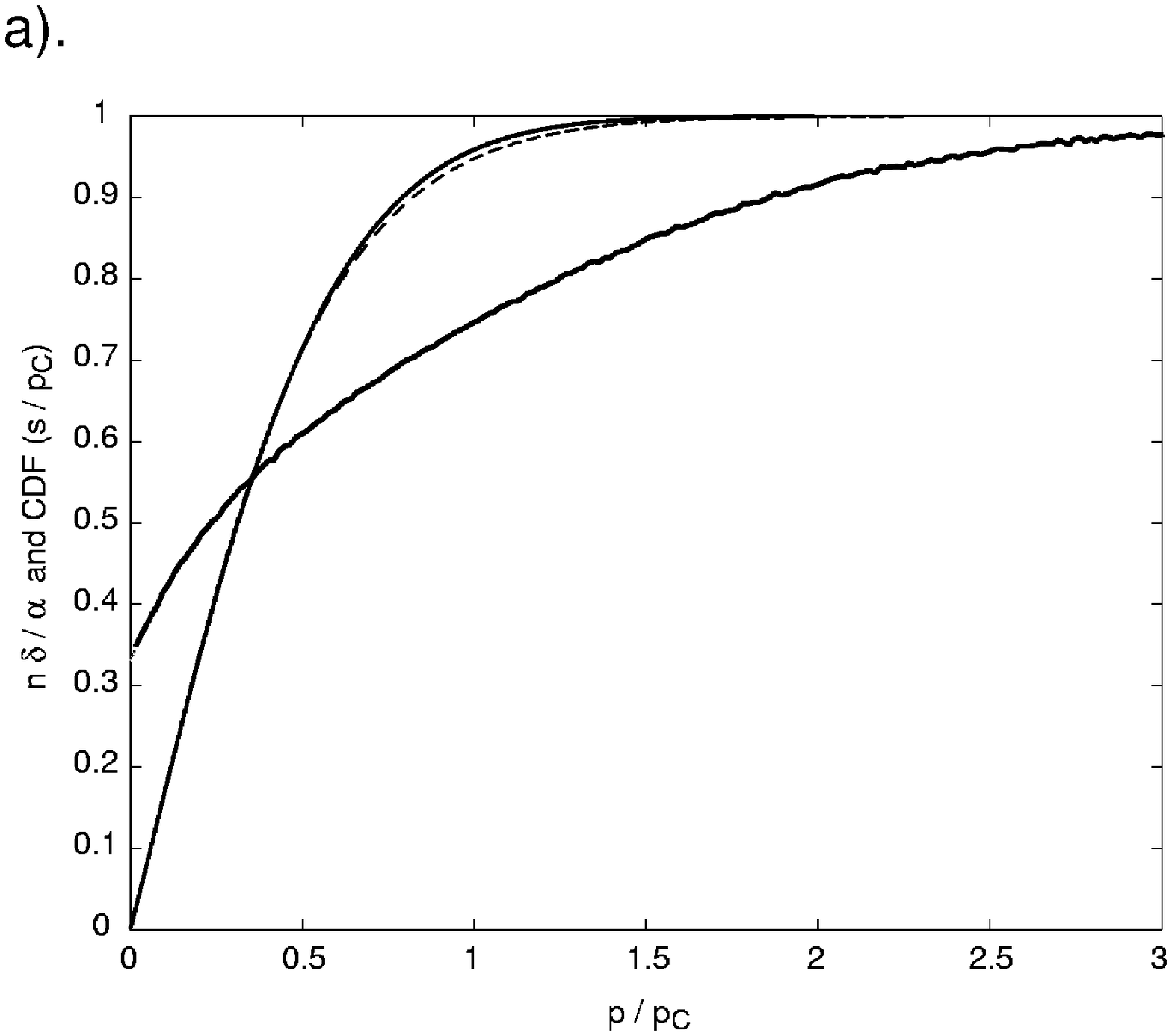}
\epsfxsize=2.5in 
\epsfbox{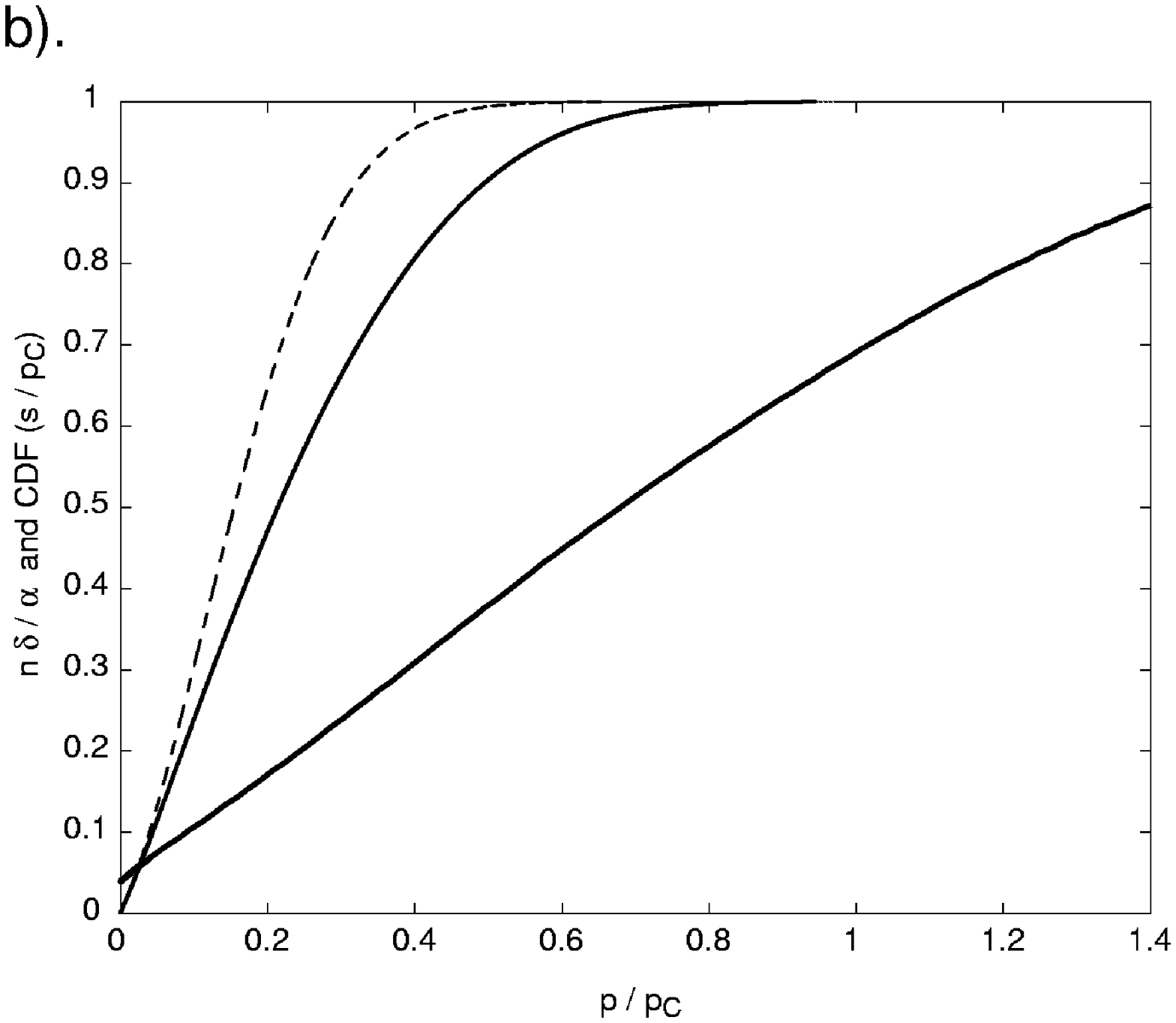}
\epsfxsize=2.5in 
\epsfbox{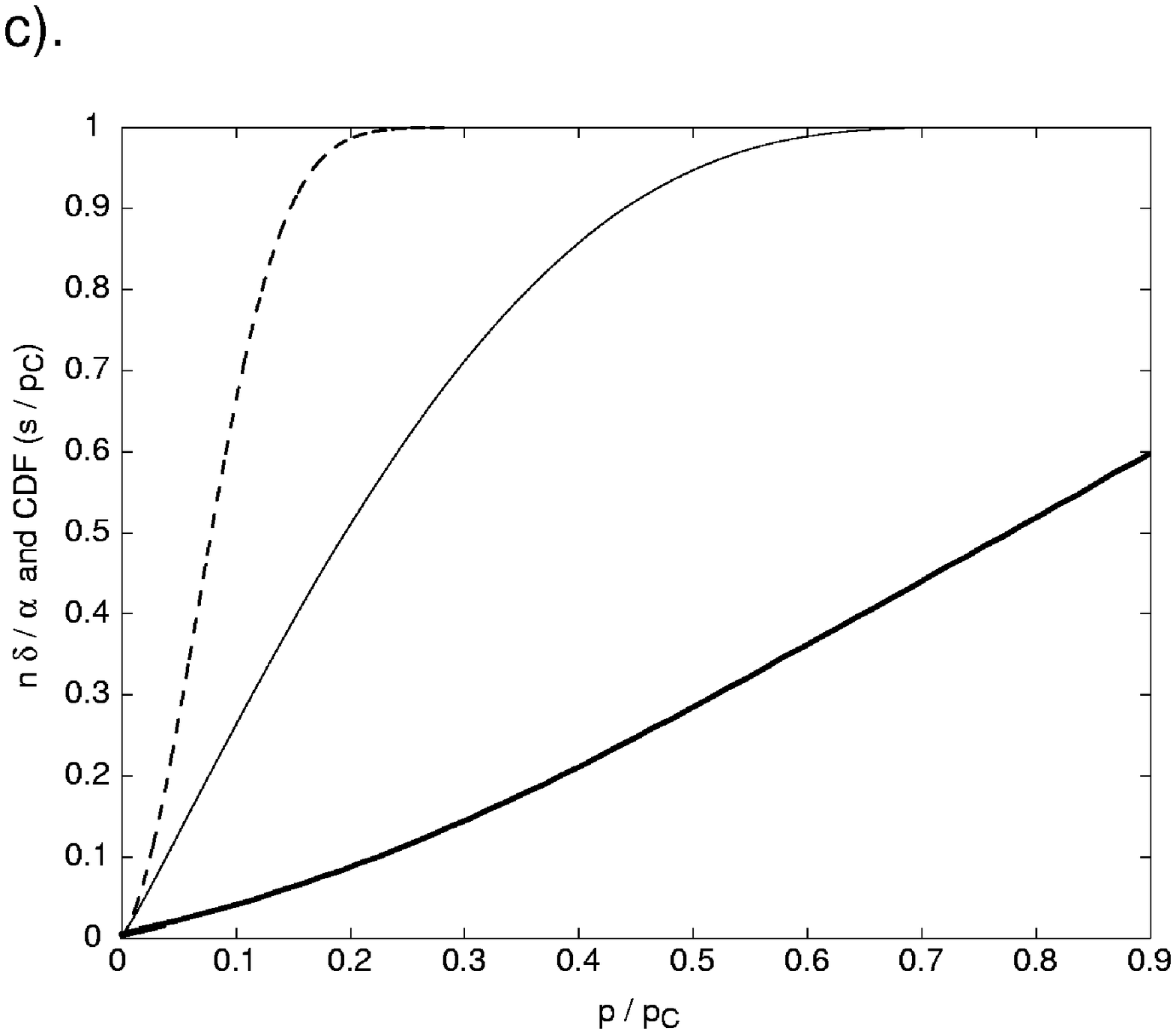}
\caption{
  CDFs $\mbox{Pr} \left( s/2 < p \right)$ from simulations (thin
  solid), mean density profile $\left< n_m \left( p \right) \right>$
  from simulations (thick solid), and computed CDF of spread (thin
  dashed) from $\left< n_m \left( p \right) \right>$, under the
  assumption of uncorrelated fluctuations, at three values of
  $\epsilon$. (a): $\epsilon = 0.2$ (low market order rate);
  approximation is very good.  (b): $\epsilon = 0.02$ (intermediate
  market order rate); approximation is marginal.  (c): $\epsilon =
  0.002$ (high market order rate); approximation is very poor.
  \label{fig:test_cdfs_and_nhat_4}
}
\end{figure}

This relation is tested against simulation results in
Fig.~\ref{fig:test_cdfs_and_nhat_4}. One can  
observe that there are three
regimes.

A high-${\epsilon}$ regime is defined when the mean density profile at the
midpoint $\left< n_m \left( 0 \right) \right> \lesssim 1$, and
strongly concave downward.  In this regime, the approximation of
independent fluctuations is excellent, and a master equation treatment
is expected to be useful.  Intermediate-${\epsilon}$ is defined by $\left<
n_m \left( 0 \right) \right> \ll 1$ and nearly linear, and the
approximation of independence is marginal.  Large-${\epsilon}$ is defined by  
$\left< n_m \left( 0 \right) \right> \ll 1$ and concave upward, and
the approximation of independent fluctuations is completely invalid.
These regimes of validity correspond also to the qualitative ranges
noted already in  Sec.~\ref{subsec:epsi_ranges}.  

In the bid centered frame however, Eq. \ref{eq:factor_test} 
never seems to be valid
for any range of parameters. We will discuss later why this might be
so.  For the present therefore, the master equation approach is
carried out in the midpoint-centered frame.  Alternatively, the mean
field theory of the separations is most convenient in the bid-centered
frame, so that frame will be studied in the dual basis.  The relation
of results in the two frames, and via the two methods of treatment,
will provide a good qualitative, and for some properties quantitative,
understanding of the depth profile and its effect on impacts.

It is possible in a modified treatment, to match certain features of
simulations at any $\epsilon$, by limited incorporation of correlated
fluctuations.  However, the general master equation will be developed
independent of these, and tested against simulation results at large
$\epsilon$, where its defining assumptions are well met.

\subsection{Comments on renormalized diffusion}
\label{subsec:ren_diff}

A qualitative understanding of why the diffusivity is different
over short and long times scales, as well as why it
may depend on $\epsilon$, may be gleaned from the
following observations.

First, global order conservation places a strong constraint on the
classically nondimensionalized density profile in the bid-centered
frame.  We have seen that at $\epsilon \ll 1$, the density profile
becomes concave upward near the bid, accounting for an increasing
fraction of the allowed ``remainder area'' as $\epsilon \rightarrow
0$ (see Figs. \ref{epsDepth} and \ref{fig:direct_x_compare_mid}).
Since this remainder area is fixed at unity, it can be conserved
only if the density profile approaches one more quickly with
increasing price.  Low density at low price appears to lead to more
frequent persistent steps in the effective short-term random walk, and
hence large short-term diffusivity.  However, increased density far
from the bid indicates less impact from market orders relative to the
relaxation time of the Poisson distribution, and thus a lower
long-time diffusivity.  

The qualitative behavior of the bid-centered density profile is the
same as that of the midpoint-centered profile, and this is expected
because the spread distribution is stationary, rather than diffusive.
In other words, the only way the diffusion of the bid or ask can
differ from that of the midpoint is for the spread to either increase
or decrease for several succeeding steps.  Such autocorrelation of the
spread cannot accumulate with time if the spread itself is to have a
stationary distribution.  Thus, the shift in the midpoint over some
time interval can only differ from that of the bid or ask by at most a
constant, as a result of a 
few correlated changes in the spread.  This difference cannot
grow with time, however, and so does not affect the diffusivity at
long times.  

Indeed, both of the predicted corrections to the classical estimate
for diffusivity are seen in simulation results for midpoint diffusion.
The simulation results, however, show that the implied
autocorrelations change the diffusivity by factors of
$\sqrt{\epsilon}$, suggesting that these corrections require a more
subtle derivation than the one attempted here.  This will be evidenced by
the difficulty of obtaining a source term ${\cal S}$ in density
coordinates (section\ref{subsec:ME}), 
which satisfied both the global order conservation law,
and the proper zero-price boundary condition, in the midpoint-centered
frame.  

An interesting speculation is that the subtlety of these correlations
also causes the density $n \left( p,t \right)$ in bid-centered
coordinates not to approximate the mean-field condition at any of the
parameters studied here, as noted in Sec.~\ref{subsec:fact_tests}.
Since short-term and long-term diffusivity corrections are related by
a hard constraint, the difficulty of producing the late-time density
profile should match that of producing the early-time profile.  The
midpoint-centered profile is potentially easier, in that the late-time
complexity must be matched by a combination of the early-time density
profile and the scaling of the expected spread.  It appears that the
complex scaling is absorbed in the spread, as per Fig.~\ref{meanSpread}
and Fig.~\ref{spread_compare}, leaving a density that can be
approximately calculated with the methods used here.

\subsection{Master equations and mean-field approximations}
\label{subsec:ME}

There are two natural limits in which functional configurations may
become simple enough to be tractable probabilistically, with analytic
methods.  They correspond to mean field theories in which fluctuations
of the dual differentials of either $N \left( p , t \right)$ or $p
\left( N , t \right)$ are independent.  In the first case,
probabilities may be defined for any density $n \left( p , t
\right)$ independently at each $p$, and in the second for the
separation intervals $x \left( N , t \right)$ at each $N$.  The mean
field theory from the first approximation will be solved in
Subsec.~\ref{subsec:number_MFT}, and that from the second in
Subsec.~\ref{subsec:IIA}.  As mentioned above,
because the fluctuation independence
approximation is only usable in a midpoint-centered frame, $n \left( p , t
\right)$ will refer always to this frame.  $x \left( N , t \right)$ is
well-defined without reference to any frame.  

\subsubsection{A number density master equation}
\label{subsec:number_MFT}

If share-number fluctuations are independent at different $p$, a
density $\pi \left( n , p , t \right)$ may be defined, which gives the
probability to find $n$ orders in bin $\left( p , p+dp \right)$, at
time $t$.  
The normalization condition defining $\pi$ as a probability
density is
\begin{equation}
  \sum_{n} 
  \pi
  \left( n , p , t \right) = 
  1 , 
\label{eq:pi_norm_def}
\end{equation}
for each bin index $p$ and at every $t$.  The index $t$ will be
suppressed henceforth 
in the notation since we are looking for time-independent
solutions.

Supposing an arbitrary density of order-book configurations $ \pi
\left( n , p \right)$ at time $t$, the stochastic dynamics of the
configurations causes probability to be redistributed according to the
master equation
\begin{eqnarray}
  \frac{
    \partial
  }{
    \partial t
  }
& & 
  \pi
  \left( 
    n , p 
  \right) = 
\nonumber \\
& & 
  \frac{
    \alpha \left( p \right) dp
  }{
    \sigma
  }
  \left[
    \pi
    \left( 
      n - \sigma , p 
    \right) - 
    \pi
    \left( 
      n , p 
    \right) 
  \right]
\nonumber \\
& & 
  \mbox{} + 
  \frac{
    \delta 
  }{
    \sigma
  }
  \left[
    \left(
      n + \sigma
    \right)
    \pi
    \left( 
      n + \sigma , p 
    \right) - 
    n 
    \pi
    \left( 
      n , p 
    \right) 
  \right]
\nonumber \\
& & 
  \mbox{} + 
  \frac{ 
    \mu \left( p \right)
  }{
    2 \sigma 
  }
  \left[
    \pi
    \left( 
      n + \sigma , p 
    \right) - 
    \pi
    \left( 
      n , p 
    \right) 
  \right] 
\nonumber \\
& & 
  \mbox{} + 
  \sum_{\Delta p}
  P_{+} 
  \left( \Delta p \right)
  \left[
    \pi
    \left( 
      n , p - \Delta p 
    \right) - 
    \pi
    \left( 
      n , p 
    \right) 
  \right] 
\nonumber \\
& & 
  \mbox{} + 
  \sum_{\Delta p}
  P_{-} 
  \left( \Delta p \right)
  \left[
    \pi
    \left( 
      n , p + \Delta p 
    \right) - 
    \pi
    \left( 
      n , p 
    \right) 
  \right] . 
\label{eq:number_density_master_eq}
\end{eqnarray}
Here $ \partial \pi \left( n , p \right) /\partial t $ is a continuum
notation for $ \left[ \pi \left( n , p , t +\delta t \right) - \pi
\left( n , p , t \right) \right] / \delta t$, where $\delta t$ is an
elementary time step, chosen short enough that at most one event
alters any typical configuration.
Eq.~(\ref{eq:number_density_master_eq}) represents a general balance
between additions and removals, without regard to the meaning of $n$.
Thus, $\alpha \left( p \right)$ is a function that must be determined
self-consistently with the choice of frame.  As an example of how this
works, in a bid-centered frame, $\alpha \left( p \right)$ takes a
fixed value $\alpha \left( \infty \right)$ at all $p$, because the
deposition rate is independent of position and frame shifts.  The
midpoint-centered frame is more complicated, because depositions below
the midpoint cause shifts that leave the deposited order above the
midpoint.  The specific consequence for $\alpha \left( p \right)$ in
this case will be considered below.  $\mu \left( p \right) / 2$ is,
similarly, the rate of market orders surviving to cancel limit orders
at price $p$.  $\mu \left( p \right) / 2$ decreases from $\mu \left( 0
\right) / 2$ at the ask (for buy market orders, because $\mu$ total
orders are divided evenly between buys and sells) to zero as $p
\rightarrow \infty$, as market orders are screened probabilistically
by intervening limit orders.  $\alpha \left( \infty \right)$ and $\mu
\left( 0 \right)$ are thus the parameters $\alpha$ and $\mu$ of the
simulation.

The lines of Eq.~(\ref{eq:number_density_master_eq}) correspond to the
following events.  The term proportional to $\alpha \left( p \right)
dp / \sigma$ describes depositions of discrete orders at that rate
(because $\alpha$ is expressed in {\em shares per price per time}),
which raise configurations from $n - \sigma$ to $n$ shares at price
$p$.  The term proportional to $\delta$ comes from deletions and has
the opposite effect, and is proportional to $n / \sigma$, the number
of {\em orders} that can independently decay.  The term proportional
to $\mu \left( p \right) / 2 \sigma$ describes market order
annihilations.  For general configurations, the preceding three
effects may lead to shifts of the origin by arbitrary intervals
$\Delta p$, and $P_{\pm}$ are for the moment unknown distributions
over the frequency of those shifts.  They must be determined
self-consistently with the configuration of the book which emerges
from any solution to Eq.~(\ref{eq:number_density_master_eq}).

A limitation of the simple product representation of frame shifts is
that it assumes that whole order-book configurations are transported
under $p \pm \Delta p \rightarrow p$, independently of the value of $n
\left( p \right)$.  As long as fluctuations are independent, this is a
good approximation for orders at all $p$ which are not either the bid
or the ask, either before or after the event that causes the shift.
The correlations are never ignorable for the bins which are the bid
and ask, though, and there is some distribution of instances in which
any $p$ of interest plays those parts.  Approximate methods to
incorporate those correlations will require replacing the product form
with a sum of products conditioned on states of the order book,
as will be derived below.

The important point is that the order-flow dependence of
Eq.~(\ref{eq:number_density_master_eq}) is independent of these
self-consistency requirements, and may be solved by use of generating
functionals at general $\alpha \left( p \right)$, $\mu \left( p
\right)$, and $P_{\pm}$.  The solution, exact but not analytically
tractable at general $dp$, will be derived in closed form in the next
subsection.  It has a well-behaved continuum limit at $dp \rightarrow
0$, however, which is analytically tractable, so that special case
will be considered in the following subsection.

\subsubsection{Solution by generating functional}

The moment generating functional for $\pi$ is defined for a parameter
$\lambda \in \left[ 0 , 1 \right]$, as
\begin{equation}
  \Pi 
  \left( 
    \lambda , p 
  \right) \equiv
  \sum_{n / \sigma = 0}^{\infty}
  {\lambda}^{n / \sigma}
  \pi
  \left( 
    n , p 
  \right) .
\label{eq:gen_funal_def}
\end{equation}
Introducing a shorthand for its value at $\lambda = 0$, 
\begin{equation}
  \Pi 
  \left( 
    0 , p 
  \right) = 
  \pi
  \left( 
    0 , p 
  \right) \equiv 
  {\pi}_0
  \left( 
    p 
  \right) , 
\label{eq:Pi_zero_term_notation}
\end{equation}
while the normalization condition~(\ref{eq:pi_norm_def}) for
probabilities gives
\begin{equation}
  \Pi 
  \left( 
    1 , p 
  \right) = 
  1 \mbox{ } , 
  \forall p . 
\label{eq:Norm_condition}
\end{equation}
By definition of the average of $n \left( p \right)$ in the
distribution $\pi$, denoted $ \left< n \left( p \right) \right>$,
\begin{equation}
  {
    \left.
      \frac{\partial}{\partial \lambda} 
      \Pi 
      \left( 
        \lambda , p 
      \right) 
    \right|
  }_{\lambda \rightarrow 1} = 
  \frac{
    \left<
      n \left( p \right)
    \right>
  }{
    \sigma 
  } , 
\label{eq:Pi_give_N_ave}
\end{equation}
and because $\Pi$ will be regular in some sufficiently small
neighborhood of $\lambda = 1$, one can expand
\begin{equation}
  \Pi 
  \left( 
    \lambda , p 
  \right) = 
  1 + 
  \left( \lambda - 1 \right)  
  \frac{
    \left<
      n \left( p \right)
    \right>
  }{
    \sigma
  } + 
  {\cal O}
  {
    \left( \lambda - 1 \right)
  }^2 . 
\label{eq:lamb_power_expn}
\end{equation}

Multiplying Eq.~(\ref{eq:number_density_master_eq}) by ${\lambda}^{n /
\sigma}$ and summing over $n$, (and suppressing the argument $p$ in
the notation everywhere; $\alpha \left( p \right)$ or $\alpha \left( 0
\right)$ will be used where the distinction of the function from its
boundary value is needed) the stationary solution for $\Pi$ must
satisfy
\begin{eqnarray} 
  0 
& = & 
  \frac{\lambda - 1}{\sigma}
  {
    \left.
      \left\{
        \alpha dp 
        \Pi - 
        \delta \, \sigma 
        \frac{
          \partial \Pi
        }{
          \partial \lambda
        } - 
        \frac{\mu}{2 \lambda}
        \left(
          \Pi - 
          {\pi}_0
        \right)
      \right\}
    \right|
  }_{\left( \lambda , p \right)}
\nonumber \\
  & + & 
  \sum_{\Delta p}
  P_{+} 
  \left( \Delta p \right)
  \left[
    \Pi
    \left( 
      \lambda , p - \Delta p 
    \right) - 
    \Pi
    \left( 
      \lambda , p 
    \right) 
  \right] 
\nonumber \\
  & + & 
  \sum_{\Delta p}
  P_{-} 
  \left( \Delta p \right)
  \left[
    \Pi
    \left( 
      \lambda , p + \Delta p 
    \right) - 
    \Pi
    \left( 
      \lambda , p 
    \right) 
  \right] . 
\label{eq:gen_fnal_master_eqn}
\end{eqnarray}

Only the symmetric case with no net drift will be considered here for
simplicity, which requires
$ P_{+} 
  \left( \Delta p \right) = 
  P_{-} 
  \left( \Delta p \right) \equiv 
  P
  \left( \Delta p \right) $. 
In a Fokker-Planck expansion, the (unrenormalized) diffusivity of
whatever reference price is used as coordinate origin, is related to
the distribution $P$ by
\begin{equation}
  D \equiv
  \sum_{\Delta p}
  P
  \left( \Delta p \right)
  {\Delta p}^2 . 
\label{eq:diffusivity_def}
\end{equation}
The rate at which shift events happen is 
\begin{equation}
  R \equiv
  \sum_{\Delta p}
  P
  \left( \Delta p \right) , 
\label{eq:rate_shift_def}
\end{equation}
and the mean shift amount appearing at linear order in derivatives
(relevant at $p \rightarrow 0$), is 
\begin{equation}
  \left< \Delta p \right> 
  \equiv
  \frac{
  \sum_{\Delta p}
    P
    \left( \Delta p \right)
    \Delta p 
  }{
    \sum_{\Delta p}
    P
    \left( \Delta p \right)
  } . 
\label{eq:mean_shift_def}
\end{equation}

Anywhere in the interior of the price range (where $p$ is not at any
stage the bid, ask, or a point in the spread),
Eq.~(\ref{eq:gen_fnal_master_eqn}) may be written
\begin{equation}
  \left\{
    \frac{\partial}{\partial \lambda} - 
    \frac{
      D
    }{
      \delta \left( \lambda - 1 \right)
    } \, 
    \frac{{\partial}^2}{\partial p^2} - 
    \frac{
      \alpha dp - \mu / 2 \lambda
    }{
      \delta \, \sigma 
    }
  \right\}
  \Pi =
  \frac{
    \mu 
  }{
    2 \delta \, \sigma \lambda
  }
  {\pi}_0 . 
\label{eq:diff_form_gen_ME}
\end{equation}
Evaluated at $\lambda \rightarrow 1$, with the use of the
expansion~(\ref{eq:lamb_power_expn}), this becomes 
\begin{equation}
  \left(
    1 - 
    \frac{D}{\delta} \, 
    \frac{\partial}{\partial p^2} 
  \right)
  \left< n  \right> = 
    \frac{
      \alpha dp 
    }{
      \delta 
    } - 
    \frac{
      \mu 
    }{
      2 \delta 
    }
  \left(
    1 - {\pi}_0
  \right) . 
\label{eq:diff_form_N_ave}
\end{equation}

At this point it is convenient to specialize to the case $dp
\rightarrow 0$, wherein the eligible values of any $\left< n \left( p
\right) \right>$ become just $\sigma$ and zero.  The expectation is
then related to the probability of zero occupancy (at each $p$) as
\begin{equation}
  \left< 
    n 
  \right> = 
  \sigma
  \left[
    1 - {\pi}_0
  \right] , 
\label{eq:ave_n_from_pi_0}
\end{equation}
yielding immediately 
\begin{equation}
  \frac{
    \alpha dp 
  }{
    \delta
  } = 
  \left[
    \frac{
      \mu 
    }{
      2 \delta \, \sigma
    } + 
    \left(
      1 - 
      \frac{D}{\delta} \, 
      \frac{d^2}{d p^2} 
    \right)
  \right]
  \left< n \right> . 
\label{eq:series_invert}
\end{equation}

Eq.~(\ref{eq:series_invert}) defines the general solution $\left< n
\left( p \right) \right>$ for the master
equation~(\ref{eq:number_density_master_eq}), in the continuum limit
$2 \alpha dp / \mu \rightarrow 0$.  The shift distribution $P
\left( \Delta p \right)$ appears only through the diffusivity $D$,
which must be solved self-consistently, along with the otherwise
arbitrary functions $\alpha$ and $\mu$.  The more general solution at
large $dp$ is carried out in App.~\ref{sec:app_genf_large_dp}.  

A first step toward nondimensionalization may be taken by writing
Eq.~(\ref{eq:series_invert}) in the form (re-introducing the indexing
of the functions)
\begin{equation}
  \frac{
    \alpha \left( p \right)
  }{
    \alpha \left( \infty \right)
  } = 
  \left[
    \frac{
      \mu \left( p \right)
    }{
      \mu \left( 0 \right)
    } + 
    \epsilon
    \left(
      1 - 
      \frac{D}{\delta} \, 
      \frac{d^2}{d p^2} 
    \right)
  \right]
  \frac{
    1
  }{
    \epsilon
  }
  \frac{
    \delta \left< n \right> . 
  }{
    \alpha dp
  } . 
\label{eq:series_invert_nondim}
\end{equation}
Far from the midpoint, where only depositions and cancellations take
place, orders in bins of width $dp$ are Poisson distributed with mean
$\alpha \left( \infty \right) dp / \delta$.  Thus, the asymptotic value of
$\delta \left< n \right> / \alpha \left( \infty \right) dp$ at large $p$ is
unity.  This is consistent with a limit for $\alpha \left( p \right) /
\alpha \left( \infty \right)$ of unity, and a limit for
the screened $\mu \left( p
\right) / \mu \left( 0 \right)$ of zero.  The reason for grouping the
nondimensionalized number density with $1 / \epsilon$, together with
the proper normalization of the characteristic price scale, will come
from examining the decay of the dimensionless function $\mu \left( p
\right) / \mu \left( 0 \right)$.

\subsubsection{Screening of the market-order rate}
\label{subsec:ctm_limit}

In the context of independent fluctuations,
Eq.~(\ref{eq:ave_n_from_pi_0}) implies a relation between the mean
density and the rate at which market orders are screened as price
increases.  The effect of a limit order, resident in the price bin $p$
when a market order survives to reach that bin, is to prevent its
arriving at the bin at $p+dp$.  Though the nature of the shift
induced, when such annihilation occurs, depends on the comoving frame
being modeled, the change in the number of orders surviving is
independent of frame, and is given by
\begin{equation}
  d \mu = 
  - \mu 
  \left(
    1 - {\pi}_0 
  \right) = 
  - \mu 
  \left< n \right> / \sigma . 
\label{eq:rate_reduction}
\end{equation}

Eq.~(\ref{eq:rate_reduction}) may be rewritten 
\begin{equation}
  \frac{
    d \log \left( \mu \left( p \right) / 
      \mu \left( 0 \right) \right) = 
  }{
    dp
  }
  - 
  \frac{1}{\epsilon}
  \left(
    \frac{
      2 \alpha \left( \infty \right)
    }{
      \mu \left( 0 \right)
    }
  \right)
  \left(
    \frac{
      \delta \left< n  \left( p \right) \right>
    }{
      \alpha \left( \infty \right) dp
    }
  \right) , 
\label{eq:rate_reduction_again}
\end{equation}
identifying the characteristic scale for prices as $p_c = \mu \left( 0
\right) / 2 \alpha \left( \infty \right) \equiv \mu / 2 \alpha$.  Writing
$\hat{p} \equiv p / p_c$, the function that screens market orders is
the same as the argument of Eq.~(\ref{eq:series_invert_nondim}), and
will be denoted
\begin{equation}
  \frac{1}{\epsilon}
  \frac{
    \delta \left< n \left( p \right) \right>
  }{
    \alpha \left( \infty \right) dp
  } \equiv 
  \psi 
  \left( \hat{p} \right)
\label{eq:psi_def}
\end{equation}

Defining a nondimensionalized diffusivity $\beta \equiv D / \delta
p_c^2$, Eq.~(\ref{eq:series_invert}) can then be put in the form
\begin{equation}
  \frac{
    \alpha \left( p \right)
  }{
    \alpha \left( \infty \right)
  } = 
  \left[
    \frac{
      \mu \left( p \right)
    }{
      \mu \left( 0 \right)
    } + 
    \epsilon
    \left(
      1 - 
      \beta
      \frac{d^2}{d {\hat{p}}^2} 
    \right)
  \right]
  \psi , 
\label{eq:series_invert_nondim_more}
\end{equation}
with 
\begin{equation}
  \frac{
    \mu \left( p \right)
  }{
    \mu \left( 0 \right)
  } \equiv 
  \varphi 
  \left( \hat{p} \right) = 
  \exp 
  \left( - 
    \int_0^{\hat{p}} d{\hat{p}'}
    \psi \left( \hat{p}' \right)
  \right) , 
\label{eq:phi_sigma_constr}
\end{equation}

\subsubsection{Verifying the conservation laws}

Since nothing about the derivation so far has made explicit use of the
frame in which $n$ is averaged, the combination of
Eq.~(\ref{eq:series_invert_nondim_more}) with
Eq.~(\ref{eq:phi_sigma_constr}) respects the conservation
laws~(\ref{eq:n_b_balance_dim}) and~(\ref{eq:n_m_balance_dim}), if
appropriate forms are chosen for the deposition rate $\alpha \left( p
\right)$.

For example, in the bid-centered frame, $\alpha \left( p \right) /
\alpha \left( \infty \right) = 1$ everywhere.  Multiplying
Eq.~(\ref{eq:series_invert_nondim_more}) by $d\hat{p}$ and integrating
over the whole range from the bid to $+\infty$, we recover the
nondimensionalized form of Eq.~(\ref{eq:n_b_balance_dim}):
\begin{equation}
  \int_0^{\infty}
  d\hat{p} 
  \left( 
    1 - \epsilon \psi 
  \right) = 1 , 
\label{eq:n_b_balance_nodim}
\end{equation}
{\em iff} we are careful with one convention.  The integral of the
diffusion term formally produces the first derivative $ { \left. d
\psi / d \hat{p} \right| }_0^{\infty}$.  We must regard this as a true
first derivative, and consider its evaluation at zero continued far
enough below the bid to capture the identically zero first derivative
of the sell order depth profile.  

In the midpoint centered frame, the correct form for
the source term should be $\alpha \left( \hat{p} \right) / \alpha
\left( \hat{\infty} \right) = 1 + \mbox{Pr} \left( \hat{s}/2 \ge \hat{p}
\right) $, whatever the expression for the cumulative distribution
function.  Recognizing that the integral of the CDF is, by parts, the
mean value of $\hat{s}/2$, the same integration of
Eq.~(\ref{eq:series_invert_nondim_more}) gives
\begin{equation}
  \int_0^{\infty}
  d\hat{p} 
  \left( 
    1 - \epsilon \psi 
  \right) = 1 - 
  \frac{\left< \hat{s} \right>}{2}, 
\label{eq:n_m_balance_nodim}
\end{equation}
the nondimensionalized form of Eq.~(\ref{eq:n_m_balance_dim}).  Again,
this works only if the surface contribution from integrating the
diffusion term vanishes.  

Neither of these results required the assumption of independent
fluctuations, though that will be used below to give a simple
approximate form for $\mbox{Pr} \left( \hat{s}/2 \ge \hat{p} \right)
\approx \varphi \left( \hat{p} \right)$.  They therefore provide a
check that the extinction form~(\ref{eq:phi_sigma_constr}) propagates
market orders correctly into the interior of the order-book
distribution, to respect global conservation.  They also check the
consistency of the intuitively plausible form for $\alpha$ in the
midpoint-centered frame.  The detailed form is then justified whenever
the assumption of independent fluctuations is checked to be valid.

\subsubsection{Self-consistent parametrization}

The assumption of independent fluctuations of $n \left( p \right)$ 
used above to derive the screening of market orders, is
equivalent to a specification of the CDF of the ask.  Market orders
are only removed between prices $p$ and $p + dp$ in those instances
when the ask is at $p$.  Therefore 
\begin{equation}
  \mbox{Pr}
  \left(
    \hat{s}/2 \ge \hat{p}
  \right) = 
  {\varphi} \left( \hat{p} \right) ,  
\label{eq:factor_test_ctm}
\end{equation}
the continuum limit of Eq.~(\ref{eq:factor_test}).  Together with the
form $\alpha \left( \hat{p} \right) / \alpha \left( \hat{\infty} \right) =
1 + \mbox{Pr} \left( \hat{s}/2 \ge \hat{p} \right) $,
Eq.~(\ref{eq:series_invert_nondim_more}) becomes
\begin{equation}
  1 + {\varphi} = 
  - 
  \left[
    \frac{d {\varphi}}{d\hat{p}} + 
    \epsilon
    \left(
      1 - 
      \beta
      \frac{d^2}{d {\hat{p}}^2} 
    \right)
    \frac{d \log {\varphi}}{d\hat{p}} 
  \right] .  
\label{eq:phi_self_con_eqn_mid}
\end{equation}
(If the assumption of independent fluctuations were valid in the
bid-centered frame, it would take the same form, but with $\varphi$
removed on the left-hand side.)

To consistently use the diffusion approximation, with the realization
that for $p = 0$, $n \pi \left( n , p - \Delta p \right) = 0$ for
essentially all $\Delta p$ in Eq.~(\ref{eq:number_density_master_eq}),
it is necessary to set the Fokker-Planck approximation to $\psi \left(
0 - \left< \Delta p \right> \right) = 0$ as a boundary condition.
Nondimensionalized, this gives
\begin{equation}
{ \left.
  \frac{\beta}{2} \, 
  \frac{d^2 \psi }{d {\hat{p}}^2} 
  \right|
}_0 = 
  \frac{R}{\delta} 
{ \left.
  \left( 
    \left< \Delta \hat{p} \right> 
    \frac{d}{d \hat{p}} - 
    1 
  \right)
  \psi 
  \right|
}_0 ,
\label{eq:diff_boundary_self_conm}
\end{equation}
where $R$ is the rate at which shifts occur (Eq. \ref{eq:rate_shift_def}).
In the solutions below, the curvature will typically be much smaller
than $\psi \left( 0 \right) \sim 1$, so it will be convenient to
enforce the simpler condition
\begin{equation}
  \left< \Delta \hat{p} \right> 
{ \left.
  \frac{d \psi}{d \hat{p}}
  \right|
}_0 - 
  \psi \left( 0 \right) \approx
  0 , 
\label{eq:diff_boundary_approx}
\end{equation}
and verify that it is consistent once solutions have been evaluated. 

Self-consistent expressions for $\beta$ and $\left< \Delta p \right>$
are then constructed as follows.  Given an ask at some position $a$
(in the midpoint-centered frame), there is a range from $-a$ to $a$ in
which sell limit orders may be placed, which will induce positive
midpoint-shifts.  The shift amount is half as great as the distance
from the bid, so the measure for shifts $dP_{+} \left( \Delta p
\right) $ from sell limit-order addition inherits a term $2 \alpha
\left( 0 \right) \left( d \Delta p \right) \mbox{Pr} \left( a \ge
\Delta p \right)$, where the last factor counts the instances with
asks large enough to admit shifts by $\Delta p$.  There is an equal
contribution to $dP_{-}$ from addition of buy limit orders.  Symmetry
requires that for every positive shift due to an addition, there is a
negative shift due to evaporation with equal measure, so the
contribution from buy limit order removal should equal that for sell
limit order addition.  When these contributions are summed, the
measures for positive and negative shifts both equal
\begin{equation}
  dP_{\pm} \left( \Delta p \right) = 
  4 \alpha \left( \infty \right)
  \left( d \Delta p \right)
  \mbox{Pr} 
  \left( a \ge \Delta p \right) . 
\label{eq:measure_shifts_dim}
\end{equation}

Eq.~(\ref{eq:measure_shifts_dim}) may be inserted into the continuum
limit of the definition~(\ref{eq:diffusivity_def}) for $D$, and then
nondimensionalized to give 
\begin{equation}
  \beta = 
  \frac{4}{\epsilon}
  \int_0^{\infty}
  d \Delta \hat{p} \, 
  {
    \left( \Delta \hat{p} \right)
  }^2
  \varphi \left( \Delta \hat{p} \right) , 
\label{eq:beta_simple_source}
\end{equation}
where the mean-field substitution of $\varphi \left( \Delta \hat{p}
\right)$ for $\mbox{Pr} \left( a \ge \Delta p \right)$ has been used.
Similarly, the mean shift amount used in
Eq.~(\ref{eq:diff_boundary_approx}) is
\begin{equation}
  \left< \Delta \hat{p} \right> = 
  \frac{
  \int_0^{\infty}
    d \Delta \hat{p} \, 
    \left( \Delta \hat{p} \right)
    \varphi \left( \Delta \hat{p} \right) 
  }{
    \int_0^{\infty}
    d \Delta \hat{p} \, 
    \varphi \left( \Delta \hat{p} \right) . 
  }
\label{eq:mean_deltp_simple_source}
\end{equation}

\begin{figure}[t]
\epsfxsize=2.75in 
\epsfbox{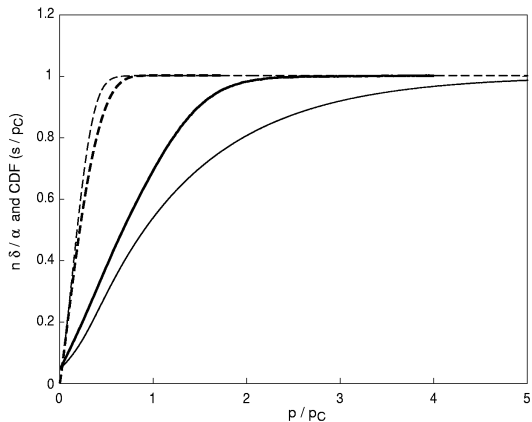}
\caption{
  Fit of the self-consistent solution with diffusivity term to
  simulation results for the midpoint-centered frame.  Thin solid line
  is the analytic solution for the mean number density, and thick
  solid line is simulation result, at $\epsilon = 0.02$.  Thin dashed
  line is the analytic prediction for the cumulative distribution
  function $\mbox{Pr} \left( \hat{s} / 2 \le \hat{p}\right)$, and
  thick dashed line is simulation result.
  \label{fig:compare_eps0p02}
}
\end{figure}

A fit of Eq.~(\ref{eq:phi_self_con_eqn_mid}) to simulations, using
these self-consistent measures for shifts, is shown in
Fig.~\ref{fig:compare_eps0p02}.  This solution is actually a
compromise between approximations with opposing ranges of validity.
The diffusion equation using the mean order depth describes nonzero
transport of limit orders through the midpoint, an approximation
inconsistent with the correlations of shifts with states of the order
book.  This approximation is a small error only at $\epsilon
\rightarrow 0$.  On the other hand, both the form of $\alpha$, and the
self-consistent solutions for $\left< \Delta \hat{p} \right>$ and
$\beta$, made use of the mean-field approximation, which we saw was
only valid for $\epsilon \lesssim 1$.  The two approximations appear
to create roughly compensating errors in the intermediate range
$\epsilon \sim 0.02$.

\subsubsection{Accounting for correlations}
\label{subsec:acc_corrs}

The numerical integral implementing the diffusion solution actually
doesn't satisfy the global conservation condition that the diffusion
term integrate to zero over the whole price range.  Thus, it describes
diffusive transport of orders through the midpoint, and as such also
doesn't have the right $\hat{p} = 0$ boundary condition.  The
effective absorbing boundary represented by the pure diffusion
solution corresponds roughly to the approximation made by Bouchaud
{\it et al.}~\cite{Bouchaud02} It differs from theirs, though, in that
their method of images effectively approximates the region of the
spread as a point, whereas Eq.~(\ref{eq:series_invert_nondim_more})
actually resolves the screening of market orders as the spread
fluctuates.

Treating the spread region -- roughly defined as the range over which
market orders are screened -- as a point is consistent with treating
the resulting coarse-grained ``midpoint'' as an absorbing boundary.
If the spread is resolved, however, it is not consistent for diffusion
to transport any finite number density through the midpoint, because
the midpoint is strictly always in the center of an open set with no
orders, in a continuous price space.  The correct behavior in a
neighborhood of the ``fine-grained midpoint'' can be obtained by
explicitly accounting for the correlation of the state of orders, with
the shifts that are produced when market or limit order additions
occur.  

We expect the problem of recovering both the global conservation law
and the correct $\hat{p} = 0$ boundary condition to be difficult, as
it should be responsible for the non-trivial corrections to
short-term and long-term diffusion mentioned earlier.
We have found, however, that by
explicitly sacrificing the global conservation law, we can incorporate
the dependence of shifts on the position of the ask, in an
interesting range around the midpoint.  At general $\epsilon$, the
corrections to diffusion reproduce the mean density over the main
support of the CDF of the spread.  While the resulting density does
not predict that CDF (due to correlated fluctuations), it closely
enough resembles the real density that the independent CDFs of the two
are similar.

\subsubsection{Generalizing the shift-induced source terms}

Nondimensionalizing the generating-functional master
equation~(\ref{eq:gen_fnal_master_eqn}) and keeping leading terms in
$dp$ at $\lambda \rightarrow 1$, get
\begin{eqnarray}
  \frac{
    \alpha \left( \hat{p} \right)
  }{
    \alpha \left( \hat{\infty} \right)
  } 
& = & 
  \left(
    \frac{
      \mu \left( \hat{p} \right)
    }{
      \mu \left( \hat{0} \right)
    } + 
    \epsilon
  \right)
  \psi \left( \hat{p} \right) 
\nonumber \\
& & 
  \mbox{} - 
  \int 
  d P_{+} 
  \left( \Delta \hat{p} \right)
  \left[
    \psi \left( \hat{p} - \Delta \hat{p} \right) - 
    \psi \left( \hat{p} \right) 
  \right] 
\nonumber \\
& & 
  \mbox{} - 
  \int 
  d P_{-} 
  \left( \Delta \hat{p} \right)
  \left[
    \psi \left( \hat{p} + \Delta \hat{p} \right) - 
    \psi \left( \hat{p} \right) 
  \right]
\label{eq:gen_fnal_master_nondim}
\end{eqnarray}
where $d P_{\pm} \left( \Delta \hat{p} \right)$ is the
nondimensionalized measure that results from taking the continuum
limit of $P_{\pm}$ in the variable $\Delta \hat{p}$.  

Eq.~(\ref{eq:gen_fnal_master_nondim}) is inaccurate because the number
of orders shifted into or out of a price bin $p$, at a given spread,
may be identically zero, rather than the unconditional mean value
$\psi$.  We take that into account by replacing the last two lines of
Eq.~(\ref{eq:gen_fnal_master_nondim}) with lists of source terms,
whose forms depend on the position of the ask, weighted by the
probability density for that ask.  Independent fluctuations are
assumed by using Eq.~(\ref{eq:factor_test_ctm}).

It is convenient at this point to denote the replacement of the last
two lines of Eq.~(\ref{eq:gen_fnal_master_nondim}) with the notation
${\cal S}$, yielding 
\begin{equation}
  \frac{
    \alpha \left( \hat{p} \right)
  }{
    \alpha \left( \hat{\infty} \right)
  } = 
  \left(
    \frac{
      \mu \left( \hat{p} \right)
    }{
      \mu \left( \hat{0} \right)
    } + 
    \epsilon
  \right)
  \psi 
  - {\cal S} . 
\label{eq:gen_mast_anon_src} 
\end{equation}
The global conservation laws for orders would be satisfied if $\int
d\hat{p} {\cal S} = 0$.

\begin{figure}[t]
\epsfxsize=2.75in 
\epsfbox{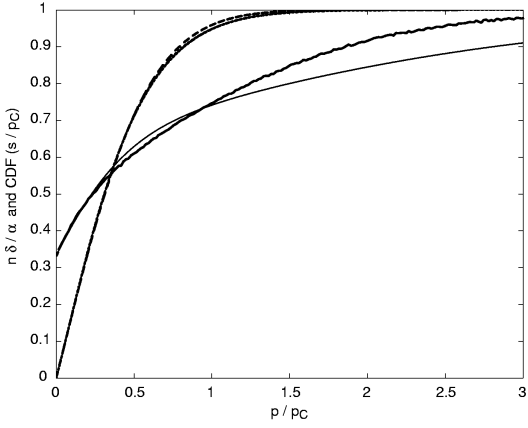}
\caption{
  Reconstruction with source terms ${\cal S}$ that approximately
  account for correlated fluctuations near the midpoint.  $\epsilon =
  0.2$.  Thick solid line is averaged order book depth from
  simulations, and thin solid is the mean field result.  Thin dotted
  line is the simulated CDF for $\hat{s} / 2$, and thick dotted line
  is the mean field result.  Thick dashed line is the CDF that would
  be produced from the simulated depth, if the mean-field
  approximation were exact.
  \label{fig:better_wrong_ind_eps_0p2}
}
\end{figure}

The source term ${\cal S}$ is derived approximately in
App.~\ref{sec:FK_expand_corrs}.  The solution to
Eq.~(\ref{eq:gen_mast_anon_src}) at $\epsilon = 0.2$, with the
simple-diffusive source term replaced by the
evaluations~(\ref{eq:detailed_nondim_src_formal} -
\ref{eq:cal_I_0_def}), is compared to the simulated order-book depth
and spread distribution in Fig.~\ref{fig:better_wrong_ind_eps_0p2}.
The simulated $\left< n \left( p \right) \right>$ satisfies
Eq.~(\ref{eq:n_m_balance_nodim}), showing what is the correct
``remainder area'' below the line $\left< n \right> \equiv 1$.  The
numerical integral deviates from that value by the incorrect integral
$\int d\hat{p} {\cal S} \neq 0$.  However, most of the probability for
the spread lies within the range where the source terms ${\cal S}$ are
approximately correct, and as a result the distribution for $\hat{s} /
2$ is predicted fairly well.

\begin{figure}[t]
\epsfxsize=2.75in 
\epsfbox{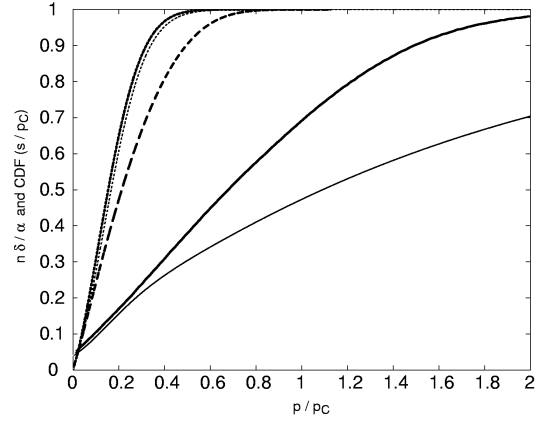}
\caption{
  Reconstruction with correlated source terms for $\epsilon = 0.02$.
  Line style and thickness are the same as in
  Fig.~\ref{fig:better_wrong_ind_eps_0p2}.
  \label{fig:better_wrong_ind_eps_0p02}
}
\end{figure}

\begin{figure}[t]
\epsfxsize=2.75in 
\epsfbox{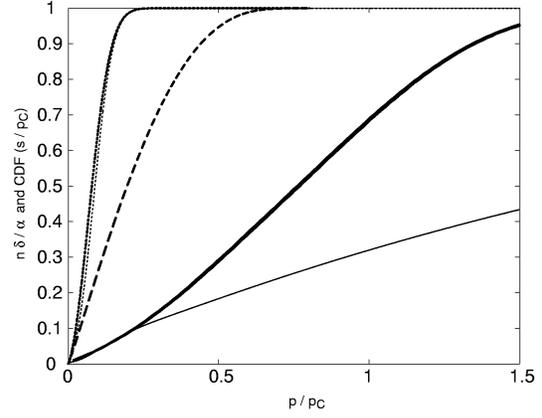}
\caption{
  Reconstruction with correlated source terms for $\epsilon = 0.002$.
  Line style and thickness are the same as in
  Fig.~\ref{fig:better_wrong_ind_eps_0p2}.
  \label{fig:better_wrong_ind_eps_0p002}
}
\end{figure}

Even where the mean-field approximation is known to be inadequate, the
source terms defined here capture most of the behavior of the
order-book distribution in the region that affects the spread
distribution.  Fig.~\ref{fig:better_wrong_ind_eps_0p02} shows the
comparison to simulations for $\epsilon = 0.02$, and
Fig.~\ref{fig:better_wrong_ind_eps_0p002} for $\epsilon = 0.002$.
Both cases fail to reproduce the distribution for the spread, and also
fail to capture the large-$\hat{p}$ behavior of $\psi$.  However, they
approximate $\psi$ at small $\hat{p}$ well enough that the resulting
distribution for the spread is close to what would be produced by the
simulated $\psi$ if fluctuations were independent.

\subsection{A mean-field theory of order separation intervals: The Independent Interval Approximation}

\label{subsec:IIA}

A simplifying assumption that is in some sense dual to independent
fluctuations of $n \left( p \right)$, is independent fluctuations
in the intervals $x \left( N \right)$ at different $N$.  Here we
develop a mean-field theory for the order separation
intervals in this model. From this,  we will also be
able to make an estimate of the depth profiles for
any value of the parameters. For convenience of notation
we will use $x_N$ to denote $x \left( N \right)$.

Limit order placements are considered to take place
strictly on sites which are not occupied. 
This is the same level of approximation as made in the 
previous section.
The time step  is normalized to unity, as above, so
that rates are equal to probabilities after one update of the whole
configuration.  The rates $\alpha$ and $\mu$ used in this
section correspond to $\alpha (\infty)$ and $\mu(0)$ 
as defined earlier.

As shown in Fig. ~\ref{fig:price_bins_label} the configuration
is entirely specified instant by instant if the instantaneous
values of the order separation intervals are known.

Consider now, how these intervals might change due to various 
processes. For the spread $x_0$, these processes
and the corresponding change in $x_0$, are listed below.

\begin{enumerate}

\item $x_0 \rightarrow x_0+x_1$ with rate $\left( \delta + \mu / 2
\sigma \right)$ (when the ask either evaporates or is deleted by a
market order).

\item $x_0 \rightarrow x_0+x_{-1}$ with rate $\left( \delta + \mu
/ 2 \sigma \right)$ (when the bid either evaporates or is deleted by
the corresponding market order).

\item $x_0 \rightarrow x^{\prime}$ for any value $1 \le x^{\prime} \le
x_0-1$, when a sell limit order is deposited anywhere in the spread.
The rate for any single deposition is $\alpha dp / \sigma $, so
the cumulative rate for {\em some} deposition is $\alpha dp \left(
x_0 - 1 \right) / \sigma $.  (The $-1$ comes from the prohibition
against depositing on occupied sites.)

\item Similarly $x_0 \rightarrow x_0- x^{\prime}$ for any $1 \le x^{\prime}
\le x_0-1$, when a buy limit order is deposited in the spread, also
with cumulative rate $\alpha dp \left( x_0 - 1 \right) / \sigma $.

\item Since the above processes describe all possible single-event
changes to the configuration, the probability that it remains
unchanged in a single time step is $1 - 2 \delta - \mu / \sigma - 2
\alpha dp \left( x_0 - 1 \right) / \sigma $.

\end{enumerate}

In all that follows, we will put $\sigma =1$ without
loss of generality.
If we know $x_0$, $x_1$, and $x_{-1}$ at time $t$, the expected value
at time $t + dt$ is then 
\begin{eqnarray}
\lefteqn{
  x_0 
    \left( t+ dt \right)
      = 
  x_0
  \left( t \right) 
  \left[
    1 - 2\delta - 
    { \mu_0 }- 
    { 2 \alpha } \left( x_0-1 \right) 
  \right] 
} & & 
\nonumber \\
& & 
  \mbox{} + 
  \left( x_0+x_1 \right) 
  \left( \delta + \frac{\mu}{2} \right) + 
  \left( x_0+x_{-1} \right) 
  \left( \delta + \frac{\mu}{2} \right)  
\nonumber \\
& & 
  \mbox{} + 
  (\alpha_0 dp) 
   x_0 \left( x_0-1 \right) 
\end{eqnarray}

Here, $x_i(t)$ represents the value of the interval averaged
over many realizations of the process evolved up to time $t$.

Again representing the finite difference as a time derivative, the
change in the expected value, given $x_0$, $x_1$, and $x_{-1}$, is 
\begin{equation}
  \frac{
    d 
     x_0 
  }{
    dt
  } =  
  \left( x_1+x_{-1} \right) 
  \left( 
    \delta + 
    \frac{ \mu }{ 2 } 
  \right) - 
  (\alpha dp) 
  x_0 \left( x_0-1 \right) . 
\label{eq:x_0_mean_from_given}
\end{equation}

Were it not for the quadratic term arising from deposition, 
Eq.~(\ref{eq:x_0_mean_from_given}) would be a linear
function of $ x_0 $, $ x_1 $, and $x_{-1} $.  
However
we now need an approximation for
$ \left< x_0^2 \right> $, where the angle brackets
represent an average over realizations as before
or equivalently a time average in the
steady state.
Let us for the moment assume that we can
approximate $\left< x_0^2 \right>$ by $a {\left< x_0 \right>}^2$, 
where $a$ is some as yet 
undetermined constant to be determined self-consistently.
We will make this approximation for all the $x_k$'s.
This is clearly not entirely accurate because the
PDF of $x_k$ could depend on $k$ (as indeed it does. We
will comment on this a little later). However
as we will see this is still a very good approximation.

We will
therefore make this approximation in
Eq.~(\ref{eq:x_0_mean_from_given})  and everywhere below, and 
look for steady state solutions when the $x_k$'s
have reached a time independent average value.

It then follows that, 
\begin{equation}
  \left(
    \delta + \mu / 2 
  \right)
  \left( 
    x_1+x_{-1}
  \right) = 
  a\alpha dp  
  x_0 
  \left( x_0-1 \right) 
\label{eq:x_0_mean_MFT}
\end{equation}

The interval $x_k$ may be though of 
as the inverse of the density at a distance $ \sum_{j=0}^{k-1} x_j$
from the bid. That is,
$x_i  \approx 1 / \left< n \left( \sum_{j =
0}^{i-1} x_j dp \right) \right>$, the dual to the mean depth, at least
at large $i$.  It therefore makes sense to introduce a normalized
interval
\begin{equation}
  {\hat{x}}_i \equiv 
  \epsilon
  \frac{\alpha }{\delta} x_i dp = 
  \frac{
    x_i dp
  }{
    p_c
  } \approx
  \frac{ 1 }{
    \psi  
    \left(  
      \sum_{j = 0}^{i-1} {\hat{x}}_j 
    \right) 
  } , 
\label{eq:hat_x_i_def}
\end{equation}
the mean-field inverse of the normalized depth $\psi$.  In this
nondimensionalized form, Eq.~(\ref{eq:x_0_mean_MFT}) becomes
\begin{equation}
  \left(
    1 + \epsilon 
  \right)
  \left( 
    {\hat{x}}_1+{\hat{x}}_{-1}
  \right) = 
  {a}{\hat{x}}_0 
  \left( 
    {\hat{x}}_0 - d\hat{p}
  \right) ,
\label{eq:x_0_mean_MFT_nondim}
\end{equation}
where $d\hat{p} = dp/p_c$.

Since the depth profile is symmetric about the
origin, ${\hat{x}}_1 = {\hat{x}}_{-1}$.
From the equations, it can be seen that this 
ansatz is self-consistent and extends to all higher ${\hat{x}}_i$.
Substituting this in Eq.~(\ref{eq:x_0_mean_MFT_nondim}) we
get
\begin{equation}
  \left(
    1 + \epsilon 
  \right)
  {\hat{x}}_1 = 
  \frac{a}{2}
  {\hat{x}}_0 
  \left( 
    {\hat{x}}_0 - d\hat{p}
  \right) = 
  \left(
    1 + \epsilon 
  \right)
  {\hat{x}}_{-1} 
\label{eq:x_0_mean_MFT_nondim_sym}
\end{equation}

Proceeding to the change of $x_1$, the events that can occur, with
their probabilities, are shown in Table~\ref{tab:x_1_change_events}, 
with the remaining probability that $x_1$ remains unchanged.  
\begin{table}
\begin{tabular}{|l|c|c|}
  case & rate & range \\ \hline
  $ x_1 \rightarrow x_2 $ &
  $ \left( \delta + \mu / 2  \right) $ & \\
  $ x_1 \rightarrow \left( x_1 + x_2 \right) $ & 
  $ \delta $ & \\
  $ x_1 \rightarrow x^{\prime} $ & 
  $ \alpha dp$ & 
  $ x^{\prime} \in \left( 1 , x_0 - 1 \right) $ \\
  $ x_1 \rightarrow x_1 - x^{\prime} $ & 
  $ \alpha dp$ & 
  $ x^{\prime} \in \left( 1 , x_1 - 1 \right) $ 
\end{tabular}
\caption{
  Events that can change the value of $x_1$, with their rates of
  occurrence.  
\label{tab:x_1_change_events}
}
\end{table}

The differential equation for the mean change of $x_1$ can be
derived along previous lines and becomes
\begin{eqnarray}
 \frac{dx_1}{dt} 
& = & 
  \left(
    2 \delta + 
    \frac{\mu}{2}
  \right) 
  x_2 - 
  \left(
    \delta + 
    \frac{\mu}{2}
  \right) 
  x_1
\nonumber \\
& & 
  \mbox{} + 
  {\alpha dp}
  \left[
    \frac{
      x_0 
      \left( x_0 - 1 \right) 
    }{2} - 
    \frac{
      x_1 
      \left( x_1 - 1 \right)
    }{2} - 
    x_1 
    \left( x_0 - 1 \right)
  \right]
\nonumber \\
& & 
\label{eq:x_1_change_MFT}
\end{eqnarray}
Note that in the above equations, the mean-field
approximation consists of assuming that
terms like $\left< x_0x_1 \right>$ are
approximated by the product $ \left< x_0 \right> \left< x_1 \right>$.
This is thus an `independent interval' approximation.

Nondimensionalizing Eq.~(\ref{eq:x_1_change_MFT}) and combining the
result with Eq.~(\ref{eq:x_0_mean_MFT_nondim_sym}) gives the
stationary value for $x_2$ from $x_0$ and $x_1$,
\begin{equation}
  \left(
    1 + 2 \epsilon 
  \right)
  {\hat{x}}_2 = 
  \frac{a}{2}
  {\hat{x}}_1 
  \left( 
    {\hat{x}}_1 - d\hat{p}
  \right) + 
  {\hat{x}}_1 
  \left( 
    {\hat{x}}_0 - d\hat{p}
  \right) . 
\label{eq:x_1_change_MFT_nondim}
\end{equation}
Following the same procedure for
general $k$, the nondimensionalized
recursion relation is 
\begin{equation}
  \left(
    1 + k \epsilon 
  \right)
  {\hat{x}}_k = 
  \frac{a}{2}
  {\hat{x}}_{k-1} 
  \left( 
    {\hat{x}}_{k-1} - d\hat{p}
  \right) + 
  {\hat{x}}_{k-1} 
  \sum_{i=0}^{k-2}
  \left( 
    {\hat{x}}_i - d\hat{p}
  \right) . 
\label{eq:x_k_change_MFT_nondim}
\end{equation}

\subsubsection{Asymptotes and conservation rules}

Far from the bid or ask, ${\hat{x}}_k$ must go to a constant value,
which we denote ${\hat{x}}_{\infty}$.  In other words, for large $k$,
${\hat{x}}_{k+1} \rightarrow {\hat{x}}_k$.  Taking the difference of
Equation~(\ref{eq:x_k_change_MFT_nondim}) for $k+1$ and $k$ in this
limit gives the identification
\begin{equation}
  \epsilon {\hat{x}}_{\infty} = 
  {\hat{x}}_{\infty}
  \left( 
    {\hat{x}}_{\infty} - d\hat{p}
  \right) , 
\label{eq:x_infty_id}
\end{equation}
or ${\hat{x}}_{\infty} = \epsilon + d\hat{p}$.  Apart from the factor
of $d\hat{p}$, arising from the exclusion of deposition on
already-occupied sites, this agrees with the limit $\psi \left( \infty
\right) \rightarrow 1 / \epsilon$ found earlier.  In the continuum
limit $d\hat{p} \rightarrow 0$ at fixed $\epsilon$, these are the
same.  

From the large-$k$ limit of Eq.~(\ref{eq:x_k_change_MFT_nondim}), one
can also solve easily for the quantity $S_{\infty} \equiv \sum_{i =
0}^{\infty} \left( {\hat{x}}_i - {\hat{x}}_{\infty} \right)$, which is
related to the bid-centered order conservation law mentioned in Section
\ref{subsec:frames_marginals}.  Dividing by a
factor of  ${\hat{x}}_{\infty}$ at large $k$, 
\begin{equation}
  \left(
    1 + k \epsilon 
  \right) =
  \frac{a}{2}
  \left( 
    {\hat{x}}_{\infty} - d\hat{p}
  \right) + 
  \sum_{i = 0}^{k-2}
  \left( 
    {\hat{x}}_i - d\hat{p}
  \right) , 
\label{eq:x_k_change_MFT_nondim_large_k}
\end{equation}
or, using Eq.~(\ref{eq:x_infty_id}) and rewriting the sum on the
righthand side as $ \sum_{i = 0}^{k-2} (\hat{x}_i - \hat{x}_{\infty})
+ \sum_{i = 0}^{k-2} \hat{x}_{\infty} - d\hat{p}$

\begin{equation}
    1 + 
  (1-\frac{a}{2})\epsilon =
  S_{\infty} . 
\label{eq:solve_S_infty}
\end{equation}

The interpretation of $S_{\infty}$ is straightforward.  There
are $k+1$ orders in the price range $\sum_{i=0}^k x_i$.  Their decay
rate is $\delta \left( k+1 \right)$, and the rate of annihilation from
market orders is $\mu / 2 $.  The rate of additions, up to
an uncertainty about what should be considered the center of the
interval, is $\left( \alpha dp \right) \sum_{i=0}^k
(x_i-1)$ in the bid-centered frame (where effective $\alpha$ is
constant and additions on top of previously occupied sites is forbidden).  
Equality of addition and removal is the bid-centered order
conservation law (again), in the form
\begin{equation}
 \frac{
    \mu 
  }{
    2 
  } + 
  \delta 
  \left( k+1 \right) = 
  {
    \alpha dp
  }
  \sum_{i=0}^k (x_i-1) .
\label{eq:bid_balance_dim}
\end{equation}
Taking $k$ large, nondimensionalizing, and using
Eq.~(\ref{eq:x_infty_id}), Eq.~(\ref{eq:bid_balance_dim}) becomes
\begin{equation}
  1 = S_{\infty} . 
\label{eq:bid_balance_nondim}
\end{equation}
This conservation law is indeed respected to a remarkable
accuracy in Monte Carlo simulations of the model as indicated in table
~\ref{tab:S_infty}.
\begin{table}
\begin{tabular}{|l|c|c|}
$\epsilon$ & $S_\infty $ from theory & $S_\infty$ from MCS \\ \hline
$ 0.66$ & $1$ & $1.000$ \\ 
$ 0.2$ & $1$ & $1.000$ \\
$ 0.04$ & $1$ & $0.998$ \\
$ 0.02$ & $1$ & $1.000$ \\

\end{tabular}
\caption{
Theoretical vs. results from simulations for $S_\infty$.
\label{tab:S_infty}
}
\end{table}

In order that the equation for the $x$'s obey this exact
conservation law, we require Eq. ~\ref{eq:solve_S_infty}
to be equal to Eq. ~\ref{eq:bid_balance_nondim}.
We can hence now self-consistently set the value of $a=2$.

\begin{figure}[t]
\epsfxsize=3.0in 
\epsfbox{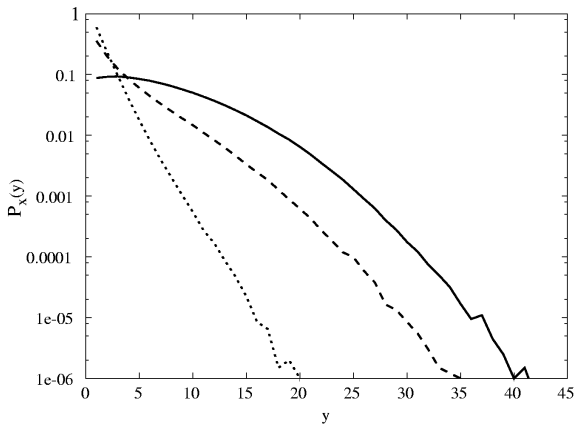}
\caption{
 The probability distribution functions $P_x(y)$ vs. $y$ for the
 intervals $x = x_0,x_1$ and $x_5$ at $\epsilon = 0.1$, on a semi-log
 scale. Solid curve is for $x_0$, dashed for $x_1$, and dot for
 $x_5$. The functional form of the distribution changes from a
 Gaussian to an exponential.
  \label{fig:prob_x}
}
\end{figure}

The value of $a$ implies that we have now set
$ \left< x_k^2 \right> \sim 2 \left< x_k \right> ^2$.
This would be strictly true if the probability distribution
function of the interval $x_k$ were exponentially distributed
for all $k$.  This is generally a good approximation for
large $k$ for any $\epsilon$. Fig ~\ref{fig:prob_x}
shows the numerical results from Monte Carlo simulations of the
model,  for the probability distribution function for three
intervals $x_0$, $x_1$ and $x_5$ at $\epsilon = 0.1$.
The functional form for $P(x_0)$ and $P(x_1)$ are
better approximated by a Gaussian than an exponential. However
$P(x_5)$ is clearly an exponential.

Eq.~(\ref{eq:bid_balance_nondim}) has an important consequence for the
short-term and long-term diffusivities, which can also be seen in
simulations, as mentioned in earlier sections.  The nondimensionalization of the diffusivity $D$ with
the rate parameters, suggests a
classical scaling of the diffusivity
\begin{equation}
  D \sim 
  p_c^2 \delta = 
  \frac{
    \mu^2
  }{
    4 \alpha^2
  }
  \delta . 
\label{eq:D_class_scaling}
\end{equation}
As mentioned earlier, it 
is observed from simulations that the locally best short-time fit
to the actual diffusivity of the midpoint is $\sim \sqrt{1 /
\epsilon}$ times the estimate~(\ref{eq:D_class_scaling}), and the
long-time diffusivity is $\sim \sqrt{\epsilon}$ times the classical
estimate. 
While we do not yet know how to derive this relation
analytically, the fact that early and late-time renormalizations must
have this qualitative relation can be argued from the conservation
law~(\ref{eq:bid_balance_nondim}).

$S_{\infty}$ is the area enclosed 
between the actual density and the asymptotic value.
Increases in $1
/ \epsilon$ (descaled market-order rate) deplete orders near the
spread, diminishing the mean depth at small $\hat{p}$, and induce the
upward curvature seen in Fig.~\ref{epsDepth}, and even more strongly
in Fig.~\ref{fig:direct_x_compare_mid} below.  As noted above, they
cause more frequent shifts (more than compensating for the slight
decrease in average step size), and increase the classically descaled
diffusivity $\beta$.  However, as a result, this increases the 
fraction of the
area in $S_{\infty}$ accumulated near the spread, requiring that the
mean depth at larger $\hat{p}$ increase to compensate (see Fig. ~\ref{fig:direct_x_compare_mid}). The resulting
steeper approach to the asymptotic depth at prices greater than the
mean spread, and the larger negative curvature of the distribution,
are fit by an effective diffusivity that {\em decreases} with
increasing $1 / \epsilon$.  Since the distribution further from the
midpoint represents the imprint of market order activity further in
the past, this effective diffusivity describes the long-term evolution
of the distribution.  The resulting anticorrelation of the
small-$\hat{p}$ and large-$\hat{p}$ effective diffusion constants
implied by conservation of the area $S_{\infty}$ is exactly consistent
with their respective $\sim \sqrt{1 / \epsilon}$ and $\sim
\sqrt{\epsilon}$ scalings. The general idea here is to 
connect diffusivities at short and long time scales
to the depth profile near the spread and far away from the spread
respectively. The conservation law for the depth
profile, then implies a connection between these two diffusivities.

\subsubsection{Direct simulation in interval coordinates}
\label{subsec:xsimulations}

The set of equations determined by the general
form~(\ref{eq:x_k_change_MFT_nondim}) is ultimately parametrized by
the single input ${\hat{x}}_0$.  The correct value for ${\hat{x}}_0$
is determined when the ${\hat{x}}_k$ are solved recursively, by
requiring convergence to ${\hat{x}}_{\infty}$.  We do this recursion
numerically, in the same manner as was done to solve the differential
equation for the normalized mean density $\psi \left( \hat{p}
\right)$. 

\begin{figure}[ptb]
  \begin{center} 
  \includegraphics[scale=0.65]{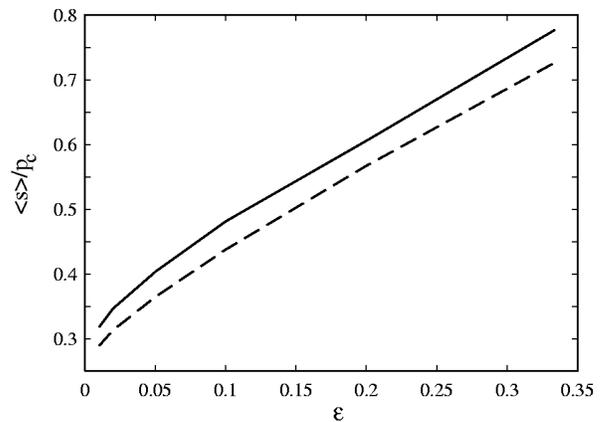}
  \caption{
    The mean value of the spread in nondimensional units $\hat{s} = s
    / p_c$ as a function of $\epsilon$. The numerical value above (solid) is
    compared with the theoretical estimate below (dash).
  \label{spread_compare}.
  }
  \end{center}
\end{figure}

In Fig.~\ref{spread_compare} we compare the numerical result for
${\hat{x}}_0$ with the analytical estimate generated as explained
above.  The results are surprisingly good throughout the entire range.
Though the theoretical value consistently underestimates the numerical
value, yet the functional form is captured accurately.

\begin{figure}[t]
\epsfxsize=3.0in 
\epsfbox{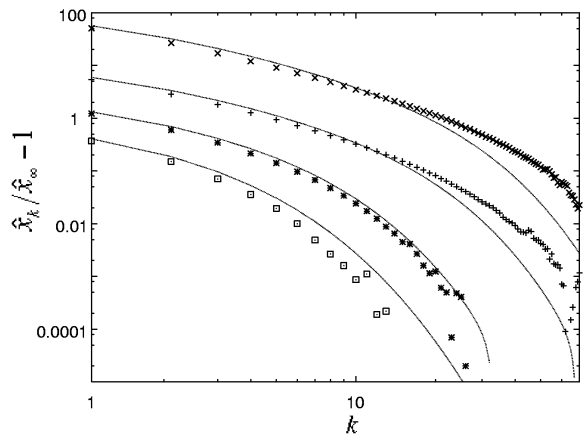}
\caption{
  Four pairs of curves for the quantity
  ${\hat{x}}_k/{\hat{x}}_{\infty} -1$ vs. $k$. The value of $\epsilon$
  increases from top to bottom ($ \epsilon = 0.02, 0.04, 0.2, 0.66$).
  In each pair of curves, the markers are obtained from simulations
  while the solid curve is the prediction of
  Eq.~\ref{eq:x_k_change_MFT_nondim} evaluated numerically. The
  difference between numerics and mean-field increases as $\epsilon$
  decreases, especially for large $k$.
  \label{fig:direct_x_compare}
}
\end{figure}

In Fig.~\ref{fig:direct_x_compare}, the values of $x_k$ for all $k$, 
are compared
to the values determined directly from
simulations.

\begin{figure}[t]
\epsfxsize=3.0in 
\epsfbox{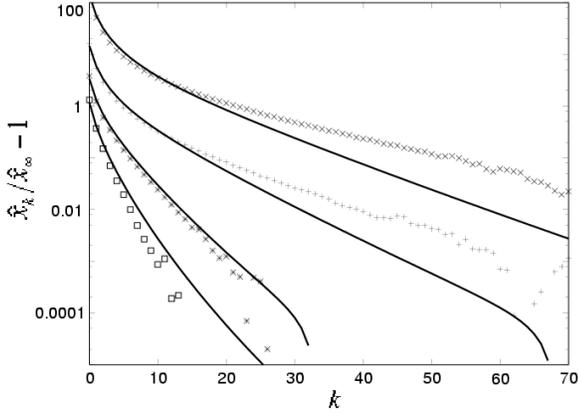}
\caption{
  Same plot as Fig.~\ref{fig:direct_x_compare} but on a semi-log scale
  to show exponential decay at large $k$.
  \label{fig:direct_x_compare_semilog}
}
\end{figure}

Fig.~\ref{fig:direct_x_compare_semilog} shows the same data on a
semilog scale for ${\hat{x}}_k/{\hat{x}}_{\infty} -1$, showing the
exponential decay at large argument characteristic of a simple
diffusion solution.  The IIA is clearly a good approximation for
large $\epsilon$. However for small $\epsilon$ it starts
deviating significantly from the simulations, especially for
large $k$.

\begin{figure}[t]
\epsfxsize=3.0in 
\epsfbox{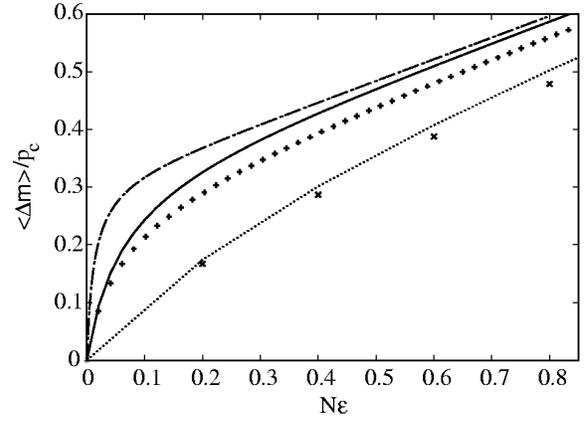}
\caption{
  Three pairs of curves for the quantity $\langle \Delta m \rangle /p_c$ vs. $N \epsilon$
  where 
  $\langle \Delta m \rangle =  1/2\sum_{k=1}^{N} x_{k}$.
  The value of $\epsilon$
  increases from top to bottom ($ \epsilon = 0.002, 0.02, 0.2$).
  In each pair of curves, the markers are obtained from simulations
  while the solid curve is the prediction of the IIA. For $\epsilon = 
  0.002$, we show only the theoretical prediction.
  The theory captures the functional form of the price impact
  curves for different $\epsilon$. Quantitatively, its
  better for larger epsilon, as remarked earlier.            
  \label{priceimpact_th}
}
\end{figure}

The values of $x_k$ computed from the IIA, can be 
very directly used to get an estimation of the
price impact. The price impact, as defined
in earlier sections, can be thought of as the 
change in the position of the midpoint (or the bid), consecutive 
to a certain number of orders being filled. 
Within the framework of the simplified model we study here, 
this is simply the
quantity $\langle \Delta m \rangle= 1/2\sum_{k^{\prime}=1}^{k} x_{k^{\prime}}$, for
$k$ orders. The factor of $1/2$ comes
from considering the change in the position of the midpoint and 
not the bid.
Fig ~\ref{priceimpact_th} shows $\langle \Delta m \rangle$ nondimensionalized
by $p_c$ plotted as a function of the number of orders (multiplied
by $\epsilon$), for three different values of $\epsilon$.
Again, the theory matches quite well with the numerics, qualitatively.
For large $\epsilon$ the agreement
is quantitative as well.

\begin{figure}[t]
\epsfxsize=3.00 in 
\epsfbox{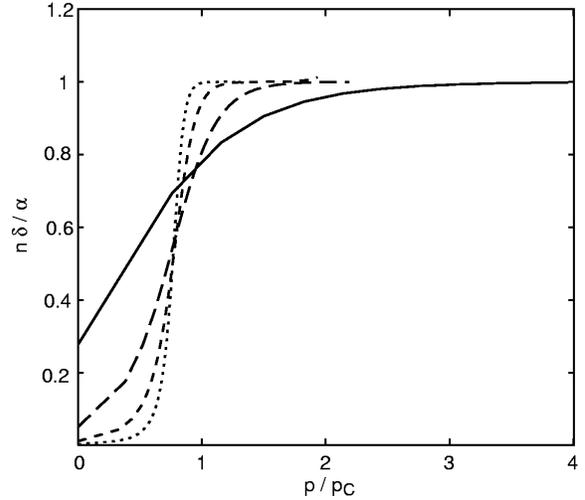}
\caption{
  Density profiles for different values of $\epsilon$ ranging over the
  values $ 0.2, 0.02, 0.004, 0.001$, obtained from the Independent
  Interval Approximation.  
  \label{fig:direct_x_compare_mid}
}
\end{figure}

The simplest approximation to the density profiles in the
midpoint-centered frame is to continue to approximate the mean density
as $1 / x_k$, but to regard that density as evaluated at position $x_0
/ 2 + \sum_{k=1}^i x_k$.  This clearly is not an adequate treatment in
the range of the spread, both because the intervals are discrete,
whereas mean $\psi$ is continuous, and because the density profiles
satisfy different global conservation laws associated with
non-constancy of $\alpha$.  For large $k$ however, this approximation 
might hold. The mean-field values (only) corresponding to a
plot of $\epsilon \psi \left( \hat{p} \right)$ versus $\hat{p}$, are
shown in Fig.~\ref{fig:direct_x_compare_mid}. Here
the theoretically estimated $x_k$'s at different parameter
values are used to generate the depth profile using the
procedure detailed above.

\begin{figure}[t]
\epsfxsize=3.00 in 
\epsfbox{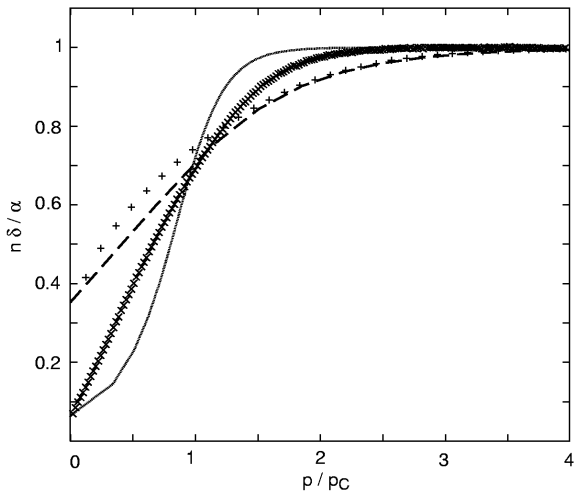}
\caption{
  Density profiles from Monte Carlo simulation (markers) and the
  Independent Interval Approximation (lines).  Pluses and dash line
  are for $\epsilon = 0.2$, while crosses and dotted line are for
  $\epsilon = 0.02$.
  \label{fig:compare}
}
\end{figure}

A comparison of the theoretically estimated profiles with the
results from Monte Carlo simulations of the model, is shown in
Fig. ~\ref{fig:compare}.
As  evident, the theoretical
estimate for the density profile is better for large $\epsilon$
rather than small $\epsilon$.

\begin{figure}[t]
\epsfxsize=3.0in 
\epsfbox{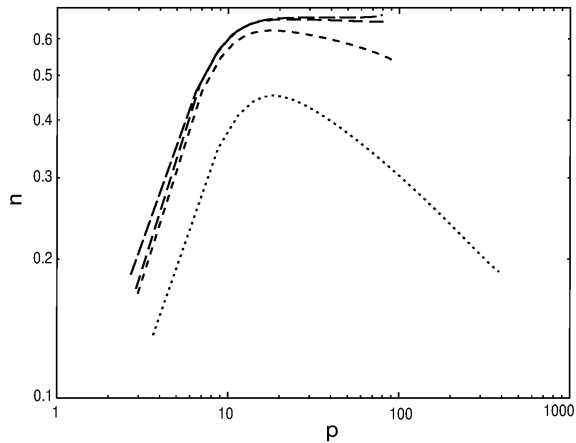}
\caption{
  Density profiles for a power-law order placement process for
  different values of $\Delta_0$.
  \label{fig:powerlaw}
}
\end{figure}

We can also generalize the above analysis to when the order placement
process is no longer uniform. In particular it has been found that a
power-law order placement process is relevant
\cite{Zovko02,Bouchaud02}. We carry out the above analysis for when
$\alpha = {\Delta_0}^{\beta}/{(\Delta +\Delta_0)}^\beta$ where
$\Delta$ is the distance from the current bid and $\Delta_0$
determines the 'shoulder' of the power-law.  We find an interesting
dependence of the existence of solutions on $\beta$. In particular we
find that for $\beta>1$, $\Delta_0$ needs to be larger than some value
(which depends on $\beta$ as well as other parameters of the model
such as $\mu$ and $\delta$) for solutions of the IIA to exist. This
might be interpreted as a market order wiping out the entire book, if
the exponent is too large. When solutions exist, we find that the the
depth profile has a peak , consistent with the findings of
~\cite{Bouchaud02}. In Fig.~\ref{fig:powerlaw} the depth profiles for
three different values of $\Delta_0$ are plotted.

\section{Concluding remarks}
\label{summa}

\subsection{Ongoing work on empirical validation}

Members of our group are working on the problem of empirically testing
this model.  We are using a dataset from the London Stock Exchange.
We have chosen this data because it contains every order and every
cancellation.  This makes it possible to measure all the parameters of
this model directly.  It is also possible to reconstruct the order
book and measure all the statistical properties we have studied in
this paper.  Our empirical work so far shows that, despite its many
limitations, our model can act as an effective guide to future
research.  We believe that the main discrepancies between the
predictions of our model and the data can be dealt with by using a
more sophisticated model of order flow.  We summarize some of the
planned improvements in the following subsection.

\subsection{Future Enhancements}
\label{subsec:futurework}

As we have mentioned above, the zero intelligence, IID order flow
model should be regarded as just a starting point from which to add
more complex behaviors.  We are considering several enhancements to
the order flow process whose effects we intend to discuss in future
papers.  Some of the enhancements include:

\begin{itemize}

\item {\it Trending of order flow.}

We have demonstrated that IID order flow necessarily leads to non-IID
prices.  The converse is also true: Non-IID order flow is necessary
for IID prices.  In particular, the order flow must contain trends,
i.e. if order flow has recently been skewed toward buying, it is more
likely to continue to be skewed toward buying.  If we
assume perfect market efficiency, in the sense that prices are a
random walk, this implies that there must be trends in order flow.

\item {\it Power law placement of limit prices}

For both the London Stock Exchange and the Paris Bourse, the
distribution of the limit price relative to the best bid or ask
appears to decay as a power-law \cite{Bouchaud02,Zovko02}.  Our
investigations of this show that this can have an important effect.
Exponents larger than one result in order books with a finite number
of orders.  In this case, depending on other parameters, there is a
finite probability that a single market order can clear the entire
book (see Section~\ref{subsec:xsimulations}).

\item {\it Power law or log-normal order size distribution.}

Real order placement processes have order size distributions that
appear to be roughly like a log-normal distribution with a power law
tail \cite{Maslov01}.  This has important effects on the 
fluctuations in liquidity. 

\item {\it Non-Poisson order cancellation process.}

When considered in real time order placement cancellation does not appear
to be Poisson \cite{Challet01}.  However, this may not be a bad
approximation in event time rather than real time.

\item {\it Conditional order placement.}

Agents may conditionally place larger market orders when the book is
deeper, causing the market impact function to grow more slowly.  We
intend to measure this effect and incorporate it into our model.

\item {\it Feedback between order flow and prices.}

In reality there are feedbacks between order flow and price movements,
beyond the feedback in the reference point for limit order placement built into this model.  This can induce bursts of trading, causing
order flow rates to speed up or slow down, and give rise to
clustered volatility.

\end{itemize}

The last item is just one of many examples of how one can surely
improve the model by making order flow conditional on available
information.  However, we believe it is important to first gain an
understanding of the properties of simple unconditional models, and
then build on this foundation to get a fuller understanding of the
problem.

\subsection{Comparison to standard models based on valuation and 
information arrival}
\label{valuation}

In the spirit of Gode and Sunder \cite{Gode93}, we assume a simple,
zero-intelligence model of agent behavior and show that the market
institution exerts considerable power in shaping the properties of
prices.  While not disputing that agent behavior might be important,
our model suggests that, at least on the short timescale many of the
properties of the market are dictated by the market institution, and
in particular the need to store supply and demand.  Our model is
stochastic and fully dynamic, and makes predictions that go beyond the
realm of experimental economics, giving quantitative predictions about
fundamental properties of a real market.  We have developed what were
previously conceptual toy models in the physics literature into a
model with testable explanatory power.  

This raises questions about the comparison to standard models based on
the response of valuations to news.  The idea that news might drive
changes in order flow rates is compatible with our model.  That is,
news can drive changes in order flow, which in turn cause the best bid
or ask price to change.  But notice that in our model there are no
assumptions about valuations.  Instead, everything depends on order
flow rates.  For example, the diffusion rate of prices increases as
the $5/2$ power of market order flow rate, and thus volatility, which
depends on the square root of the diffusion rate, increases as the
$5/4$ power.  Of course, order flow rates can respond to information;
an increase in market order rate indicates added impatience, which
might be driven by changes in valuation.  But changes in long-term
valuation could equally well cause an increase in limit order flow
rate, which {\it decreases} volatility.  Valuation {\it per se} does
not determine whether volatility will increase or decrease.  Our model
says that volatility does not depend directly on valuations, but
rather on the urgency with which they are felt, and the need for
immediacy in responding to them.

Understanding the shape of the price impact function was one of the
motivations that originally set this project into motion.  The price
impact function is closely related to supply and demand functions,
which have been central aspects of economic theory since the 19th
century.  Our model suggests that the shape of price impact functions
in modern markets is significantly influenced not so much by strategic
thinking as by an economic fundamental: The need to store supply and
demand in order to provide liquidity.  {\it A priori} it is surprising
that this requirement alone may be sufficient to dictate at least the
broad outlines of the price impact curve.

Our model offers a ``divide and conquer'' strategy to understanding
fundamental problems in economics.  Rather than trying to ground our
approach directly on assumptions of utility, we break the problem into
two parts.  We provide an understanding of how the statistical
properties of prices respond to order flow rates, and leave the
problem open of how order flow rates depend on more fundamental
assumptions about information and utility.  Order flow rates have the
significant advantage that, unlike information, utility, or the
cognitive powers of an agent, they are directly measurable.  We hope
that by breaking the problem into two pieces, and partially solving
the second piece, we can ultimately help provide a deeper
understanding of how markets work.

\appendix

\section{Relationship of Price impact to cumulative depth}

An important aspect of markets is the immediate liquidity, by which we
mean the immediate response of prices to incoming market orders.  When
a market order enters, its execution range depends both on the spread
and on the depth of the orders in the book.  These determine the
sequence of transaction prices produced by that order, as well as the
instantaneous market impact.  Long term liquidity depends on the
longer term response of the limit order book, and is characterized by
the price impact function $\phi(\omega, \tau)$ 
for values of $\tau > 0$.  Immediate
liquidity affects short term volatility, and long term liquidity
affects volatility measured over longer timescales.  In this section
we address only short term liquidity.
We address volatility on longer timescales in section \ref{subsec:longterm}.

We characterize liquidity in terms of either the depth profile or the
price impact.  The \textit{depth profile} $n(p,t)$ is the number of
shares $n$ at price $p$ at time $t$.  For many purposes it is
convenient to think in terms of the {\it cumulative depth profile}
$N$, which is the sum of $n$ values up (or down) to some price.  For
convenience we establish a reference point at the center of the book
where we define $p \equiv 0$ and $N(0) \equiv 0$).  The reference
point can be either the midpoint quote, or the best bid or ask. We
also study the {\it price impact function} $\Delta p = \phi(\omega,
\tau, t)$, where $\Delta p$ is the shift in price at time $t + \tau$
caused by an order of size $\omega$ placed at time $t$.  Typically we
define $\Delta p$ as the shift in the midpoint price, though it is
also possible to use the best bid or ask (Eq. ~\ref{volumeConstraint}).

The price impact function and the depth profile are closely related,
but the relationship is not trivial. $N (\Delta p)$ gives us the average
total number of orders upto  a distance $\Delta p$ away from the origin.
Whereas, in order to calculate the price impact, what we need
is the average shift $\Delta p$ caused by 
a {\it fixed} number of orders.
Making the identifications $p =
\Delta p$, $N = \omega$, and choosing a common reference point, the
instantaneous price impact is the inverse of the instantaneous
cumulative depth,
i.e. $\phi(\omega, 0, t) = N^{-1}(\omega, t)$.  This relationship is
clearly true instant by instant.
However it is not true for averages, since 
the mean of the inverse is not in general equal to the inverse of the
means, i.e. $\langle \phi \rangle \neq {\langle N \rangle}^{-1}$.  This
is highly relevant here, since because the fluctuations in these
functions are huge, our interest is primarily in their statistical
properties, and in particular the first few moments.

A relationship between the moments can be derived as follows:

\subsection{ Moment expansion}
\label{momentExpansion}

There is some subtlety in how we relate the market impact to the
cumulative order count $N \left( p,t \right)$.
One eligible definition of market impact $\Delta p$ is the
movement of the midpoint, following the placement of an order of size
$\omega$.  If we define the reference point so that
$N \left( a,t \right) \equiv 0$, and the market order is a buy, this
definition puts $\omega \left( \Delta p ,t \right) = N \left( a + 2
\Delta p,t \right) - N \left( a,t \right)$.  In words, the midpoint
shift is half the shift in the best offer.  An alternative choice
would be to let $\omega \left( \Delta p ,t \right) = N \left( \Delta
p,t \right)$, which would include part of the instantaneous spread in
the definition of impact in midpoint-centered coordinates, or none of
it in ask-centered coordinates.  The issue of how impact is related to
$N \left( p,t \right)$ is separate from whether the best ask is set
equal to the reference point for prices, and may be chosen differently
to answer different questions.

Under any such definition, however, the impact $\Delta p$ is a
monotonic function of $\omega$ in every instance, so either may be
taken as the independent variable, along with the index $t$ that
labels the instance.  We wish to account for the differences in
instance averages $\left< \right>$ of $\omega$ and $\Delta p$,
regarded respectively as the dependent variables, in terms of the
fluctuations of either.  

In spite of the fact that the density $n \left( p,t \right)$ is a
highly discontinuous variable in general, monotonicity of the
cumulative $N \left( p,t \right)$ enables us to picture a power series
expansion for $\omega \left( p,t \right)$ in $p$, with coefficients
that fluctuate in time.  The simplest such expansion that captures
much of the behavior of the simulated output is 
\begin{equation}
  \omega \left( p,t \right) = 
  a \left( t \right) + 
  b \left( t \right)
  p + 
  \frac{
    c \left( t \right)
  }{
    2
  }
  p^2 , 
\label{eq:impact_inst_expn}
\end{equation}
if $p$ is regarded as the independent variable, or 
\begin{equation}
  p \left( \omega ,t \right) = 
  \frac{
    -b \left( t \right) + 
    \sqrt{
      b^2 \left( t \right) + 
      2
      c \left( t \right)
      \left(
        \omega - 
        a \left( t \right)
      \right)
    }
  }{
    c \left( t \right)
  } , 
\label{eq:inv_impact_inst_expn}
\end{equation}
if $\omega$ is.  While the variable $a \left( t \right)$ would seem
unnecessary since $\omega$ is zero at
$p=0$, empirically we find that simultaneous fits to both  $\omega$ and
${\omega}^2$ at low order can be made better by incorporating the
additional freedom of fluctuations in $a$. 

We imagine splitting each $t$-dependent coefficient into its mean, and
a zero-mean fluctuation component, as
\begin{equation}
  a \left( t \right) \equiv
  \bar{a} + 
  \delta a \left( t \right) , 
\label{eq:a_fluct_expand}
\end{equation}
\begin{equation}
  b \left( t \right) \equiv
  \bar{b} + 
  \delta b \left( t \right)
\label{eq:b_fluct_expand}
\end{equation}
and 
\begin{equation}
  c \left( t \right) \equiv
  \bar{c} + 
  \delta c \left( t \right) . 
\label{eq:c_fluct_expand}
\end{equation}
The fluctuation components will in general depend on $\epsilon$.
The values of the mean and second moment of the fluctuations can be extracted 
from the  mean distributions $\left< \omega \right>$ and $\left< {\omega}^2
\right>$.  The mean values come from the linear expectation: 
\begin{equation}
  \left< \omega \left( 0 \right) \right>
  = 
  \bar{a} , 
\label{eq:bar_a_extract}
\end{equation}
\begin{equation}
  {
    \left.
      \frac{
        \partial 
        \left< \omega \left( p \right) \right>
      }{
        \partial p 
      } 
    \right|
  }_{p = 0}
  = 
  \bar{b} , 
\label{eq:bar_b_extract}
\end{equation}
and 
\begin{equation}
  {
    \left.
      \frac{
        {\partial}^2 
        \left< \omega \left( p \right) \right>
      }{
        {\partial p}^2 
      } 
    \right|
  }_{p = 0}
  = 
  \bar{c} . 
\label{eq:bar_c_extract}
\end{equation}
Given these, the fluctuations then come from the quadratic expectation
as
\begin{equation}
  \left< {\omega}^2 \left( 0 \right) \right>
  = 
  {\bar{a}}^2 + 
  \left< {\delta a}^2 \right> , 
\label{eq:delt_aa_extract}
\end{equation}
\begin{equation}
  {
    \left.
      \frac{
        \partial 
        \left< {\omega}^2 \left( p \right) \right>
      }{
        \partial p 
      } 
    \right|
  }_{p = 0}
  = 
  2 \bar{a} \bar{b} + 
  2 \left< \delta a \delta b \right> , 
\label{eq:delt_ab_extract}
\end{equation}
\begin{equation}
  {
    \left.
      \frac{
        {\partial}^2 
        \left< {\omega}^2 \left( p \right) \right>
      }{
        {\partial p}^2 
      } 
    \right|
  }_{p = 0}
  = 
  2 \left( {\bar{b}}^2 + \bar{a} \bar{c} \right) + 
  2 \left< {\delta b}^2 + \delta a \delta c \right>, 
\label{eq:delt_bb_extract}
\end{equation}
\begin{equation}
  {
    \left.
      \frac{
        {\partial}^3 
        \left< {\omega}^2 \left( p \right) \right>
      }{
        {\partial p}^3 
      } 
    \right|
  }_{p = 0}
  = 
  6 
  \left(
    \bar{b} \bar{c} + 
    \left< \delta b \delta c \right>
  \right) , 
\label{eq:delt_bc_extract}
\end{equation}
and
\begin{equation}
  {
    \left.
      \frac{
        {\partial}^4 
        \left< {\omega}^2 \left( p \right) \right>
      }{
        {\partial p}^4 
      } 
    \right|
  }_{p = 0}
  = 
  6 
  \left(
    {\bar{c}}^2 + 
    \left< {\delta c}^2 \right>
  \right) .  
\label{eq:delt_cc_extract}
\end{equation}
When $\omega$ is given a specific definition in terms of the
cumulative distribution, its averages become averages over the density
in the order book.

The values of the moments as obtained above may then be
used in a derivative expansion of
the inverse function~(\ref{eq:inv_impact_inst_expn}), making the
prediction for the averaged impact 
\begin{eqnarray}
  \left< p  \left( \omega \right) \right> 
& = & 
  \bar{p} + 
  \frac{1}{2}
  \overline{
    \frac{
      {\partial}^2 
      p
    }{
      {\partial a}^2 
    }
  }
  \left< {\delta a}^2 \right> + 
  \frac{1}{2}
  \overline{
    \frac{
      {\partial}^2 
      p
    }{
      {\partial b}^2 
    }
  }
  \left< {\delta b}^2 \right> + 
  \frac{1}{2}
  \overline{
    \frac{
      {\partial}^2 
      p
    }{
      {\partial c}^2 
    }
  }
  \left< {\delta c}^2 \right>
\nonumber \\
& & 
  \mbox{} + 
  \overline{
    \frac{
      {\partial}^2 
      p
    }{
      \partial a \partial b
    }
  }
  \left< \delta a \delta b \right> + 
  \overline{
    \frac{
      {\partial}^2 
      p
    }{
      \partial a \partial c
    }
  }
  \left< \delta a \delta c \right> + 
  \overline{
    \frac{
      {\partial}^2 
      p
    }{
      \partial b \partial c
    }
  }
  \left< \delta b \delta c \right> ,
\nonumber \\
& &  
\label{eq:inverse_fluct_expn}
\end{eqnarray}
where overbar denotes the evaluation of the
function~(\ref{eq:inv_impact_inst_expn}) or its indicated derivative
at $b \left( t \right) = \bar{b}$, $c \left( t \right) = \bar{c}$, and
$\omega$.  The fluctuations $\left< {\delta b}^2 \right>$ and $\left<
\delta a \delta c \right>$ cannot be determined independently from
Eq.~(\ref{eq:delt_bb_extract}).  However, in keeping with this fact,
their coefficient functions in Eq.~(\ref{eq:inverse_fluct_expn}) are
identical, so the inversion remains fully specified.

\begin{figure}[ptb]
  \begin{center} 
\includegraphics[scale=0.37]{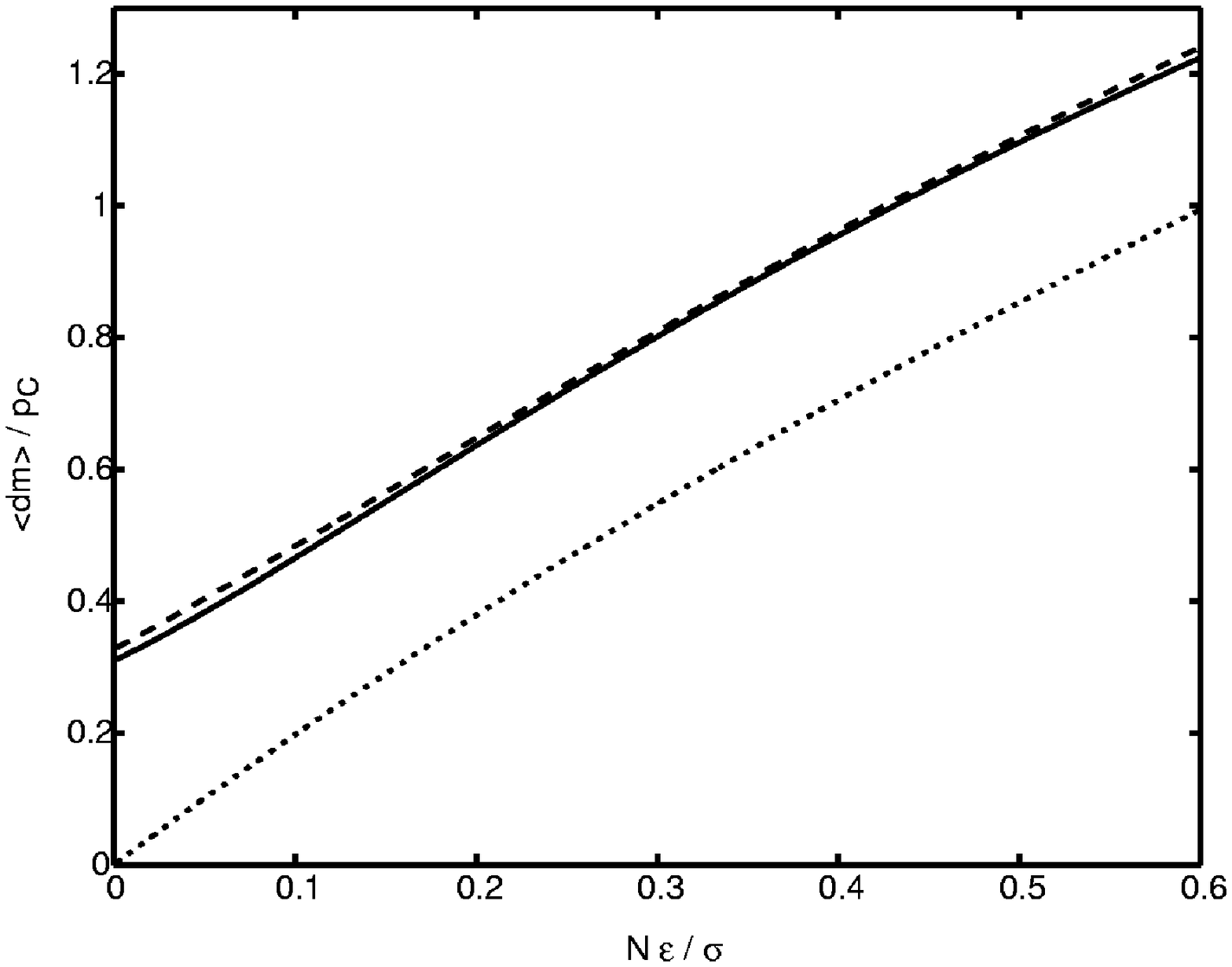}
\includegraphics[scale=0.37]{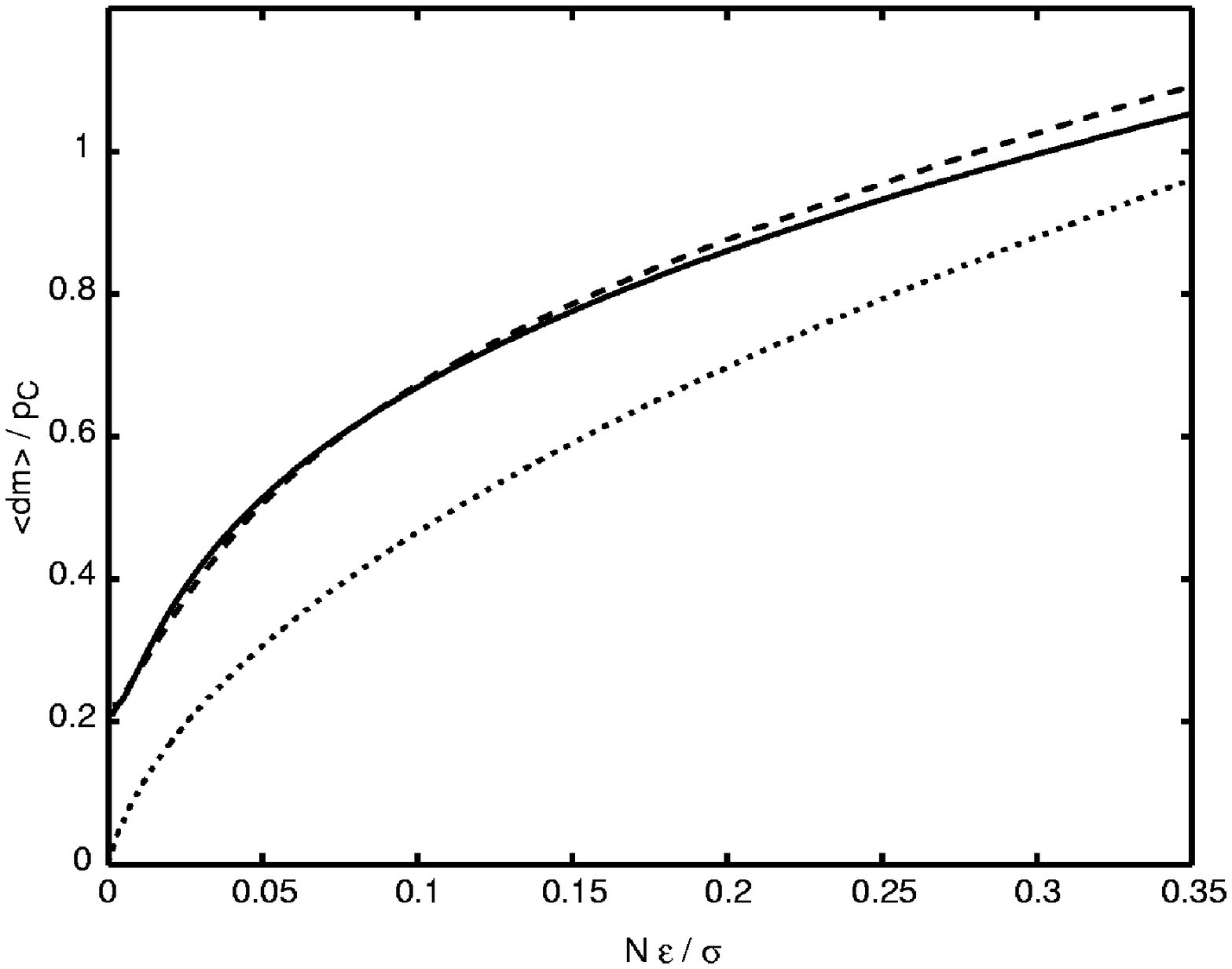}
\includegraphics[scale=0.37]{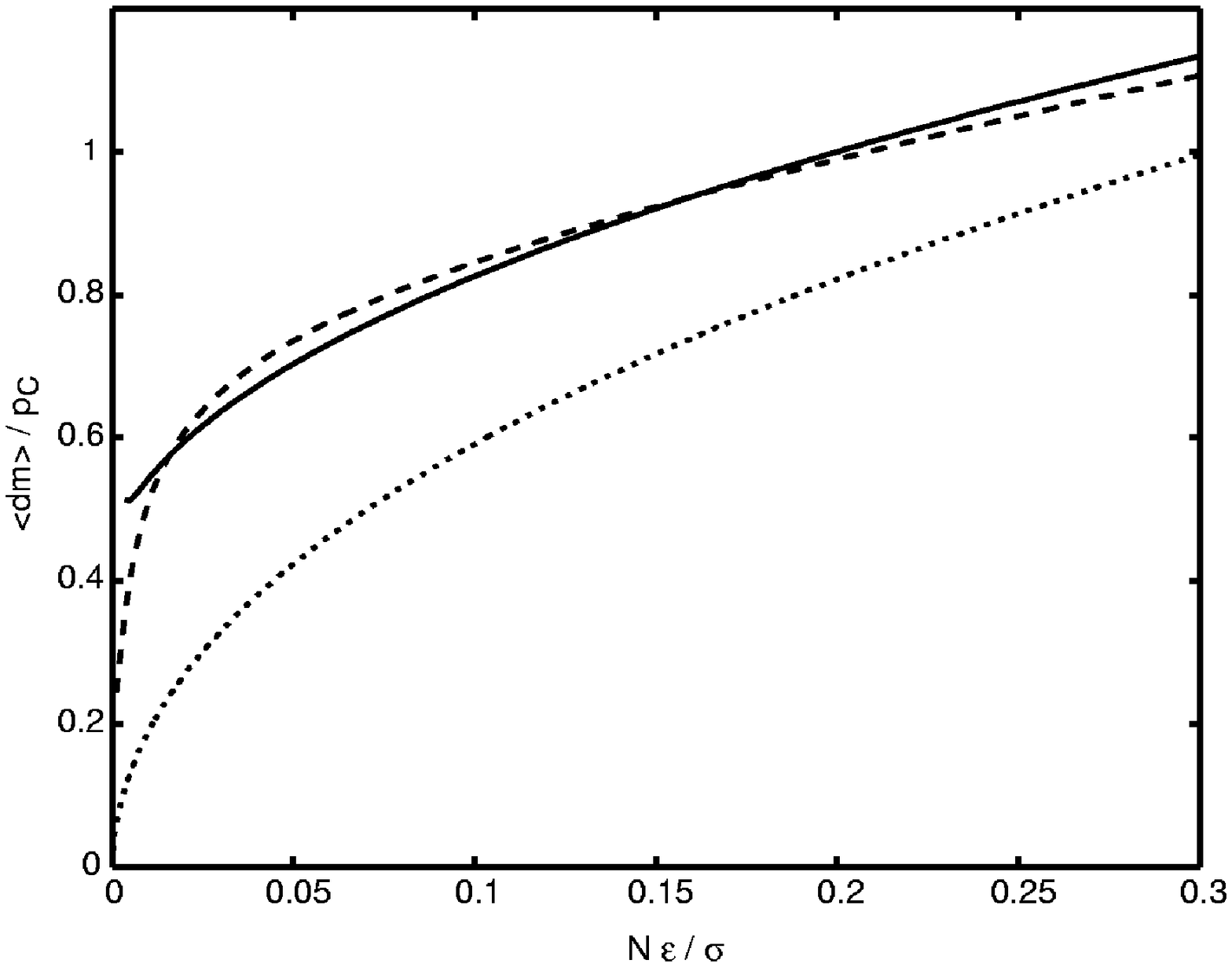}
    \caption{ 
      Comparison of the inverse mean cumulative order distribution
      $\bar{p}$ (dot), to the actual mean impact (solid), and the
      second-order fluctuation expansion~(\ref{eq:inverse_fluct_expn},
      dash).  (a) : $\epsilon = 0.2$.  (b): $\epsilon = 0.02$.  (c):
      $\epsilon = 0.002$.
    \label{fig:moment_expn}
    }  
  \end{center}
\end{figure}

If we denote by $\bar{Z}$ the radical
\begin{equation}
  \bar{Z} \equiv 
  \sqrt{
   {\bar{b}}^2 + 
    2
    \bar{c} 
    \left(
      \omega - \bar{a} 
    \right) 
  } , 
\label{eq:bar_Z_def}
\end{equation}
the various partial derivative functions in
Eq.~(\ref{eq:inverse_fluct_expn}) evaluate to 
\begin{equation}
  \frac{1}{2}
  \overline{
    \frac{
      {\partial}^2 
      p
    }{
      {\partial a}^2 
    }
  } = 
  \frac{
    -\bar{c}
  }{
    {2 \bar{Z}}^3
  } , 
\label{eq:d2p_daa}
\end{equation}
\begin{equation}
  \overline{
    \frac{
      {\partial}^2 
      p
    }{
      \partial a \partial b
    }
  } = 
  \frac{
    \bar{b}
  }{
    {\bar{Z}}^3
  } , 
\label{eq:d2p_dab}
\end{equation}
\begin{equation}
  \frac{1}{2}
  \overline{
    \frac{
      {\partial}^2 
      p
    }{
      {\partial b}^2 
    }
  } = 
  \overline{
    \frac{
      {\partial}^2 
      p
    }{
      \partial a \partial c
    }
  } = 
  \frac{
    \omega - \bar{a}
  }{
    {\bar{Z}}^3
  } , 
\label{eq:d2p_dbb}
\end{equation}
\begin{equation}
  \overline{
    \frac{
      {\partial}^2 
      p
    }{
      \partial b \partial c
    }
  } = 
  \frac{
    1
  }{
    {\bar{c}}^2
  } - 
  \frac{
    \bar{b}
  }{
    \bar{Z} {\bar{c}}^2
  } - 
  \frac{
    \left( 
      \omega - \bar{a}
    \right)
    \bar{b}
  }{
    \bar{c} {\bar{Z}}^3
  } , 
\label{eq:d2p_dbc}
\end{equation}
and 
\begin{equation}
  \frac{1}{2}
  \overline{
    \frac{
      {\partial}^2 
      p
    }{
      {\partial c}^2 
    }
  } = 
  \frac{
    \bar{Z} - \bar{b}
  }{
    {\bar{c}}^3
  } - 
  \frac{
    \omega - \bar{a}
  }{
    2 {\bar{c}}^2 \bar{Z}
  } - 
  \frac{
    {
      \left(
        \omega - \bar{a}
      \right)
    }^2
  }{
    2 \bar{c} {\bar{Z}}^3
  } . 
\label{eq:d2p_dcc}
\end{equation}
Plugging these into Eq.~(\ref{eq:inverse_fluct_expn}) gives the
predicted mean price impact, compared to actual mean in
Fig.~\ref{fig:moment_expn}.  Here the measure used for price impact is
the movement of the ask from buy market orders.  The cumulative order
distribution is computed in ask-centered coordinates, eliminating the
contribution from the half-spread in the $p$ coordinate.  The inverse
of the mean cumulative distribution (dotted), which corresponds to
$\bar{p}$ in Eq.~(\ref{eq:inverse_fluct_expn}), clearly underestimates
the actual mean impact (solid).  However, the corrections from only
second-order fluctuations in $a$, $b$, and $c$ account for much of the
difference at all values of $\epsilon$.  

\subsection{Quantiles}

Another way to characterize the relationship between depth profile and
market impact is in terms of their quantiles (the fraction greater
than a given value, for example the median is the 0.5 quantile).
Interestingly, the relationship between quantiles is trivial.  Letting
$Q_r(x)$ be the $r^{th}$ quantile of $x$, because the 
the cumulative depth $N(p)$ is a non-decreasing function with inverse
$p = \phi (N)$, we have the relation
\begin{equation}
Q_r(\phi) = (Q_{1-r}(N))^{-1}
\end{equation}
This provides an easy and accurate way to compare depth and price
impact when the tick size is sufficiently small.  However, when the
tick size is very coarse, the quantiles are in general not very
useful, because unlike the mean, the quantiles do not vary
continuously, and only take on a few discrete values.

As we have argued in the previous section, in nondimensional
coordinates all of the properties of the limit order book are
described by the two dimensionless scale factors $\epsilon$ and
$dp/p_c$ (see table (\ref{discreteParams}).  When expressed in
dimensionless coordinates, any property, such as depth, spread, or
price impact, can only depend on these two parameters.  This reduces
the search space from five dimensions to two, which greatly simplifies
the analysis.  Any results can easily be re-expressed in dimensional
coordinates from the definitions of the dimensionless parameters.

\section{Supporting calculations in density coordinates}

The following two subsections provide details for the master equation
solution in density coordinates.  The first provides the generating
functional solution for the density at general $dp$, and the second
the approximate source term for correlated fluctuations.  

\subsection{Generating functional at general bin width}
\label{sec:app_genf_large_dp}

As in the main text, $\alpha$ and $\mu$ represent the functions of $p$
everywhere in this subsection, because the boundary values do not
propagate globally.  Eq.~(\ref{eq:diff_form_gen_ME}) can be solved by
assuming there is a convergent expansion in (formal) small $D /
\delta$,
\begin{equation}
  \Pi \equiv
  \sum_j
  {
    \left(
      \frac{D}{\delta}
    \right)
  }^j
  {\Pi}_j , 
\label{eq:series_exp_Pi}
\end{equation}
and it is convenient to embellish the shorthand notation as well, with
\begin{equation}
  {\Pi}_j
  \left( 
    0 , p 
  \right) \equiv
  {{\pi}_0}_j 
  \left( p \right) . 
\label{eq:pi0_notation}
\end{equation}
It follows that expected number also expands as 
\begin{equation}
  \left< n \right> \equiv
  \sum_j
  {
    \left(
      \frac{D}{\delta}
    \right)
  }^j
  {\left< n \right>}_j . 
\label{eq:series_exp_N_ave}
\end{equation}
Order by order in $D / \delta$, Eq.~(\ref{eq:diff_form_gen_ME})
requires 
\begin{equation}
  \left(
    \frac{\partial}{\partial \lambda} - 
    \frac{
      \alpha dp - \mu / 2 \lambda
    }{
      \delta \, \sigma 
    }
  \right)
  {\Pi}_j = 
  \frac{
    \mu 
  }{
    2 \delta \, \sigma \lambda
  }
  {{\pi}_0}_j  + 
  \frac{\partial}{\partial p^2} 
  \frac{
    {\Pi}_{j-1} 
  }{
    \left( \lambda - 1 \right)
  } . 
\label{eq:Pi_master_eq_terms}
\end{equation}
Because ${\Pi}_j$ have been introduced in order to be chosen
homogeneous of degree zero in $D / \delta$, the normalization
condition requires that
\begin{equation}
  {\Pi}_0
  \left( 
    1 , p 
  \right) = 
  1 \mbox{ } , \mbox{ } 
  {\Pi}_{j \neq 0}
  \left( 
    1 , p 
  \right) = 
  0 \mbox{ } , \mbox{ }
  \forall p 
\label{eq:Pi0_limit_conditions}
\end{equation}
The implied recursion relations for expected occupation numbers are 
\begin{equation}
  {\left< n \right>}_0 = 
  \frac{
    \alpha dp 
  }{
    \delta 
  } - 
  \frac{
    \mu 
  }{
    2 \delta 
  }
  \left(
    1 - {{\pi}_0}_0 
  \right) , 
\label{eq:N0_ave_recursion}
\end{equation}
at $j = 0$, and 
\begin{equation}
  {\left< n \right>}_j = 
  \frac{
    \mu 
  }{
    2 \delta 
  }
  {{\pi}_0}_j + 
  \frac{\partial}{\partial p^2} 
  {\left< n  \right>}_{j-1} 
\label{eq:Nj_ave_recursion}
\end{equation}
otherwise.  

Eq.~(\ref{eq:Pi_master_eq_terms}) is solved immediately by use of an
integrating factor, to give the recursive integral relation
\begin{eqnarray} 
  {\Pi}_j 
  \left( \lambda \right) 
& = & 
  {{\pi}_0}_j
  \left[
    1 + 
    \frac{
      \alpha dp
    }{
      \delta \, \sigma
    }
    {\cal I} 
    \left( \lambda \right) 
  \right] 
\nonumber \\
& + & 
  {\cal I} 
  \left( \lambda \right) 
  {
    \left< \! \! \left<
      \frac{\partial}{\partial p^2} \, 
      \frac{{\Pi}_{j-1}}{\lambda - 1} 
    \right> \! \! \right>  
  }_{\lambda} , 
\label{eq:Pi_soln_recursion}
\end{eqnarray}
where 
\begin{equation}
  {\cal I} 
  \left( \lambda \right) \equiv 
  \lambda
  \int_0^1
  dz 
  e^{
    \left(
      \alpha dp / 
      \delta \, \sigma
    \right)
    \left( 1 - z \right)
  }  
  z^{
    \left( \mu / 2 \delta \, \sigma \right)
  } , 
\label{eq:I_of_lambda_def}
\end{equation}
and 
\begin{eqnarray}
  {
    \left< \! \! \left<
      \frac{\partial}{\partial p^2} 
      \frac{{\Pi}_{j-1}}{\lambda - 1} 
    \right> \! \! \right>  
  }_{\lambda} 
& & 
  \equiv
\nonumber \\
  \frac{
    \lambda
  }{
    {\cal I} 
    \left( \lambda \right) 
  }
  \int_0^1
  dz 
& & 
  e^{
    \left(
      \alpha dp / 
      \delta \, \sigma
    \right)
    \left( 1 - z \right)
  }  
  z^{
    \left( \mu / 2 \delta \, \sigma \right)
  }
  \frac{{\partial}^2}{\partial p^2} \, 
  \frac{
    {\Pi}_{j-1} \left( \lambda z \right)
  }{
    \lambda z - 1 
  } . 
\nonumber \\
& & 
\label{eq:lambda_double_ave_def}
\end{eqnarray}
The surface condition~(\ref{eq:Pi0_limit_conditions}) provides the
starting point for this recursion, by giving at $j = 0$
\begin{equation}
  {{\pi}_0}_0 = 
  \frac{
    1
  }{
    1 + 
    \left( \alpha dp / \delta \, \sigma \right)
    {\cal I} \left( 1 \right) 
  } . 
\label{eq:pi_00_eval_general}
\end{equation}
Given forms for $\alpha$ and $\mu$, Eq.~(\ref{eq:N0_ave_recursion})
may be solved directly from Eq.~(\ref{eq:pi_00_eval_general}), and
extended by Eq.~(\ref{eq:Nj_ave_recursion}) to solve for $\left< n
\left( p \right) \right>$.  More generally,
equations~(\ref{eq:I_of_lambda_def}),
(\ref{eq:lambda_double_ave_def}), and~(\ref{eq:pi_00_eval_general})
may be solved to any desired order numerically, to obtain the
fluctuation characteristics of $n \left( p \right)$.  Finding the
solution becomes difficult, however, when $\alpha$ and $\mu$ must be
related self-consistently to the solutions for $\Pi$.
The special case $dp \rightarrow 0$ admits a drastic simplification,
in which the whole expansion for $\left< n \left( p \right) \right>$
may be directly summed, to recover the result in the main text.  
In this limit, one gets a single differential equation in $p$ which
is solvable by numerical integration.  The existence and regularity of
this solution demonstrates the existence of a continuum limit on the
price space, and can be simulated directly by allowing orders to be
placed at arbitrary real-valued prices.

\subsubsection{Recovering the continuum limit for prices}

In the limit that the dimensionless quantity $\alpha dp / \delta \,
\sigma \rightarrow 0$, Eq.~(\ref{eq:I_of_lambda_def}) simplifies to
\begin{equation}
  {\cal I} 
  \left( \lambda \right) \rightarrow 
  \frac{
    \lambda
  }{
    1 + \mu / 2 \delta \, \sigma
  } + 
  {\cal O}
  \left( dp \right) , 
\label{eq:I_lambda_simplifies}
\end{equation}
from which it follows that 
\begin{equation}
  {\left< n \right>}_0 \rightarrow 
  \frac{
    \alpha dp / \delta 
  }{
    1 + \mu / 2 \delta \, \sigma
  } + 
  {\cal O}
  \left( {dp}^2 \right) . 
\label{eq:N0_simplifies}
\end{equation}

The important simplification given by vanishing $dp$, as will be seen
below, is that the expansion~(\ref{eq:lamb_power_expn}) collapses, at
leading order in $dp$, to 
\begin{equation}
  {\Pi}_0 
  \left( \lambda \right) \rightarrow 
  1 + 
  \left( \lambda - 1 \right)
  \frac{
    {\left< n \right>}_0 
  }{
    \sigma 
  } + 
  {\cal O}
  \left( {dp}^2 \right) . 
\label{eq:Pi0_simplifies}
\end{equation}
Eq.~(\ref{eq:Pi0_simplifies}) is used as the input to an inductive
hypothesis 
\begin{equation}
  {\Pi}_{j-1} 
  \left( \lambda \right) \rightarrow 
  {\Pi}_{j-1} 
  \left( 1 \right) + 
  \left( \lambda - 1 \right)
  \frac{
    {\left< n \right>}_{j-1}
  }{
    \sigma 
  } + 
  {\cal O}
  \left( {dp}^2 \right) , 
\label{eq:Pij_hypothesis}
\end{equation}
(n.~b.~$ {\left< n \right>}_{j-1} \sim {\cal O} \left( dp \right)$,
${\Pi}_{j-1} \left( 1 \right) = \mbox{ either } 1 \mbox{ or } 0$), 
which with Eq.~(\ref{eq:lambda_double_ave_def}), then recovers the
condition at $j$:
\begin{equation}
  {\Pi}_j 
  \left( \lambda \right) \rightarrow 
  \left( \lambda - 1 \right)
  {\cal I} 
  \left( 1 \right) 
  \frac{d^2}{d p^2} \, 
  \frac{
    {\left< n \right>}_{j-1} 
  }{
    \sigma 
  } + 
  {\cal O}
  \left( {dp}^2 \right) . 
\label{eq:Pij_consequence}
\end{equation}
Using Eq.~(\ref{eq:lamb_power_expn}) at $\lambda \rightarrow 1$, and
Eq.~(\ref{eq:I_lambda_simplifies}) for ${\cal I}$, gives the recursion
for the number density
\begin{equation}
  {\left< n \right>}_{j \neq 0} \rightarrow 
  \frac{
    1
  }{
    1 + \mu / 2 \delta \, \sigma
  } \, 
  \frac{d^2}{d p^2} 
  {\left< n \right>}_{j-1} + 
  {\cal O}
  \left( {dp}^2 \right) . 
\label{eq:N_ave_j_form}
\end{equation}
The sum~(\ref{eq:series_exp_N_ave}) for $\left< n \right>$ is then
\begin{equation}
  \left< n \right> = 
  \sum_j
  {
    \left(
      \frac{D}{\delta} \, 
      \frac{
        1
      }{
        1 + \mu / 2 \delta \, \sigma
      } \, 
      \frac{d^2}{d p^2} 
    \right)
  }^j
  {\left< n \right>}_0 .  
\label{eq:N_ave_formal_expn}
\end{equation}
Using Eq.~(\ref{eq:N0_simplifies}) for ${\left< n \right>}_0$ and
re-arranging terms, Eq.~(\ref{eq:N_ave_formal_expn}) is equivalent to 
\begin{equation}
  \left< n \right> = 
  \frac{
    1
  }{
    1 + \mu / 2 \delta \, \sigma
  }  
  \sum_j
  {
    \left(
      \frac{D}{\delta} \, 
      \frac{d^2}{d p^2} \, 
      \frac{
        1
      }{
        1 + \mu / 2 \delta \, \sigma
      } 
    \right)
  }^j
  \frac{
    \alpha dp 
  }{
    \delta
  } . 
\label{eq:N_reverse_series}
\end{equation}

The series expansion in the price Laplacian is formally the geometric
sum 
\begin{equation}
  \left(
    1 + \mu / 2 \delta \, \sigma
  \right)
  \left< n \right> = 
  {
    \left[
      1 - 
      \frac{D}{\delta} \, 
      \frac{d^2}{d p^2} \, 
      \frac{
        1
      }{
        1 + \mu / 2 \delta \, \sigma
      } 
    \right]
  }^{-1} 
  \frac{
    \alpha dp 
  }{
    \delta
  } , 
\label{eq:N_sum_reversed_series}
\end{equation}
which can be inverted to give Eq.~(\ref{eq:series_invert}), a relation
that is local in derivatives.

\subsection{Cataloging correlations}
\label{sec:FK_expand_corrs}

A correct source term ${\cal S}$ must correlate the incidences of zero
occupation with the events producing shifts.  It is convenient to
separate these into the four independent types of deposition and
removal.  

First we consider removal of buy limit orders, which generates a
negative shift of the midpoint.  Let $\hat{a}'$ denote the position of
the ask after the shift.  Then all possible shifts $\Delta \hat{p}$
are related to a given price bin $\hat{p}$ and $\hat{a}'$ in one of
three ordering cases, shown in Table~\ref{tab:buy_lim_ord_rem}.  For
each case, the source term corresponding to $\left[ \psi \left(
\hat{p} - \Delta \hat{p} \right) - \psi \left( \hat{p} \right)
\right]$ in Eq.~(\ref{eq:gen_fnal_master_nondim}) is given, together
with the measure of order-book configurations for which that case
occurs.  The mean-field assumption~(\ref{eq:factor_test_ctm}) is used
to estimate these measures.

\begin{table}
\begin{tabular}{|c|c|c|}
  \mbox{case} & \mbox{source} & \mbox{prob} \\ \hline
  $ \Delta \hat{p} \le 
  \hat{a}' <
  \hat{p} $ & 
  $ \psi \left( \hat{p} - \Delta \hat{p} \right) - 
  \psi \left( \hat{p} \right) $ & 
  $ \varphi \left( \Delta \hat{p} \right) - 
  \varphi \left( \hat{p} \right) $ \\ \hline
  $ \Delta \hat{p} \le \hat{p} < 
  \hat{a}' \le
  \hat{p} + \Delta \hat{p} $ & 
  $ 0 - 
  \psi \left( \hat{p} \right) $ & 
  $ \varphi \left( \hat{p} \right) - 
  \varphi \left( \hat{p} + \Delta \hat{p}\right) $ \\ \hline
  $ \hat{p} < \Delta \hat{p} < 
  \hat{a}' \le
  \hat{p} + \Delta \hat{p} $ & 
  $ 0 - 
  \psi \left( \hat{p} \right) $ & 
  $ \varphi \left( \Delta \hat{p} \right) - 
  \varphi \left( \hat{p} + \Delta \hat{p}\right) $ \\ \hline
\end{tabular}
\caption{
  Contributions to ``effective $P_{-}$'' from removal of a buy limit
  order, conditioned on the position of the ask relative to $p$. 
  \label{tab:buy_lim_ord_rem}
}
\end{table}

As argued when defining $\beta$ in the simpler diffusion approximation
for the source terms, the measure of shifts from removal of either buy
or sell limit orders should be symmetric with that of their addition
within the spread, which is is $2 d \Delta \hat{p}$ for either type,
in cases when the shift $\pm \Delta \hat{p}$ is consistent with the
value of the spread.  The only change in these more detailed source
terms is replacement of the simple $\mbox{Pr} \left( a \ge \Delta p
\right)$ with the entries in the third column of
Table~\ref{tab:buy_lim_ord_rem}.  When the $\Delta \hat{p}$ cases are
integrated over their range as specified in the first column and
summed, the result is a contribution to ${\cal S}$ of 
\begin{eqnarray}
& & 
  \int_0^{\hat{p}}
  2 d \Delta \hat{p} \, 
  \psi \left( \hat{p} - \Delta \hat{p} \right) 
  \left[
    \varphi \left( \Delta \hat{p} \right) - 
    \varphi \left( \hat{p} \right) 
  \right]
\nonumber \\
& & 
  \mbox{} - 
  \int_0^{\infty}
  2 d \Delta \hat{p} \, 
  \psi \left( \hat{p} \right) 
  \left[
    \varphi \left( \Delta \hat{p} \right) - 
    \varphi \left( \hat{p} + \Delta \hat{p} \right) 
  \right]
\label{eq:buy_lim_ord_rem_src}
\end{eqnarray}

\begin{table}
\begin{tabular}{|c|c|c|}
  \mbox{case} & \mbox{source} & \mbox{prob} \\ \hline
  $ \Delta \hat{p} \le 
  \hat{a}' <
  \hat{p} $ & 
  $ \psi \left( \hat{p} + \Delta \hat{p} \right) - 
  \psi \left( \hat{p} \right) $ & 
  $ \varphi \left( \Delta \hat{p} \right) - 
  \varphi \left( \hat{p} \right) $ \\ \hline
  $ \Delta \hat{p} \le \hat{p} < 
  \hat{a}' \le
  \hat{p} + \Delta \hat{p} $ & 
  $ \psi \left( \hat{p} + \Delta \hat{p} \right) - 
  0 $ & 
  $ \varphi \left( \hat{p} \right) - 
  \varphi \left( \hat{p} + \Delta \hat{p}\right) $ \\ \hline
  $ \hat{p} < \Delta \hat{p} < 
  \hat{a}' \le
  \hat{p} + \Delta \hat{p} $ & 
  $ \psi \left( \hat{p} + \Delta \hat{p} \right) - 
  0 $ & 
  $ \varphi \left( \Delta \hat{p} \right) - 
  \varphi \left( \hat{p} + \Delta \hat{p}\right) $ \\ \hline
\end{tabular}
\caption{
  Contributions to ``effective $P_{+}$'' from removal of a sell limit
  order, conditioned on the position of the ask relative to $p$.
  \label{tab:sell_lim_ord_rem}
}
\end{table}

Sell limit-order removals generate another sequence of cases,
symmetric with the buys, but inducing positive shifts.  The cases,
source terms, and frequencies are given in
Table~\ref{tab:sell_lim_ord_rem}.  Their contribution to ${\cal S}$,
after integration over $\Delta \hat{p}$, is then
\begin{eqnarray}
& & 
  \int_0^{\infty}
  2 d \Delta \hat{p} \, 
  \psi \left( \hat{p} + \Delta \hat{p} \right) 
  \left[
    \varphi \left( \Delta \hat{p} \right) - 
    \varphi \left( \hat{p} + \Delta \hat{p} \right) 
  \right]
\nonumber \\
& & 
  \mbox{} - 
  \int_0^{\hat{p}}
  2 d \Delta \hat{p} \, 
  \psi \left( \hat{p} \right) 
  \left[
    \varphi \left( \Delta \hat{p} \right) - 
    \varphi \left( \hat{p} \right) 
  \right]
\label{eq:sell_lim_ord_rem_src}
\end{eqnarray}

\begin{table}
\begin{tabular}{|c|c|c|}
  \mbox{case} & \mbox{source} & \mbox{prob} \\ \hline
  $ \Delta \hat{p} \le 
  \hat{a} <
  \hat{p} - \Delta \hat{p}$ & 
  $ \psi \left( \hat{p} - \Delta \hat{p} \right) - 
  \psi \left( \hat{p} \right) $ & 
  $ \varphi \left( \Delta \hat{p} \right) - 
  \varphi \left( \hat{p} - \Delta \hat{p} \right) $ \\ \hline
  $ \Delta \hat{p} \le \hat{p} - \Delta \hat{p} < 
  \hat{a} \le
  \hat{p} $ & 
  $ 0 - 
  \psi \left( \hat{p} \right) $ & 
  $ \varphi \left( \hat{p} - \Delta \hat{p} \right) - 
  \varphi \left( \hat{p} \right) $ \\ \hline
  $ \hat{p} - \Delta \hat{p} < \Delta \hat{p} < 
  \hat{a} \le
  \hat{p} $ & 
  $ 0 - 
  \psi \left( \hat{p} \right) $ & 
  $ \varphi \left( \Delta \hat{p} \right) - 
  \varphi \left( \hat{p} \right) $ \\ \hline
\end{tabular}
\caption{
  Contributions to ``effective $P_{-}$'' from addition of a sell limit
  order, conditioned on the position of the ask relative to $p$.
  \label{tab:sell_lim_ord_add}
}
\end{table}

Order addition is treated similarly, except that $\hat{a}$ denotes the
position of the ask before the event.  Sell limit-order additions
generate negative shifts, with the cases shown in
Table~\ref{tab:sell_lim_ord_add}.  Integration over $\Delta \hat{p}$
consistent with these cases gives the negative-shift contribution to
${\cal S}$
\begin{eqnarray}
& & 
  \int_0^{\hat{p}/2}
  2 d \Delta \hat{p} \, 
  \psi \left( \hat{p} + \Delta \hat{p} \right) 
  \left[
    \varphi \left( \Delta \hat{p} \right) - 
    \varphi \left( \hat{p} - \Delta \hat{p} \right) 
  \right]
\nonumber \\
& & 
  \mbox{} - 
  \int_0^{\hat{p}}
  2 d \Delta \hat{p} \, 
  \psi \left( \hat{p} \right) 
  \left[
    \varphi \left( \Delta \hat{p} \right) - 
    \varphi \left( \hat{p} + \Delta \hat{p} \right) 
  \right]
\label{eq:sell_lim_ord_add_src}
\end{eqnarray}

\begin{table}[t]
\begin{tabular}{|c|c|c|}
  \mbox{case} & \mbox{source} & \mbox{prob} \\ \hline
  $ \Delta \hat{p} \le 
  \hat{a}' <
  \hat{p} $ & 
  $ \psi \left( \hat{p} + \Delta \hat{p} \right) - 
  \psi \left( \hat{p} \right) $ & 
  $ \varphi \left( \Delta \hat{p} \right) - 
  \varphi \left( \hat{p} \right) $ \\ \hline
  $ \Delta \hat{p} \le \hat{p} < 
  \hat{a}' \le
  \hat{p} + \Delta \hat{p} $ & 
  $ \psi \left( \hat{p} + \Delta \hat{p} \right) - 
  0 $ & 
  $ \varphi \left( \hat{p} \right) - 
  \varphi \left( \hat{p} + \Delta \hat{p} \right) $ \\ \hline
  $ \hat{p} < \Delta \hat{p} < 
  \hat{a}' \le
  \hat{p} + \Delta \hat{p} $ & 
  $ \psi \left( \hat{p} + \Delta \hat{p} \right)  - 
  0 $ & 
  $ \varphi \left( \Delta \hat{p} \right) - 
  \varphi \left( \hat{p} + \Delta \hat{p}\right) $ \\ \hline
\end{tabular}
\caption{
  Contributions to ``effective $P_{+}$'' from addition of a buy limit
  order, conditioned on the position of the ask relative to $p$.
  \label{tab:buy_lim_ord_add}
}
\end{table}

The corresponding buy limit-order addition cases are given in
Table~\ref{tab:buy_lim_ord_add}, and their positive-shift contribution
to ${\cal S}$ turns out to be the same as that from removal of sell
limit orders~(\ref{eq:sell_lim_ord_rem_src}).

Writing the source as a sum of two terms ${\cal S} \equiv {\cal
S}_{\mbox{\scriptsize buy}} + {\cal S}_{\mbox{\scriptsize sell}}$, the
combined contribution from buy limit-order additions and removals is
\begin{eqnarray}
\lefteqn{
  {\cal S}_{\mbox{\scriptsize buy}}
  \left( \hat{p} \right) = 
  \int_0^{\hat{p}}
  2 d \Delta \hat{p} \, 
  \left[
    \psi \left( \hat{p} - \Delta \hat{p} \right) - 
    \psi \left( \hat{p} \right) 
  \right]
  \left[
    \varphi \left( \Delta \hat{p} \right) - 
    \varphi \left( \hat{p} \right) 
  \right]
} & & 
\nonumber \\
& & 
  \mbox{} - 
  \int_0^{\infty}
  2 d \Delta \hat{p} \, 
  \left[
    \psi \left( \hat{p} + \Delta \hat{p} \right) - 
    \psi \left( \hat{p} \right) 
  \right]
  \left[
    \varphi \left( \Delta \hat{p} \right) - 
    \varphi \left( \hat{p} + \Delta \hat{p} \right) 
  \right]
\nonumber \\
& & 
\label{eq:buy_lim_ord_both_src}
\end{eqnarray}
The corresponding source term from sell order addition and removal is 
\begin{eqnarray}
  {\cal S}_{\mbox{\scriptsize sell}}
  \left( \hat{p} \right)
& = & 
  \int_0^{\hat{p}/2}
  2 d \Delta \hat{p} \, 
  \psi \left( \hat{p} - \Delta \hat{p} \right) 
  \left[
    \varphi \left( \Delta \hat{p} \right) - 
    \varphi \left( \hat{p} - \Delta \hat{p} \right) 
  \right]
\nonumber \\
& & 
  \mbox{} - 2 
  \int_0^{\hat{p}}
  2 d \Delta \hat{p} \, 
  \psi \left( \hat{p} \right) 
  \left[
    \varphi \left( \Delta \hat{p} \right) - 
    \varphi \left( \hat{p} \right) 
  \right]
\nonumber \\
& & 
  \mbox{} + 
  \int_0^{\infty}
  2 d \Delta \hat{p} \, 
  \psi \left( \hat{p} + \Delta \hat{p} \right) 
  \left[
    \varphi \left( \Delta \hat{p} \right) - 
    \varphi \left( \hat{p} + \Delta \hat{p} \right) 
  \right]
\nonumber \\
& & 
\label{eq:sell_lim_ord_both_src}
\end{eqnarray}

The forms~(\ref{eq:buy_lim_ord_both_src})
and~(\ref{eq:sell_lim_ord_both_src}) do not lead to $\int d\hat{p}
{\cal S} = 0$, and correcting this presumably requires distributing
the orders erroneously transported through the midpoint by the
diffusion term, to interior locations where they then influence
long-time diffusion autocorrelation.  These source terms manifestly
satisfy ${\cal S} \left( 0 \right) = 0$, though, and that determines
the intercept of the average order depth.

\subsubsection{Getting the intercept right}

Evaluating Eq.~(\ref{eq:gen_mast_anon_src}) with $\alpha \left(
\hat{p} \right) / \alpha \left( \hat{\infty} \right) = 1 + \mbox{Pr} \left(
\hat{s}/2 \ge \hat{p} \right) $, at $\hat{p} = 0$ gives the boundary
value of the nondimensionalized, midpoint-centered, mean order density
\begin{equation}
  {\psi} \left( 0 \right) = 
  \frac{2}{1 + \epsilon} , 
\label{eq:nondim_ord_mid}
\end{equation}
which dimensionalizes to 
\begin{equation}
  \frac{
    \left< 
      n \left( 0 \right) 
    \right> 
  }{\sigma dp} = 
  \frac{
    2 \alpha \left( \infty \right) / \sigma 
  }{
    \mu \left( 0 \right) / 2 \sigma  + \delta
  } .  
\label{eq:dim_ord_mid}
\end{equation}
Eq.~(\ref{eq:dim_ord_mid}), for the {\em total} density, is the same
as the form~(\ref{eq:N0_ave_recursion}) produced by the diffusion
solution for the {\em zeroth order} density, as should be the case if
diffusion no longer transports orders through the midpoint.  This form
is verified in simulations, with midpoint-centered averaging.  

Interestingly, the same argument for the bid-centered frame would
simply omit the $\varphi$ from $\alpha \left( 0 \right) / \alpha
\left( \infty \right) $, predicting that
\begin{equation}
  {\psi} \left( 0 \right) = 
  \frac{1}{1 + \epsilon} , 
\label{eq:nondim_ord_bid}
\end{equation}
a result which is {\em not} confirmed in simulations.  Thus, in
addition to not satisfying the mean-field approximation, the
bid-centered density average appears to receive some diffusive
transport of orders all the way down to the bid.

\subsubsection{Fokker-Planck expanding correlations}

Equations~(\ref{eq:buy_lim_ord_both_src})
and~(\ref{eq:sell_lim_ord_both_src}) are not directly easy to use in a
numerical integral.  However, they can be Fokker-Planck expanded to
terms with behavior comparable to the diffusion equation, and the
correct behavior near the midpoint.  Doing so gives the nondimensional
expansion of the source term ${\cal S}$ corresponding to the diffusion
contribution in Eq.~(\ref{eq:series_invert_nondim_more}): 
\begin{equation}
  {\cal S} = 
  {\cal R} \left( \hat{p} \right)
  \psi \left( \hat{p} \right) + 
  {\cal P} \left( \hat{p} \right)
  \frac{
    d
    \psi \left( \hat{p} \right)
  }{
    d \hat{p}
  } + 
  \epsilon \beta \left( \hat{p} \right)
  \frac{
    {
      d
    }^2
    \psi \left( \hat{p} \right)
  }{
    {
      d \hat{p}
    }^2
  } . 
\label{eq:detailed_nondim_src_formal}
\end{equation}
The rate terms in Eq.~(\ref{eq:detailed_nondim_src_formal}) are
integrals defined as 
\begin{eqnarray}
  {\cal R} \left( \hat{p} \right)
& = & 
  \int_0^{\infty}
  2d \Delta \hat{p}
  \left[
    \varphi 
    \left( \Delta \hat{p} \right) - 
    \varphi 
    \left( \hat{p} + \Delta \hat{p} \right)
  \right]
\nonumber \\
& & 
  \mbox{} - 
  2 \int_0^{\hat{p}}
  2d \Delta \hat{p}
  \left[
    \varphi 
    \left( \Delta \hat{p} \right) - 
    \varphi 
    \left( \hat{p} \right)
  \right]
\nonumber \\
& & 
  \mbox{} + 
  \int_0^{\hat{p} / 2}
  2d \Delta \hat{p}
  \left[
    \varphi 
    \left( \Delta \hat{p} \right) - 
    \varphi 
    \left( \hat{p} - \Delta \hat{p} \right)
  \right] , 
\label{eq:cal_R_express}
\end{eqnarray}
\begin{eqnarray}
  {\cal P} \left( \hat{p} \right)
& = & 
  2 \int_0^{\infty}
  2d \Delta \hat{p} \, 
  \Delta \hat{p}
  \left[
    \varphi 
    \left( \Delta \hat{p} \right) - 
    \varphi 
    \left( \hat{p} + \Delta \hat{p} \right)
  \right]
\nonumber \\
& & 
  \mbox{} - 
  \int_0^{\hat{p}}
  2d \Delta \hat{p} \, 
  \hat{p}
  \left[
    \varphi 
    \left( \Delta \hat{p} \right) - 
    \varphi 
    \left( \hat{p} \right)
  \right]
\nonumber \\
& & 
  \mbox{} - 
  \int_0^{\hat{p} / 2}
  2d \Delta \hat{p} \, 
  \Delta \hat{p}
  \left[
    \varphi 
    \left( \Delta \hat{p} \right) - 
    \varphi 
    \left( \hat{p} - \Delta \hat{p} \right)
  \right] , 
\label{eq:cal_P_express}
\end{eqnarray}
and 
\begin{eqnarray}
  \epsilon \beta \left( \hat{p} \right)
& = & 
  \int_0^{\infty}
  2d \Delta \hat{p} \, 
  {
    \left( \Delta \hat{p} \right)
  }^2
  \left[
    \varphi 
    \left( \Delta \hat{p} \right) - 
    \varphi 
    \left( \hat{p} + \Delta \hat{p} \right)
  \right]
\nonumber \\
& & 
  \mbox{} + 
  \frac{1}{2}
  \int_0^{\hat{p}} \, 
  {
    \left( \Delta \hat{p} \right)
  }^2
  2d \Delta \hat{p}
  \left[
    \varphi 
    \left( \Delta \hat{p} \right) - 
    \varphi 
    \left( \hat{p} \right)
  \right]
\nonumber \\
& & 
  \mbox{} + 
  \frac{1}{2}
  \int_0^{\hat{p} / 2}
  2d \Delta \hat{p} \, 
  {
    \left( \Delta \hat{p} \right)
  }^2
  \left[
    \varphi 
    \left( \Delta \hat{p} \right) - 
    \varphi 
    \left( \hat{p} - \Delta \hat{p} \right)
  \right] . 
\nonumber \\ 
\label{eq:cal_D_express}
\end{eqnarray}

All of the coefficients ~(\ref{eq:cal_R_express} -
\ref{eq:cal_P_express}) vanish manifestly as $\hat{p} \rightarrow 0$,
and at large $\hat{p}$, ${\cal R}, {\cal P} \rightarrow 0$, while
$\epsilon \beta \left( \hat{p} \right) \rightarrow 4 \int_0^{\infty} d \Delta
\hat{p} \, { \left( \Delta \hat{p} \right) }^2 \varphi \left( \Delta
\hat{p} \right) $, recovering the diffusion
constant~(\ref{eq:beta_simple_source}) of the simplified source term.
However, they are still not convenient for numerical integration,
being nonlocal in $\varphi$.  

The exponential form~(\ref{eq:phi_sigma_constr}) is therefore
exploited to approximate $\varphi$, in the region where its value is
largest, with the expansion
\begin{equation}
  \varphi
  \left( \hat{p} \pm \Delta \hat{p} \right) \approx
  \varphi
  \left( \hat{p} \right) 
  \varphi
  \left( \Delta \hat{p} \right) 
  e^{
    \pm \hat{p} \, \Delta \hat{p} 
    {
      \left. 
        \partial \psi / 
        \partial \hat{p}
      \right|
    }_0
  }
\label{eq:varphi_approx}
\end{equation}
In the range where the mean-field approximation is valid, $\varphi$ is
dominated by the constant term $\psi \left( 0 \right)$, and even the
factors $ e^{ \pm \hat{p} \, \Delta \hat{p} { \left.  \partial \psi /
\partial \hat{p} \right| }_0 }$ can be approximated as unity.  This
leaves the much-simplified expansions
\begin{eqnarray}
  {\cal R} \left( \hat{p} \right)
& = & 
  \left[
    1 - 
    \varphi 
    \left( \hat{p} \right)
  \right]
  {\cal I}_0 \left( \infty \right)
\nonumber \\
& & 
  \mbox{} - 
  2 
  \left[
    {\cal I}_0 \left( \hat{p} \right) - 
    2 \hat{p} 
    \varphi 
    \left( \hat{p} \right)
  \right]
\nonumber \\
& & 
  \mbox{} + 
  \left[
    1 - 
    \varphi 
    \left( \hat{p} \right)
  \right]
  {\cal I}_0 \left( \hat{p} / 2 \right) , 
\label{eq:cal_R_simpl}
\end{eqnarray}
\begin{eqnarray}
  {\cal P} \left( \hat{p} \right)
& = & 
  2 \left[
    1 - 
    \varphi 
    \left( \hat{p} \right)
  \right]
  {\cal I}_1 \left( \infty \right)
\nonumber \\
& & 
  \mbox{} - 
  \left[
    {\cal I}_1 \left( \hat{p} \right) - 
    { \hat{p} }^2
    \varphi 
    \left( \hat{p} \right)
  \right]
\nonumber \\
& & 
  \mbox{} - 
  \left[
    1 - 
    \varphi 
    \left( \hat{p} \right)
  \right]
  {\cal I}_1 \left( \hat{p} / 2 \right) , 
\label{eq:cal_P_simpl}
\end{eqnarray}
and 
\begin{eqnarray}
  \epsilon \beta \left( \hat{p} \right)
& = & 
  \left[
    1 - 
    \varphi 
    \left( \hat{p} \right)
  \right]
  {\cal I}_2 \left( \infty \right)
\nonumber \\
& & 
  \mbox{} +  
  \frac{1}{2}
  \left[
    {\cal I}_2 \left( \hat{p} \right) - 
    \frac{2}{3}
    { \hat{p} }^3
    \varphi 
    \left( \hat{p} \right)
  \right]
\nonumber \\
& & 
  \mbox{} + 
  \left[
    1 - 
    \varphi 
    \left( \hat{p} \right)
  \right]
  {\cal I}_2 \left( \hat{p} / 2 \right) . 
\label{eq:cal_D_simpl}
\end{eqnarray}
In Equations~(\ref{eq:cal_R_simpl} - \ref{eq:cal_D_simpl}), 
\begin{equation}
  {\cal I}_j \left( \hat{p} \right) \equiv 
  \int_0^{\hat{p}}
  2d \Delta \hat{p} \, 
  {
    \left( \Delta \hat{p} \right)
  }^j
  \varphi 
  \left( \Delta \hat{p} \right) , 
\label{eq:cal_I_0_def}
\end{equation}
for $j = 0, 1 , 2$.  These forms~(\ref{eq:cal_R_simpl} -
\ref{eq:cal_D_simpl}) are inserted in
Eq.~(\ref{eq:detailed_nondim_src_formal}) for ${\cal S}$ to produce
the mean-field results compared to simulations in
Fig.~\ref{fig:better_wrong_ind_eps_0p2} -
Fig.~\ref{fig:better_wrong_ind_eps_0p002}.

\begin{acknowledgments}

We would like to thank the McKinsey Corporation, Credit Suisse First
Boston, the McDonnel Foundation, Bob Maxfield, and Bill Miller for
supporting this research.  We would also like to thank Paolo Patelli,
R. Rajesh, Spyros Skouras, and Ilija Zovko for helpful discussions,
and Marcus Daniels for valuable technical support.

\end{acknowledgments}

\end{document}